\documentclass{article}

\usepackage{arxiv}

\usepackage[utf8]{inputenc} 
\usepackage[T1]{fontenc}    
\usepackage[colorlinks=true,linkcolor=blue,citecolor=blue]{hyperref}       
\usepackage{url}            
\usepackage{booktabs}       
\usepackage{amsfonts}       
\usepackage{nicefrac}       
\usepackage{microtype}      
\usepackage{lipsum}

\usepackage{amsmath,amssymb,amsthm}
\usepackage{graphicx}
\usepackage{grffile}
\usepackage[caption=true,font=footnotesize]{subfig}
\usepackage{xcolor}
\usepackage{enumitem}

\def\R{\mathbb{R}}              		
\def\C{\mathbb{C}}              		
\def\Rp{\mathbb{R}_{+}}         		
\def\1i{\msf{i}} 						
\def\eps{\varepsilon} 					
\def\real{\operatorname{Re}} 			
\def\imag{\operatorname{Im}}		    
 
\newcommand{\vc}[1]{{\boldsymbol #1}}   
\newcommand{\mc}[1]{\mathcal{#1}} 		
\newcommand{\mbb}[1]{\mathbb{#1}} 		
\newcommand{\mbf}[1]{\mathbf{#1}} 		
\newcommand{\msf}[1]{\mathsf{#1}} 		

\def\Exp{\mathrm{exp}} 										
\def\ind{\mbb{I}} 											

\newcommand{\nrm}[1]{\|#1\|} 								
\newcommand{\Lnrm}[1]{\left\|#1\right\|} 					
\newcommand{\set}[1]{\{#1\}} 								
\newcommand{\Lset}[1]{\left\{#1\right\}} 					

\def\ker{  \operatorname{\mbf{ker}}} 		
\def\supp{ \operatorname{\mbf{supp}}} 		

\def\diag{  \operatorname{\mbf{diag}}}		

\def\ev{    \operatorname{\mbf{E}}} 		
\def\var{   \mbf{V}} 						
\def\cov{   \operatorname{\mbf{Cov}}}       
\def\prob{  \mbf{P}} 						

\theoremstyle{plain}		\newtheorem{theorem}{Theorem}
\theoremstyle{plain}		\newtheorem{lemma}{Lemma}
\theoremstyle{plain}		
\theoremstyle{plain}		\newtheorem{proposition}{Proposition}
\theoremstyle{definition}	
\theoremstyle{remark}       
\def\whx{\widehat{x}}
\def\whxO{\widehat{x}_\Omega}
\def\vx{\vc{x}}
\def\whvx{\widehat{\vc{x}}}
\def\whvxO{\widehat{\vc{x}}_\Omega}
\def\whvxOdb{\widehat{\vc{x}}_\Omega^{\text{db}}}
\def\vz{\vc{z}}

\def\vy{\vc{y}}

\def\mF{\vc{F}}
\def\mFO{\vc{F}_\Omega}
\def\mPO{\vc{P}_\Omega}
\def\mSO{\vc{S}_\Omega}
\def\mW{\vc{W}}
\def\mPhO{\vc{\Phi}_\Omega}

\def\whaO{\widehat{\alpha}_\Omega}
\def\va{\vc{\alpha}}
\def\whva{\widehat{\vc{\alpha}}}
\def\whvaO{\widehat{\vc{\alpha}}_\Omega}
\def\whvaOdb{\widehat{\vc{\alpha}}_\Omega^{\text{db}}}
\def\Argmin{\operatorname{\mathrm{argmin}}}

\def\vr{\vc{r}}

\def\arg{\operatorname{\mbf{arg}}}

\def\atan2{\operatorname{atan2}}
\def\snr{\text{SNR}}

\definecolor{lh}{RGB}{255,0,0}
\definecolor{ds}{RGB}{0,255,0}
\definecolor{csl}{RGB}{0,0,255}

\def\vo{\vc{\omega}}
\def\vf{\vc{f}}
\def\vh{\vc{h}}
\def\ovz{\omega(\vz)}
\def\vI{\vc{I}}
\def\Sph{\C\mbb{S}}
\newcommand{\bset}[1]{[#1]}
\newcommand{\Lbset}[1]{\left[#1\right]}
\newcommand{\ol}[1]{\overline{#1}}
\newcommand{\wh}[1]{\widehat{#1}}

\def\whth{\wh{\theta}}
\def\whv{\wh{v}}
\def\venc{v_{\text{enc}}}
\def\vecp{\vec{p}}
\def\vecq{\vec{q}}

\def\vv{\vc{v}}
\def\vrho{\vc{\rho}}
\def\vtho{\vc{\theta}_0}
\def\tho{\theta_0}
\def\Tf{T_{\text{4DF}}}
\def\Tfinv{\Tf^{\dag}}

\allowdisplaybreaks

\title{An analysis of reconstruction noise from undersampled 4D flow MRI}

\author{
  Lauren Partin \\
  Department of Applied and Computational Mathematics and Statistics\\
  University of Notre Dame\\
  Notre Dame, IN \\
  \texttt{lhensley@nd.edu} \\
   \And
  Daniele E. Schiavazzi \\
  Department of Applied and Computational Mathematics and Statistics\\
    University of Notre Dame\\
    Notre Dame, IN \\
  \texttt{dschiavazzi@nd.edu} \\
   \And
  Carlos A. Sing Long\thanks{Corresponding author. C.~A.~Sing~Long is also with ANID – Millennium Science Initiative Program – Millennium Nucleus Center for the Discovery of Structures in Complex Data, and with ANID – Millennium Science Initiative Program – Millennium Nucleus Center for Cardiovascular Magnetic Resonance.}\\
  Institute for Mathematical and Computational Engineering and\\
  Institute for Biological and Medical Engineering\\
  Pontificia Universidad Cat\'olica de Chile\\
  Santiago, Chile \\
  \texttt{casinglo@uc.cl}
}

\begin{document}
\maketitle

\begin{abstract}
Novel Magnetic Resonance (MR) imaging modalities can quantify hemodynamics but require long acquisition times, precluding its widespread use for early diagnosis of cardiovascular disease. To reduce the acquisition times, reconstruction methods from undersampled measurements are routinely used, that leverage representations designed to increase image compressibility.

Reconstructed anatomical and hemodynamic images may present visual artifacts. Although some of these artifact are essentially reconstruction errors, and thus a consequence of undersampling, others may be due to measurement noise or the random choice of the sampled frequencies. Said otherwise, a reconstructed image becomes a random variable, and both its bias and its covariance can lead to visual artifacts; the latter leads to spatial correlations that may be misconstrued for visual information. Although the nature of the former has been studied in the literature, the latter has not received as much attention.

In this study, we investigate the theoretical properties of the random perturbations arising from the reconstruction process, and perform a number of numerical experiments on simulated and MR aortic flow. Our results show that the correlation length remains limited to two to three pixels when a Gaussian undersampling pattern is combined with recovery algorithms based on \(\ell_{1}\)-norm minimization. However, the correlation length may increase significantly for other undersampling patterns, higher undersampling factors (i.e., 8\(\times\) or 16\(\times\) compression), and different reconstruction methods.
\end{abstract}

\keywords{4D flow MRI \and Compressed Sensing \and MRI noise characterization}

\section{Introduction}
\label{sec:introduction}

4D flow is a Magnetic Resonance (MR) imaging technique used to acquire a three-dimensional velocity field representing moving particles within a region of interest~\cite{Markl2012,Stankovic2014}. This enables the measurement of cardiovascular flow and thus the detailed study of the cardiovascular system and its diseases. Unfortunately, the amount of measurements required to form the velocity field is substantial.  Typically four complex images~\cite{Pelc1991,Markl2003,Markl2012} are needed at each instant to resolve the velocity field throughout the cardiac cycle. This increases data acquisition times and renders the technique sensitive to patient movement. To accelerate this process, parallel imaging~\cite{Hutchinson1988,Larkman2007,Deshmane2012} is routinely used. In parallel imaging, multiple coils with known sensitivities are used to increase the effective measurements from separate but overlapping spatial volumes. This allows a significant reduction in acquisition times. In addition to parallel imaging, reconstruction procedures from undersampled data have been used extensively in MR imaging in the past decade. These methods aim to reconstruct an image from an incomplete set of \(k\)-space measurements by leveraging the fact that images are compressible. In other words, images have a sparse representation under suitable bases, e.g. wavelets~\cite{Daubechies1992,Ong2015}, curvelets~\cite{Candes2006a,Yazdanpanah2017}, or shearlets~\cite{Kutyniok2012,Ma2018} among others. As a consequence, the information they contain is proportional to the number of non-zero coefficients in this representation, and recovery should be possible if the number of measurements is proportional to this number. This argument is rigorously articulated by the theory of Compressed Sensing~\cite{Candes2006,Candes2006b,Donoho2006} which has been successfully applied for MR image reconstruction from highly undersampled data, far below the Nyquist rate, without compromising their overall quality~\cite{Lustig2007}. Compressed Sensing relies on the fact that the data acquisition process is linear and that the reconstruction involves solving a convex optimization problem.

Reconstructed images may present visual artifacts. In contrast to those produced by linear reconstruction methods, which are mainly smoothing and ringing artifacts, those produced by non-linear reconstruction methods are less understood. Among other factors, they may depend on the sparse representation, the undersampling ratio, and the reconstruction algorithm. To our knowledge, there are no theoretical studies about the nature of these visual artifacts. However, there are empirical studies in the clinical literature that assess the impact of visual artifacts produced by Compressed Sensing methods on clinical diagnosis. These assessments typically involve a blind test where experts rate the image quality of a fully-sampled image compared to a reconstructed one~\cite{Sharma2013,Zhang2014,Delattre2020}. Some of these studies point to the choice of undersampling pattern~\cite{Jaspan2015,Ozturk-Isik2017}, of undersampling rate~\cite{Sharma2013,Ozturk-Isik2017,Delattre2020}, or of the regularization parameter~\cite{Jaspan2015,Ozturk-Isik2017} as some of the parameters that have an impact on image quality. In addition, the optimal undersampling rate~\cite{Sharma2013,Delattre2020} and the artifacts themselves~\cite{Sartoretti2018} may depend on the organ being imaged or the imaging modality. In the case of 4D flow, the choice of velocity encoding may also play a role~\cite{Tariq2013}. Although there is a consensus that the clinical information in the reconstructed images is no different from that in a fully-sampled image when the parameters are properly selected~\cite{Sharma2013,Tariq2013,Jaspan2015,Delattre2020}, sometimes even improving image quality~\cite{Zhang2014}, these studies suggest that there are artifacts unique to non-linear reconstruction techniques. This has a direct impact on 4D flow. Since it relies on the acquisition of complex images to resolve the flow velocity, any image artifact is a potential source of a flow artifact. Note it may be significantly harder for a practitioner to determine if the observed flow structure contains artifacts or not when the flow itself is complex~(see, e.g.~\cite{Petersson2010}). 

Another potential source of visual artifacts is {\it randomness}. For instance, measurement noise in MR is typically modeled as additive Gaussian noise~\cite[Ch.~7]{Nishimura2010} and its effects are well-understood for fully-sampled, single coil acquisitions; in particular, the distribution of the complex image is complex Gaussian, whereas the magnitude image is either Rayleigh or Riccian. In 4D flow, the distribution of the velocity~\cite{Firoozabadi2018} can be also characterized explicitly for this type of acquisitions. Critically, the image or velocity values at distinct pixels are statistically uncorrelated and the noisy images are unbiased estimates of the true images. The absence of correlations between image pixels ensures there is no coherent structure due to correlations that could be misconstrued as visual information. 

This may no longer be the case when a reconstruction method is applied. In fact, any reconstruction method interpolates the missing samples in \(k\)-space and every reconstructed sample becomes a function of the measurements. As a consequence, the noise on the measured samples may introduce correlations on \(k\)-space and hence on the pixel values of the reconstructed image. This effect is independent on whether the reconstruction method is linear or non-linear, and it should become apparent for imaging modalities with low signal-to-noise ratio (SNR). Although it is possible to characterize the correlations produced by linear methods using standard arguments, this is no longer the case for non-linear methods. For instance, there is a large literature on robustness guarantees for Compressed Sensing~\cite{Candes2006c,Candes2006d,Candes2011}. In contrast, there are no results about the structure of the correlations in imaging applications. Note that in addition to measurement noise, random sampling is another possible cause of correlations. Random sampling ensures the information contained in the image is preserved, and it is crucial for both exact recovery and robustness guarantees~\cite{Candes2005,Candes2006,Candes2006b}. In principle, different undersampling patterns could be used to image the same object several times. Even if the quality of the recovered image is excellent in all cases, minor differences between them could appear as a consequence of the choice of undersampling patterns. More broadly, the distribution of the sampled frequencies in \(k\)-space may have an impact on the artifacts observed between multiple reconstructions. This may become noticeable at higher undersampling, where the variability of the sampling set and the sparse representation selected may generate unique types of artifacts. To our knowledge, the study of these effects has not attracted much attention in the literature.

In contrast to the image artifacts commonly studied in the literature, those introduced by randomness are statistical in nature, and they may not be noticeable in a single imaging experiment. However, they might explain artifacts appearing across a large number of reconstructed images. Although this may not directly reflect the experience of an expert assessing a single image, we believe its characterization, in particular for 4D flow, is important for the following reasons:

\begin{enumerate}[leftmargin=*,label=\roman*.]
    \item{{\bf Uncertainty propagation:} Some biomarkers for cardiovascular diseases proposed in the literature, e.g. vorticity~\cite{Schafer2018,Contijoch2020} and wall shear stress~\cite{VanOoij2015,Rodriguez-Palomares2018}, are computed from flow velocity measurements. The validation of these biomarkers typically involves the acquisition of flow velocity data from patients and/or volunteers; the biomarkers are then computed and suitable statistical analyses, including training classifiers, are performed afterwards. However, if the flow velocity data is computed from images reconstructed with varying SNR and/or varying undersampling rates, then the correlations {\it within} images and {\it across} images may create a confounding effect. Furthermore, the functional relation between the value of the biomarker and the flow data may increase or decrease the strength of the correlations within the image. For instance, computing the wall shear stress involves taking a derivative of the flow velocity, which decreases the regularity, whereas computing the pressure involves solving a Poisson equation with the divergence of the velocity field as a source, which has a regularizing effect. Consequently, characterizing the artifacts introduced by random effects is crucial to assess potential confounding effects in such studies.
    }    

    \item{{\bf Reproducibility:} Clinical validation studies typically involve the assessment of a fully-sampled image and a reconstructed image by an expert. However, any artifact due to random effects may not be observed on such a validation experiment and the assessment may change, even if slightly, with a different reconstruction. The artifacts introduced by correlations {\em within} images and {\em across} images can be assessed only by evaluating several reconstructed images of the same anatomy and volunteer. This would determine if the artifacts reported on a single reconstruction are consistent across several reconstructions; in other words, whether they appear on the {\it expected} reconstructed image or if the artifacts are only fluctuations around this expectation. We believe such an experiment would better reflect the experience of an aggregate of several experts assessing fully-sampled images versus reconstructed images.
    }
\end{enumerate}

A rigorous characterization of the artifacts introduced by randomness and reconstruction methods in 4D flow in the most general setting is challenging. For this reason, our goal is to analyze a simplified model where the structure of the correlations can be studied. In particular, we neglect any artifacts that may arise due to acquisition protocols, patient movement, field inhomogeneities, multiple coils, and others, and we focus on artifacts due only to reconstruction from undersampled data. Furthermore, we focus on the setting in which each one of the complex images required to resolve the flow are reconstructed independently. We evaluate \(\ell_2\)-norm reconstruction methods, \(\ell_1\)-norm reconstruction methods, and iterative methods, and in each case, the sampling set is selected at random according to different probability distributions, including Bernoulli and Gaussian. 

In addition, we consider two regimes for the measurement noise. The first regime is the {\em noiseless} (or {\em infinite SNR}) limit. In this regime we can provide a characterization of the spatial correlations under suitable regularity conditions of the reconstruction method. Under these conditions, the structure of the correlations is similar to that obtained for a linear reconstruction method. The second regime, which we call {\em finite SNR}, considers increasing values for the noise variance. In this case, we provide numerical evidence that correlations do appear in all cases. Interestingly, the structure of these correlations is also similar to the one we find in the noiseless limit. In particular, the correlation length of the noise generated by non-linear reconstruction algorithms remains limited to two-three pixels and is significantly affected by the low-frequency content of the selected undersampling mask and by the chosen algorithm.

The paper is structured as follows. 
In Section~\ref{sec:mathematicalModel} we introduce the basic mathematical model for phase-contrast 4D flow, reconstruction methods from undersampled data.
In Section~\ref{sec:noiselessLimit} we study the noiseless limit and we provide expressions for the covariance of the reconstructed density and velocity in this limit.
In Section~\ref{sec:reconstructionMethods} we study the reconstruction methods used in our numerical tests, focusing on the image artifacts they may introduce.
In Section~\ref{sec:recoveryOrthogonalProjection} we analyze {\em least \(\ell_2\)-norm recovery} as its linear dependence on the measurements allows us to provide a detailed analysis, even though this approach is seldom used in practice.
In Section~\ref{sec:recoveryCompressedSensing} we review the theory of Compressed Sensing, discuss the shrinkage effect generated by approaches based on $\ell_{1}$-norm minimization, and mention alternative approaches producing unbiased estimates.
The results of three numerical experiments are presented in Section~\ref{sec:results}, followed by a discussion in Section~\ref{sec:discussion}.

\section{Preliminaries}\label{sec:preliminaries}

We use the following notation throughout this work. Lowercase boldface denotes vectors, e.g., \(\vx\), and uppercase boldface denotes matrices, e.g. \(\vc{A}\). Plainface denotes either scalars or scalar-valued functions, e.g. \(t, \rho, x\), and plainface with an overhead arrow denotes a function taking values in \(\R^3\), e.g. \(\vec{v}\). The methods we develop are applicable in 2D or 3D, and the location of a pixel or voxel is typically denoted by \(\vec{r}\), e.g. \(x(\vec{r})\) denotes the value of the function \(x\) at \(\vec{r}\). An image may represent a 2D section or a volume, and we use pixel to indicate a position in the image regardless of the dimension. The number of pixels along each direction is denoted by a subscript, e.g. \(n_x,n_y, n_z\), and the total number of pixels as \(n\).

\section{Mathematical model}
\label{sec:mathematicalModel}

In this section we present the mathematical model for 4D flow that we use in this work. 
We introduce the basic ideas behind the forward and inverse problems in phase-contrast 4D flow in Section~\ref{sec:mathematicalModel:4dFlow} with a focus on the relation between the density and velocity, and the complex images. 
In Section~\ref{sec:mathematicalModel:noise}, we present the standard noise model in the fully-sampled case, emphasizing the absence of correlations.
Section~\ref{sec:mathematicalModel:phaseRobustness} focuses on the different mechanisms by which noise affects the complex images or the density and velocity.
Finally, in Section~\ref{sec:mathematicalModel:undersampling} we discuss the model for random undersampling and in Section~\ref{sec:mathematicalModel:noiseUndersampling} we adapt the noise model to the undersampled case in a way that the sampling set and the noise can be regarded as statistically independent.

\subsection{Phase-Contrast 4D Flow}
\label{sec:mathematicalModel:4dFlow}

4D flow quantifies the density \(\rho\) and the velocity field \(\vec{v}\) from a set of complex images typically acquired using phase contrast (PC) cardiac Magnetic Resonance (MR)~\cite{Moran1982,Bryant1984,Pelc1991}. In PC Cardiac MR four complex images are acquired to resolve the velocity at each sampled time. The relation between the density and the velocity, and the four acquired complex images at a given time is
\begin{align}
\label{eq:4dFlow}
\begin{split}
    x_0(\vec{r})     &= \rho(\vec{r}) e^{ i \theta_0(\vec{r})} \\
    x_k(\vec{r})     &= \rho(\vec{r}) e^{ i \theta_0(\vec{r})} \Exp\left(\pi i\frac{v_k(\vec{r})}{v_{\text{enc}}}\right)\qquad k \in \set{1,2,3}
\end{split}
\end{align}
where \(v_k\) represents the \(k\)-th component of the velocity, \(\theta_0\) is a nuisance parameter that accounts for the field inhomogeneity and other acquisition artifacts, and \(v_{\text{enc}}\) is the {\em velocity encoding}. This parameter is selected so that
\[
    \max_{\vec{r}}\,\frac{|v_k(\vec{r})|}{v_{\text{enc}}}< 1\qquad k\in\set{1,2,3}
\]
to avoid phase-wrapping artifacts. If \(\vtho, \vrho, \vv_1, \vv_2\) and \(\vv_3\) represent the vectorized nuisance parameter, the magnitude image, and the velocity images, the expressions in~\eqref{eq:4dFlow} define implicitly a non-linear map
\[
    \Tf : (\vtho, \vrho, \vv_1, \vv_2, \vv_3) \mapsto (\vx_0,\vx_1,\vx_2,\vx_3)
\]
that relates the four complex images to the density and velocities. When the four complex images are known, the density and velocity components can be recovered from the expressions
\begin{align}
\label{eq:4dFlow:recovery}
\begin{split}
    \rho(\vec{r}) &= |x_0(\vec{r})| \\
    v_k(\vec{r}) &= \frac{v_{\text{enc}}}{\pi}(\arg(x_k(\vec{r})) - \arg(x_0(\vec{r}))) \qquad k\in\set{1,2,3}
\end{split}
\end{align}
where \(\arg\) denotes the complex argument taking values on \([-\pi, \pi)\); we assume the convention \(\arg(0) = 0\). The expressions in~\eqref{eq:4dFlow:recovery} along with \(\tho(\vec{r}) = \arg(x_0(\vec{r}))\) define a map
\[
    \Tfinv :  (\vx_0,\vx_1,\vx_2,\vx_3) \mapsto(\vtho, \vrho, \vv_1, \vv_2, \vv_3).
\]
Although \(\Tf\) is not invertible, e.g. there is no way to recover \(\theta_0(\vec{r})\) nor \(v_k(\vec{r})\) when \(\rho(\vec{r}) = 0\), the map \(\Tfinv\) defines a right-inverse for \(\Tf\) in the sense that
\[
    \Tf\circ \Tfinv(\vx_0,\vx_1,\vx_2,\vx_3) = (\vx_0,\vx_1,\vx_2,\vx_3).
\]
Although there are other ways in which a left-inverse for \(\Tf\) can be defined, e.g. by taking the average magnitude at each \(\vec{r}\), throughout this work we use~\eqref{eq:4dFlow:recovery} for its simplicity.

In the noiseless case, the MR acquisition process measures the Fourier coefficients \(\vy_k\) of each complex image \(\vx_k\). Therefore, the acquisition process is modeled as
\begin{equation}
\label{eq:4dFlow:fullySampledNoiseless}
\vy_k = \mF\vx_k\qquad k\in\set{0,\ldots, 3},
\end{equation}
where \(\vc{F}\) denotes the Fourier transform. The complex images can be then computed by applying the inverse Fourier transform, and the density and velocity can be recovered by applying \(\Tfinv\).

\subsection{Noise in the fully-sampled case}
\label{sec:mathematicalModel:noise}

The measurements \(\vy_k\) in~\eqref{eq:4dFlow:fullySampledNoiseless} are typically corrupted by additive Gaussian noise~\cite[Ch.~7]{Nishimura2010}. The acquisition model is
\begin{equation}
\label{eq:4dFlow:fullySampledNoisy}
    \vy_k = \mF \vx_k + \sigma\vz_k\quad\mbox{where}\quad \vz_0,\ldots,\vz_3\stackrel{\text{iid}}{\sim} N(0,\vc{I}_n) + i N(0,\vc{I}_n).
\end{equation}
Since the Fourier transform is unitary, the images are reconstructed as
\begin{equation}
\label{eq:defFullySampledEstimate}
    \whvx_k = \mF^* \vy_k = \vx_k + \sigma\vz'_k\quad\mbox{where}\quad \vz_k' \stackrel{d}{=} \vz_k.
\end{equation}
These are unbiased estimates of the true images. Furthermore, they are normally distributed, and their {\em complex covariance} is
\[
    \var_z\bset{\whvx_k} := \ev_z\bset{(\whvx - \ev(\whvx))(\whvx - \ev(\whvx))^*} = 2\sigma^2 \vc{I}_n
\]
where \(^*\) denotes the conjugate transpose. 

The random variables \(\whx_k(\vec{r}_1)\) and \(\whx_k(\vec{r}_2)\) are uncorrelated when \(\vec{r}_1\neq \vec{r}_2\) and, in particular, they are independent. 
In fact, \(\whvx_k\) is a perturbation of \(\vx_k\) by additive Gaussian noise and there are no spatial artifacts introduced due to measurement noise. 

The density and the velocity can be reconstructed from \(\whvx_0,\ldots,\whvx_3\). Since
\[
    \widehat{\theta}_k(\vec{r}) = \arg(\whx_k(\vec{r})) = \arg(x_k(\vec{r}) + \sigma z'_k(\vec{r}))
\]
the probability distribution of \(\widehat{\theta}_k(\vec{r})\) is not Gaussian. However, \(\widehat{\theta}_k\) is unbiased~\cite{Firoozabadi2018}. Furthermore, \(\widehat{\theta}_k(\vec{r}_1)\) and \(\widehat{\theta}_k(\vec{r}_2)\) are independent when \(\vec{r}_1\neq \vec{r}_2\), and the values of \(\widehat{\theta}_k\) are independent of those of \(\widehat{\theta}_\ell\) everywhere for \(k\neq \ell\). Therefore, the estimates
\[
    \widehat{v}_k(\vec{r}) = \frac{v_{\text{enc}}}{\pi}(\widehat{\theta}_k(\vec{r}) - \widehat{\theta}_0(\vec{r}))\qquad k\in\set{1,2,3},
\]
are unbiased, and \(\widehat{v}_k(\vec{r}_1)\) and \(\widehat{v}_\ell(\vec{r}_2)\) are also independent when \(\vec{r}_1 \neq \vec{r}_2\). As a consequence, we can interpret \(\widehat{v}_k\) as the corruption of \(v_k\) by uncorrelated noise, and any coherent spatial structure must arise from the heteroscedasticity of the noise at each pixel, but not from the correlation of the noise at different pixels. The same argument applies for \(\widehat{\rho} := |\whx_0|\) with minimal changes.

\subsubsection{Effects of noise on the complex phase}
\label{sec:mathematicalModel:phaseRobustness}

To recover the density and the velocity we need to compose the non-linear map \(\Tfinv\) with a linear function of the measurements as in~\eqref{eq:4dFlow:fullySampledNoisy}. As a consequence, bounds on the image reconstruction error do not imply similar bounds for the density of the velocity. To illustrate this, consider for some \(\eps > 0\) a bound of the form
\begin{equation}
\label{eq:4dFlow:reconstructionpixelAbsoluteErrorBound}
    |\whx_k(\vec{r}) - x_k(\vec{r})| \leq \eps
\end{equation}
which may hold with high-probability. Although the bound on the reconstruction error for the density is the same, this is not the case for the complex argument. For instance, when~\(x_0(\vec{r}) = 0\) the bound~\eqref{eq:4dFlow:reconstructionpixelAbsoluteErrorBound} implies
\[
    |\whx_0(\vec{r}) - x_0(\vec{r})| = |\whx_0(\vec{r})| \leq \eps,
\]
from where no constraint on the estimated complex argument can be deduced. This typically leads to artifacts in the velocity on regions where the density is close to zero. More generally, by squaring~\eqref{eq:4dFlow:reconstructionpixelAbsoluteErrorBound} and rearranging terms, we deduce
\[
    \frac{1}{2}\left(\frac{|\whx_k(\vec{r})|}{|x_k(\vec{r})|} + \frac{|x_k(\vec{r})|}{|\whx_k(\vec{r})|}\right) - \frac{1}{2} \frac{\eps^2}{|\whx_k(\vec{r})||x_k(\vec{r})|} \leq \cos(\arg(\whx_k(\vec{r})) - \arg(x_k(\vec{r}))).
\]
The right-hand side can be controlled only with a relative error bound
\begin{equation}
\label{eq:4dFlow:reconstructionpixelRelativeErrorBound}
    |\whx_k(\vec{r}) - x_k(\vec{r})| \leq \eps |x_k(\vec{r})|.
\end{equation}
In this case, 
\[
    \frac{1}{2}\left(\frac{|\whx_k(\vec{r})|}{|x_k(\vec{r})|} + (1-\eps^2)\frac{|x_k(\vec{r})|}{|\whx_k(\vec{r})|}\right) \leq \cos(\arg(\whx_k(\vec{r})) - \arg(x_k(\vec{r})))
\]
and, since~\eqref{eq:4dFlow:reconstructionpixelRelativeErrorBound} also implies
\[
    \left|\frac{|\whx_k(\vec{r})|}{|x_k(\vec{r})|} - 1\right| \leq \eps,
\]
we obtain
\[
    1-\eps \leq \cos(\arg(\whx_k(\vec{r})) - \arg(x_k(\vec{r}))).
\]
Therefore, even though the reconstruction error for the complex images may be small, the error in the velocities may be quite large. In particular, even when the noise variance \(\sigma\) is small, the artifacts in the velocities may be amplified in regions where the density is small. When the complex images are reconstructed from undersampled data, the reconstruction artifacts may also be amplified by this same effect. 

\subsection{Undersampling}
\label{sec:mathematicalModel:undersampling}

To improve data acquisition rates in MR, reconstruction methods from undersampled measurements are routinely applied. Their goal is to produce an accurate reconstruction of each one of the underlying images \(\vx_k\) from a subset \(\Omega_k\) of its Fourier coefficients.

In what follows \(m_k\) denotes the size of the sampling set \(\Omega_k\) and \(\mF_{\Omega_k}\) represents the undersampled (or partial) Fourier transform. When undersampling, the measurement model~\eqref{eq:4dFlow:fullySampledNoiseless} becomes
\[
    \vy_k = \mF_{\Omega_k}\vx_k\qquad k\in\set{0,\ldots, 3}.
\]
In 4D flow, a reconstruction method computes \(\wh{\vrho}, \wh{\vv}_1,\ldots, \wh{\vv}_3\) from the undersampled measurements \(\vy_0,\ldots, \vy_3\). This can be done in several ways. For instance, the complex images can be reconstructed first~\cite{Kwak2013,Sun2017} to then apply the map \(\Tfinv\) or equivalent, or the density and velocity may be directly reconstructed from undersampled measurements~\cite{FengZhao2012,Santelli2016}. 

Regardless of the approach, a common feature of most reconstruction methods is that the sampling set must be chosen (pseudo-)randomly~\cite{Candes2006,Lustig2007}. A typical approach is to draw \(\Omega_k\) independently from a Bernoulli process as follows. Let \(\vec{\xi}\) denote the frequency variable, let \(\mu\) be a function such that \(0 \leq \mu \leq 1\), and let \(\set{B(\vec{\xi})}\) be a collection of independent Bernoulli random variables such that \(B(\vec{\xi}) = 1\) with probability \(\mu(\vec{\xi})\). Then,
\[
    \Omega_k = \set{\vec{\xi}:\, B(\vec{\xi}) = 1}\quad\Rightarrow\quad \ev\bset{|\Omega_k|} = \sum\nolimits_{\vec{\xi}} \mu(\vec{\xi}).
\]
The parameter \(\mu\) determines the rate at which each frequency is sampled and its choice has a direct impact on the performance of the reconstruction method. For instance, sampling low-frequencies more densely leads to better results empirically, reducing visual artifacts. However, as we have discussed, some of these artifacts, although small, may result in noticeable artifacts in the reconstructed velocities. 

\subsection{Noise and undersampling}
\label{sec:mathematicalModel:noiseUndersampling}

Suppose the sampling set \(\Omega_k\) is fixed. In this case, the noise model~\eqref{eq:4dFlow:fullySampledNoisy} becomes
\begin{equation}
\label{eq:4dFlow:undersampledNoisyBasic}
    \vy_k = \vc{F}_{\Omega_k}\vx_k + \sigma \vz_k\quad\mbox{where}\quad \vz_0,\ldots,\vz_3\stackrel{\text{iid}}{\sim} N(0,\vc{I}_{m_k}) + i N(0,\vc{I}_{m_k})
\end{equation}
When the noise model is formulated this way, the noise cannot be independent of \(\Omega_k\) as the length of \(\vz_k\) is precisely \(m_k\). To avoid some technical difficulties that arise due to this dependence, we formulate this model instead as
\begin{equation}
\label{eq:4dFlow:undersampledNoisy}
    \vy_k = \vc{S}_{\Omega_k}(\mF\vx_k + \sigma \vz_k) \quad\mbox{where}\quad \vz_0,\ldots,\vz_3\stackrel{\text{iid}}{\sim} N(0,\vc{I}_n) + i N(0,\vc{I}_{n})
\end{equation}
where \(\vc{S}_{\Omega_k} \in \R^{m_k\times n}\) is a {\it sampling matrix} for which \(\mF_{\Omega_k} = \vc{S}_{\Omega_k}\vc{F}\). By considering~\eqref{eq:4dFlow:undersampledNoisy} we can assume \(\Omega_k\) and \(\vz_k\) are independent.

From~\eqref{eq:4dFlow:undersampledNoisy} we see that a reconstruction method from undersampled data is subject to the random fluctuations of both the measurement noise and the sampling set. Our goal is to understand the role of each of them in the spatial correlations in the density and the velocity in 4D flow.

The deviations of \(\vy_k\) from the noiseless measurements can be characterized directly in terms of the \(\ell_2\)-norm using classical arguments. Since
\[
    \nrm{\vy_k - \vc{F}_{\Omega_k}\vx_k}_2^2 \sim \sigma^2 \chi^2_{2m_k},
\]
where \(\chi^2_{2m_k}\) denotes a chi-squared variable with \(2m_k\) degrees of freedom, we have for \(\delta> 0\) that
\[
    \prob\set{\nrm{\vy_k - \vc{F}_{\Omega_k}\vx_k}_2 > t} \leq \delta\quad\Rightarrow\quad t \geq \sigma F_{\chi^2_{2m_k}}^{-1}(1 - \delta)^{1/2}.
\]
where \(F_{\chi^2_{2m}}\) denotes the cumulative density function (CDF). The approximation~\cite{Johnson1994}
\[
    \prob\Lset{\sigma^2 \chi^2_{2m_k} \leq t^2} \approx \prob\Lset{G \leq \sqrt{2}\frac{t}{\sigma} - \sqrt{4m_k - 1}}\qquad (G\sim N(0, 1))
\]
yields
\begin{equation}
\label{eq:defEtaThreshold}
    \frac{t(m_k,\delta)}{\sigma} := \sqrt{2m_k - \frac{1}{2}} + \frac{1}{\sqrt{2}} F_G^{-1}(1-\delta)
\end{equation}
Therefore, with probability \(\approx 1-\delta\) the acquired measurements is at most at a distance \(t(m_k,\delta)\) from the noiseless measurements.

In engineering applications the standard deviation of the noise is typically deduced from the signal-to-noise ratio (SNR). Here we consider the {\em signal power} to define the variance in terms of the SNR. In other words
\[
    \snr_x := \frac{\nrm{\vx}_2^2}{2 n \sigma^2}\quad\Rightarrow\quad \sigma = \frac{1}{\snr_x^{1/2}}\frac{\nrm{\vx}_2}{\sqrt{2 n}}
\]
where \(n\) is the total number of pixels in the image. Once the noise variance has been determined, we can compute the SNR relative to the {\em measured signal power} defined by
\[
    \snr_{y_k} = \frac{\nrm{\vc{F}_{\Omega_k} \vx}_2^2}{2m_k \sigma^2} = \frac{n}{m_k} \frac{\nrm{\mF_{\Omega_k}\vc{x}}_2^2}{\nrm{\vx}_2^2} \snr_x.
\]

\section{The covariance in the noiseless limit}
\label{sec:noiselessLimit}

The effect of random sampling and measurement noise depends on the specifics of the reconstruction method and, in particular, on whether the density and the velocity are reconstructed directly from undersampled data or the complex images are reconstructed first. To simplify our analysis, we consider the latter. Hence, the reconstruction proceeds by first reconstructing each one of the complex images from its undersampled measurements; the density and velocity are then reconstructed by applying \(\Tfinv\). 

In this section we analyze the covariance of the reconstructions in the noiseless limit, i.e., in the limit \(\sigma^2\to 0\). In Section~\ref{sec:noiselessLimit:complexImages} we provide results for the leading term of the expectation and the covariance of the reconstructed images, both with respect to the sampling set and measurement noise, in this limit. In Section~\ref{sec:noiselessLimit:densityVelocity} we show how this covariance manifests in the reconstructed density and velocity. 
Finally, in Section~\ref{sec:noiselessLimit:extensions} we discuss how our results can be extended to other reconstruction methods, e.g. to those that exploit redundancies between the images \(\vx_0,\ldots, \vx_3\) or that instead reconstruct the densities and velocities directly.

\subsection{The covariance in reconstructed images}
\label{sec:noiselessLimit:complexImages}

To simplify the notation, we drop the subscript \(k\) associated to each complex image. A reconstruction method for \(\vx\in \C^n\) from measurements \(\vy\in \C^m\) depends on the sampling set \(\Omega\). Hence, we interpret \(\Omega\) as a choice of reconstruction method \(\whxO:\C^m\to \C^n\) from a family indexed by all possible sampling sets. The reconstructed image becomes
\[
    \whx = \whx(\Omega,\vz) := \whxO(\vc{F}_{\Omega} \vc{x} + \sigma\vc{S}_{\Omega} \vc{z}).
\]
Even when the components of \(\whx\) are not statistically independent, the covariance matrix could still be diagonal and the fluctuations {\it uncorrelated}; this is the case, for instance, of a random vector uniformly distributed on the sphere. To understand if this is the case, we determine first the structure of the covariance of \(\whx\) as \(\sigma^2 \to 0\) when \(\Omega\) is fixed. This captures the influence of the measurement noise on the reconstructed image conditional on \(\Omega\). These fluctuations capture the artifacts that arise between two experiments when {\em the same} object is imaged repeatedly by measuring always the same frequencies. The following result assumes differentiability of \(\whxO\) at the noiseless measurements \(\mFO\vx\). It is a particular instance of Theorems~\ref{apx:thm:noiselessLimit:expectation} and~\ref{apx:thm:noiselessLimit:covariance} which provide similar results under weaker regularity assumptions on \(\whxO\). The proof is deferred to Appendix~\ref{apx:noiselessLimit}.

\begin{theorem}
\label{thm:noiselessLimit:momentsNoise}
    Fix \(\Omega\) and \(\vx\in \C^n\). Let \(\vy := \mFO\vx\) and suppose \(\whx_\Omega\) is differentiable at \(\vy\) with
    \[
        \lim_{t\to 0^+} t^{-2}\ev\bset{\nrm{\whxO(\vy+ t\vo) - \whxO(\vy) - t D\whxO(\vy)\vo}_2^2} = 0
    \]
    for \(\vo \sim \textsc{Uniform}(\Sph^m)\) where \(\Sph^m\) is the unit sphere in \(\C^m\), and suppose that there exist \(C, r, \alpha > 0\) such that  \(\nrm{\vz}_2 > r\) implies \(\nrm{\whxO(\vz)}_2 \leq C \nrm{\vz}_2^{\alpha}\). Then
    \begin{align*}
        \lim_{\sigma\to 0} \ev_z\bset{\whxO(\vc{y} + \sigma\vc{z})} &= \whxO(\vy),\\
        \lim_{\sigma\to 0} \sigma^{-2} \var_z\bset{\whxO(\vc{y} + \sigma\vc{z})} &= 2 D\whxO(\vy)D\whxO(\vy)^*,
    \end{align*}
    for \(\vc{z} \sim N(0,\vc{I}_m) + iN(0,\vc{I}_m)\). 
\end{theorem}

For repeated reconstructions with a fixed sampling set, the covariance behaves as
\begin{equation}
\label{eq:limitCovariance:noiseOnly}
    \var_{\vz}\bset{\whxO(\vc{F}_{\Omega} \vc{x} + \sigma\vc{S}_{\Omega} \vc{z})} = 2\sigma^2 D\whxO(\mFO\vx)D\whxO(\mFO\vx)^* + o(\sigma^2).
\end{equation}
for \(\sigma^2 \ll 1\). Under the hypotheses of the theorem, the leading term will be zero if \(\whxO\) is constant near the noiseless measurements \(\mFO\vx\). More generally, for piecewise constant reconstruction methods the leading term will be zero as long as the noiseless measurements are away from the discontinuities. Otherwise, it will be non-zero if \(\whxO\) is {\em non-constant} near the noiseless measurements \(\mFO\vx\); this is typically the case with the reconstruction methods we review in Section~\ref{sec:reconstructionMethods}.

Furthermore, in this limit the moments of the reconstruction method behave as those of the affine method
\begin{equation}
\label{eq:noiselessLimit:affineOracle}
    \whxO^{A}(\vy) = \whxO(\mFO\vx) + D\whxO(\mFO\vx) (\vy - \mFO\vx).
\end{equation}
Consequently, the covariance of a piecewise linear reconstruction method will behave similarly as that of a linear or affine method when \(\sigma^2\) is small. 

As a consequence of this result, not only are the values \(\whxO(\vec{r}_1)\) and \(\whxO(\vec{r}_2)\) correlated, but their values are constrained to lie close to a subspace; this is due to the covariance matrix becoming singular in the limit. Since MR imaging modalities, among other engineering applications, are well-approximated by the high SNR regime, the leading term may dominate the structure of the covariance of the reconstructed images. 

As discussed before, another source of variability in the reconstruction is the random choice of \(\Omega\). The covariance due to both \(\Omega\) and \(\vz\) reflects the artifacts that arise between experiments where {\em the same} object is imaged repeatedly. In this case, it is of interest to determine what is the effect of the statistical independence between \(\Omega\) and \(\vz\) in the reconstructed images. Typically, one would expect that the correlations are {\em smoothed} due to the randomness of \(\Omega\) and that the covariance matrix may no longer be singular in the limit. The following theorem is a consequence of Theorem~\ref{apx:thm:noiselessLimit:overSamplingSet}. Its proof is deferred to Appendix~\ref{apx:noiselessLimit} where we provide a similar result under weaker assumptions.

\begin{theorem}
\label{thm:noiselessLimit:momentsSamplingNoise}
    Let \(\vx\in \C^n\). Suppose for every \(\Omega\) with positive probability the hypotheses of Theorem~\ref{thm:noiselessLimit:momentsNoise} hold for \(\whxO\) near \(\mFO\vx\). Then 
    \begin{align*}
        \lim_{\sigma\to 0} \ev_{\Omega,\vz}\bset{\whxO(\mFO\vx + \sigma\mSO\vz)} &= \ev_{\Omega}\bset{\whxO(\mFO\vx)},\\
        \lim_{\sigma\to 0} \var_{\Omega,\vz}\bset{\whxO(\mFO\vx + \sigma\mSO\vz)} &= \var_{\Omega}\bset{\whxO(\mFO\vx)}.
    \end{align*}
\end{theorem}

Therefore, sampling randomly does not necessarily reduce the correlations. In particular, from
\[
    \lim_{\sigma\to 0} \var_{\Omega,\vz}\bset{\ev_{\vz}\bset{\whxO(\mFO\vx + \sigma\mSO\vz)} - \ev_{\Omega}\bset{\ev_{\vz}\bset{\whxO(\mFO\vx + \sigma\mSO\vz)}}} = \var_{\Omega}\bset{\whxO(\mFO\vx)}
\]
we see the covariance behaves as
\begin{equation}
\label{eq:limitCovariance:samplingNoise}
    \var_{\Omega,\vz}\bset{\whxO(\mFO\vx + \sigma\mSO\vz)} = \var_{\Omega}\bset{\whxO(\mFO\vx)} + 2\sigma^2\ev_{\Omega}\bset{\mSO^* D\whxO(\mFO\vx)D\whxO(\mFO\vx)^*\mSO} + o(\sigma^2).
\end{equation}
Once again, this is the same structure we would obtain from the method~\eqref{eq:noiselessLimit:affineOracle}. Therefore, piecewise linear reconstruction methods would induce correlations with this structure.

\subsection{The covariance in 4D flow}
\label{sec:noiselessLimit:densityVelocity}

From the reconstructed complex images we can obtain the reconstructed density and velocity by applying \(\Tfinv\) to \(\whvx_1,\ldots, \whvx_3\). As a consequence of the discontinuities of \(\arg\) the map \(T^{\dag}\) is not continuous and the hypotheses of Theorem~\ref{thm:noiselessLimit:momentsNoise} are not satisfied everywhere. However, away from the discontinuities the map \(\Tfinv\) is differentiable. At these points we can characterize the covariance in the noiseless limit. We provide only the form of the covariance due to measurement noise. The proof of the following result is deferred to Appendix~\ref{apx:noiselessLimit:4dFlow}.

\begin{theorem}
\label{thm:noiselessLimit:4dflow}
    Suppose the complex images \(\vx_0,\ldots, \vx_3\) are reconstructed independently using measurements on sampling sets \(\Omega_0,\ldots, \Omega_3\) that are corrupted by independent measurement noise. Let
    \[
        \whx_k^{\sigma}(\vecp) := \whx_{\Omega_k}(\vecp, \mF_{\Omega_k}\vx_k + \sigma\vc{S}_{\Omega_k}\vz_k)
    \]
    be the reconstructed \(\vx_k\) and let \(\wh{\rho}^{\sigma}, \whth_0^{\sigma}, \whv_1^{\sigma},\ldots, \whv_3^\sigma\) be the reconstructions obtained by applying the map \(\Tfinv\) defined by the expressions in~\eqref{eq:4dFlow:recovery}. Finally, suppose the hypotheses of Theorem~\ref{thm:noiselessLimit:momentsNoise} hold for each \(\whx_{\Omega_k}\) at the noiseless measurements \(\mF_{\Omega_k}\vx_k\) and that \(\whx_k^0(\vecp),\whx_k^0(\vecq)\neq 0\). Then,
    \begin{align*}
        \lim_{\sigma\to 0} \sigma^{-2}\cov(\wh{\rho}^{\sigma}(\vecp), \wh{\rho}^{\sigma}(\vecq)) &= \frac{1}{2}\real\Lbset{e^{-i\whth_0^0(\vecp)+i\whth_0^0(\vecq)} \lim_{\sigma\to 0} \sigma^{-2} \cov(\whx_0^{\sigma}(\vecp),\whx_0^{\sigma}(\vecq))} \\
        &\quad+\:\frac{1}{2}\real\Lbset{e^{i\whth_0^0(\vecp)+i\whth_0^0(\vecq)} \lim_{\sigma\to 0} \sigma^{-2}\cov(\whx_0^{\sigma}(\vecp)^*,\whx_0^{\sigma}(\vecq))},\\
        \lim_{\sigma\to 0} \sigma^{-2}\cov(\wh{\rho}^{\sigma}(\vecp), \wh{v}_k^{\sigma}(\vecq)) &= \frac{\venc}{2\pi \wh{\rho}^0(\vecq)}\imag\Lbset{e^{-i\whth_0^0(\vecp) + i\whth_0^0(\vecq)}\lim_{\sigma\to 0}\sigma^{-2}\cov(\whx_0^{\sigma}(\vecp),\whx_0^{\sigma}(\vecq)))} \\
        &\quad+\: \frac{\venc}{2\pi \wh{\rho}^0(\vecq)}\imag\Lbset{e^{i\whth_0^0(\vecp) + i\whth_0^0(\vecq)}\lim_{\sigma\to 0}\sigma^{-2}\cov(\whx_0^{\sigma}(\vecp),\whx_0^{\sigma}(\vecq)))},\\
        \lim_{\sigma\to 0}\sigma^{-2} \cov(\whv_k^{\sigma}(\vecp), \whv_\ell^{\sigma}(\vecq)) &= \frac{\venc^2}{2\pi^2\wh{\rho}^0(\vecp)\wh{\rho}^0(\vecq)}\real\Lbset{e^{-i\whth_0^0(\vecp) + i\whth_0^0(\vecq)} \lim_{\sigma\to 0} \sigma^{-2}\cov(\whx_0^{\sigma}(\vecp),\whx_0^{\sigma}(\vecq))}\\
        &\quad -\: \frac{\venc^2}{2\pi^2\wh{\rho}^0(\vecp)\wh{\rho}^0(\vecq)}\real\Lbset{e^{i\whth_0^0(\vecp) - i\whth_0^0(\vecq)} \lim_{\sigma\to 0} \sigma^{-2}\cov(\whx_0^{\sigma}(\vecp)^*,\whx_0^{\sigma}(\vecq))},
    \end{align*}
    for \(k\neq \ell\) and
    \begin{align*}
    \lim_{\sigma\to 0} \sigma^{-2}\cov(\whv_k^{\sigma}(\vecp), \whv_k^{\sigma}(\vecq)) &= \frac{\venc^2}{2\pi^2\wh{\rho}^0(\vecp)\wh{\rho}^0(\vecq)}\real\Lbset{e^{-i\whth_0^0(\vecp) + i\whth_0^0(\vecq)}\lim_{\sigma\to 0} \sigma^{-2}\cov(\whx_0^{\sigma}(\vecp),\whx_0^{\sigma}(\vecq))} \\
    &\quad -\:\frac{\venc^2}{2\pi^2\wh{\rho}^0(\vecp)\wh{\rho}^0(\vecq)}\real\Lbset{e^{i\whth_0^0(\vecp) - i\whth_0^0(\vecq)}\lim_{\sigma\to 0} \sigma^{-2}\cov(\whx_0^{\sigma}(\vecp)^*,\whx_0^{\sigma}(\vecq))} \\
    &\quad +\:\frac{\venc^2}{2\pi^2|\whx_k^0(\vecp)\whx_k^0(\vecq)|}\real\Lbset{e^{-i\whth_k^0(\vecp) + i\whth_k^0(\vecq)}\lim_{\sigma\to 0} \sigma^{-2}\cov(\whx_k^{\sigma}(\vecp),\whx_k^{\sigma}(\vecq))} \\
    &\quad -\:\frac{\venc^2}{2\pi^2|\whx_k^0(\vecp)\whx_k^0(\vecq)|}\real\Lbset{e^{i\whth_k^0(\vecp) - i\whth_k^0(\vecq)}\lim_{\sigma\to 0} \sigma^{-2}\cov(\whx_k^{\sigma}(\vecp)^*,\whx_k^{\sigma}(\vecq))}.
    \end{align*}
\end{theorem}

The correlations in the reconstructed images induce correlations on the reconstructed magnitude and velocity. Furthermore, as discussed in Section~\ref{sec:mathematicalModel:phaseRobustness}, the covariance in the complex images increases substantially on regions where the magnitude is small. Furthermore, the density and the velocities are statistically dependent, even when the density and velocity may not be physically related. Interestingly, the covariance between the density and the velocity scales linearly in \(\venc\) whereas the covariance between the components of the velocity scales quadratically in \(\venc\).

\subsection{Covariance structure for other reconstruction methods}
\label{sec:noiselessLimit:extensions}

Although we focus on recovering the individual complex images from acquired data separately to then estimate the density and the velocity, in the literature there are several recovery methods that combine together the information obtained for each complex image, and produce an estimate either for the complex images simultaneously~\cite{Kwak2013,Sun2017} or directly for the density and velocity~\cite{FengZhao2012,Santelli2016}. Our arguments still apply in this case. In general, we can consider a \(n\times 4\) complex matrix \(\vc{X}\) where the \(k\)-th column is the vectorized complex image \(\vc{x}_k\). If we let \(\mc{F}\) be the map that associates to every column of \(\vc{X}\) its Fourier transform, then~\eqref{eq:4dFlow:undersampledNoisyBasic} can be represented \(\vc{Y} = \mc{F}(\vc{X}) + \sigma \vc{Z}\) for a \(n\times 4\) random matrix \(\vc{Z}\) where \(Z_{i,j}\) are i.i.d. \(N(0,1) + iN(0,1)\).

The undersampling can be written in terms of the matrix \(\vc{X}\) as follows. To sampling sets \(\Omega_0, \ldots, \Omega_3\) we can associate a map \(\mc{S}_{\Pi_\Omega}\) that extracts the entries indexed on a suitable chosen set \(\Pi_\Omega\). It is clear the Bernoulli model described in Section~\ref{sec:mathematicalModel:undersampling} can be adapted to a Bernoulli model for \(\mc{S}_{\Pi_\Omega}\). Therefore, the model~\eqref{eq:4dFlow:undersampledNoisy} becomes \(\vc{y} = \mc{S}_{\Pi_\Omega}(\mc{F}(\vc{X}) + \sigma \vc{Z})\). Consequently, to estimate a quantity \(f\) from undersampled data, we have
\[
    \widehat{\vc{f}}_{\Pi_\Omega} = \widehat{f}_{\Pi_\Omega}(\mc{S}_{\Pi_\Omega}(\mc{F}(\vc{X}) + \sigma \vc{Z})).
\]
We see then that Theorem~\ref{thm:noiselessLimit:momentsNoise} can be applied to this function. Furthermore, if \(\widehat{\vc{f}}_{\Pi_\Omega}\) represents the complex images, Theorem~\ref{thm:noiselessLimit:4dflow} can be readily extended to this case.

\section{Image reconstruction methods}
\label{sec:reconstructionMethods}

When the complex images are reconstructed from undersampled measurements first, standard image reconstruction methods are typically used. We briefly discuss some of them. We evaluate least \(\ell_2\)-norm recovery, Compressed Sensing, and stOMP. Since we assume that the method used to recover each one of the four complex images is the same, in this section we drop the subscript \(k\) and refer simply to an image \(\vx\) to be reconstructed from incomplete Fourier measurements.

\subsection{Reconstruction by least \(\ell_2\)-norm recovery}
\label{sec:recoveryOrthogonalProjection}

Reconstruction by least \(\ell_2\)-norm is a linear method that is amenable to analysis with standard techniques. Although it has been superseded by non-linear methods, its analysis leads to closed-form expressions that can be directly compared to those presented in Section~\ref{sec:noiselessLimit}. Although some of the results we present are classical, our emphasis is on the covariance of the estimator and the visual artifacts this may create.

\begin{figure}[ht!]
    \centering
    \subfloat[]{\includegraphics[width=0.35\textwidth]{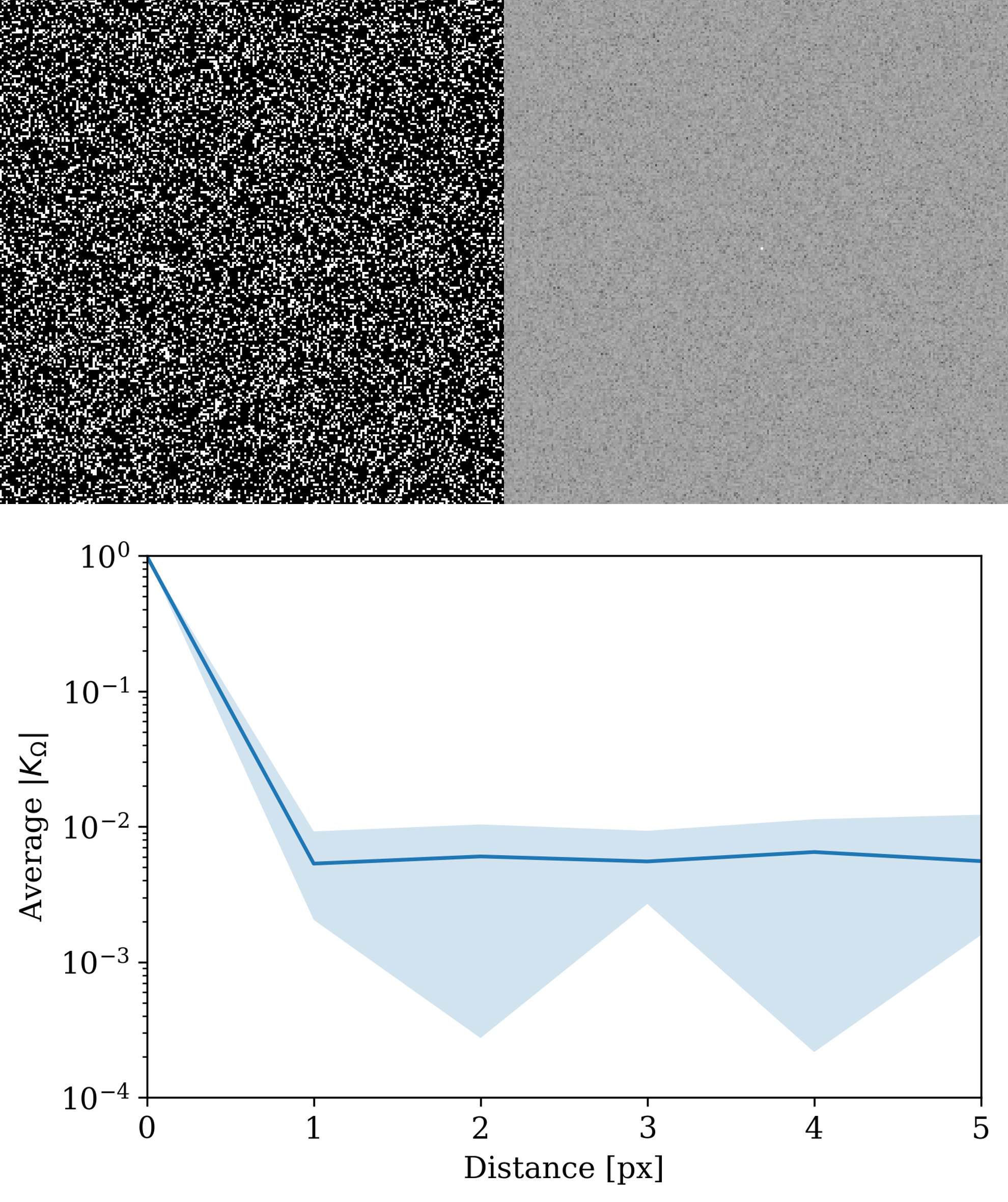}}\hspace{12pt}
    \subfloat[]{\includegraphics[width=0.35\textwidth]{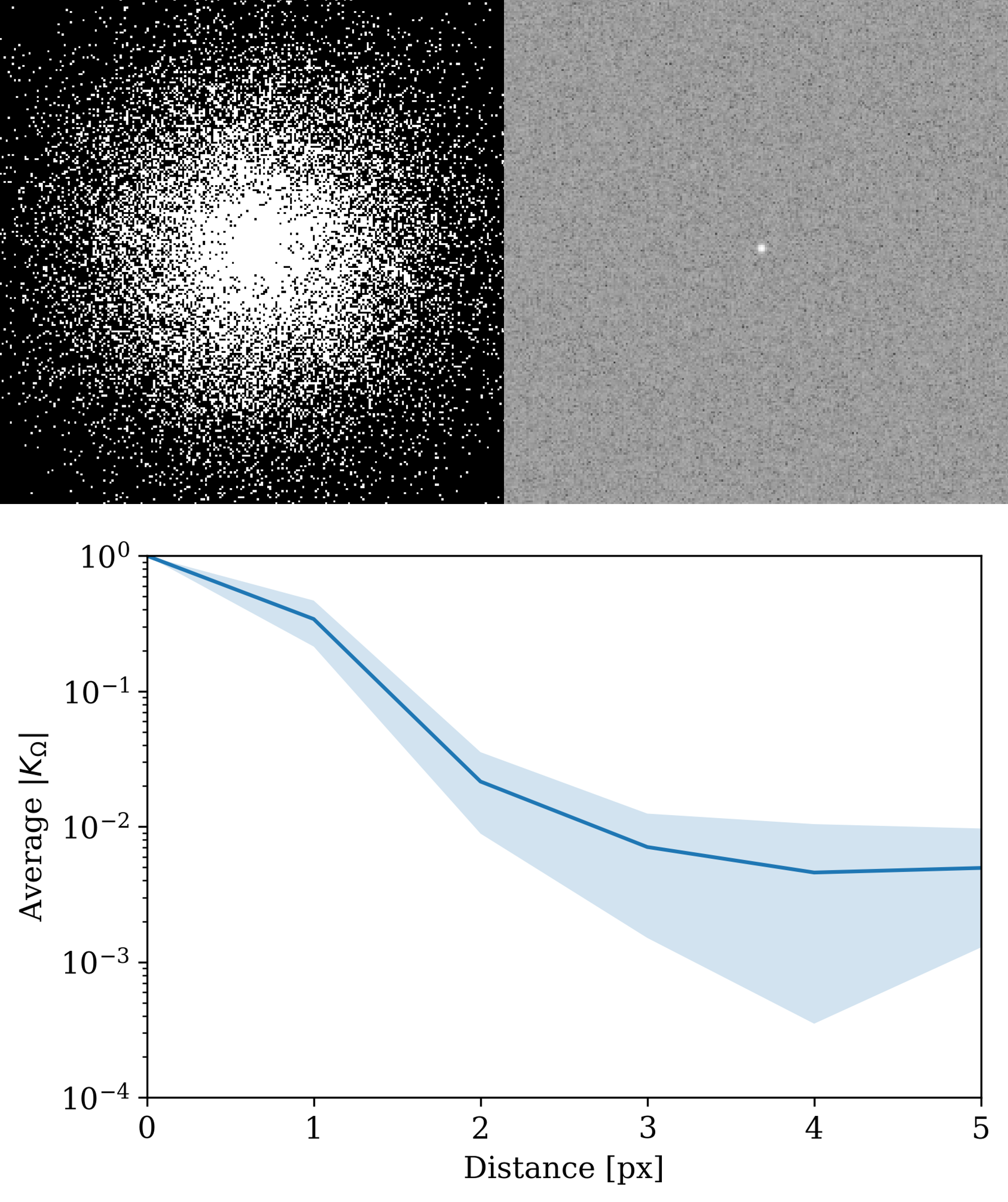}}\\
    \caption{Random samples for (a) Bernoulli and (b) Gaussian sampling for 75\% undersampling. In each case the upper left figure shows a random sampling set. The upper right figure shows the \(|K_\Omega|\) in logarithmic scale. Observe the kernel is much more concentrated at the origin for Bernoulli sampling than for Gaussian sampling. The lower figure shows the average decay of \(|K_\Omega|\). We observe that for Bernoulli sampling the kernel decays very fast. In contrast, the Gaussian kernel has a slower decay, reaching the minimum in about 3 pixels.}
    \label{fig:l2Recon:kernel}
\end{figure}

This reconstruction method involves solving the convex optimization problem
\begin{equation}
    \label{opt:l2NormRecovery}
    \begin{aligned}
        & \underset{\whvx}{\text{minimize}} & & \nrm{\whvx}_2
        & \text{subject to} & & \mFO\whvx = \vy. 
    \end{aligned}
\end{equation}
Since the Fourier transform is unitary, the optimal solution has the closed-form expression \(\whvxO = \mFO^*\vy\); in particular, it is linear and differentiable everywhere. Combining this expression with the noise model~\eqref{eq:4dFlow:undersampledNoisyBasic} we obtain
\begin{equation}
\label{eq:l2NormRecovery:estimate}
   \whvxO = \mFO^*\mSO(\mF\vx + \sigma \vz) = \mPO\vx + \sigma \mF^*\mSO\vz
\end{equation}
where \(\mPO := \mFO^*\mFO\) is the orthogonal projector onto \(\ker(\mFO)^\perp\). From this, we deduce
\[
    \ev_{z}\bset{\whvxO} = \mPO\vx = \whvxO(\mFO\vx)\quad\mbox{and}\quad\var_{z}\bset{\whvxO} = 2\sigma^2 \mFO^*\mFO = 2\sigma^2 \mPO = 2\sigma^2 D\whvx(\mFO\vx) D\whvx(\mFO\vx)^*.
\]
Hence, the covariance of a linear reconstruction method is the same as that in the noiseless limit. These closed-form expressions imply the covariance between the pixel values is
\begin{equation}
\label{eq:l2NormRecovery:covariance}
\begin{split}
    \cov_{z}(\whxO(\vec{r}_1),\whxO(\vec{r}_2)) = \frac{2\sigma^2}{n}\sum\nolimits_{\vec{\xi}\in \Omega} e^{2\pi i \vec{\xi}\cdot (\vec{r}_1-\vec{r}_2)} := 2\sigma^2 \frac{m}{n} K_\Omega(\vec{r}_1  - \vec{r}_2).
\end{split}
\end{equation}
The covariance is completely characterized by \(K_\Omega\). Since \(|K_\Omega| \leq 1\) the covariance has a magnitude proportional to the measured fraction of the variance. Fig.~\ref{fig:l2Recon:kernel} shows \(K_\Omega\) for Bernoulli and Gaussian sampling. Observe that even when only 25\% of the coefficients are sampled, \(K_\Omega\) is concentrated around the origin or, equivalently, \(\vc{P}_\Omega\) is close to diagonal. This is consistent with a trade-off between bias and variance as the covariance matrix is close to diagonal and small in norm, but the bias is, in general, large. Since the kind of images we consider have a large low-frequency content, the bias is larger for Bernoulli sampling than for Gaussian sampling even when \(K_\Omega\) is more concentrated for the former. 

\begin{figure}[ht!]
    \centering
    \subfloat[]{\includegraphics[width=0.4\textwidth]{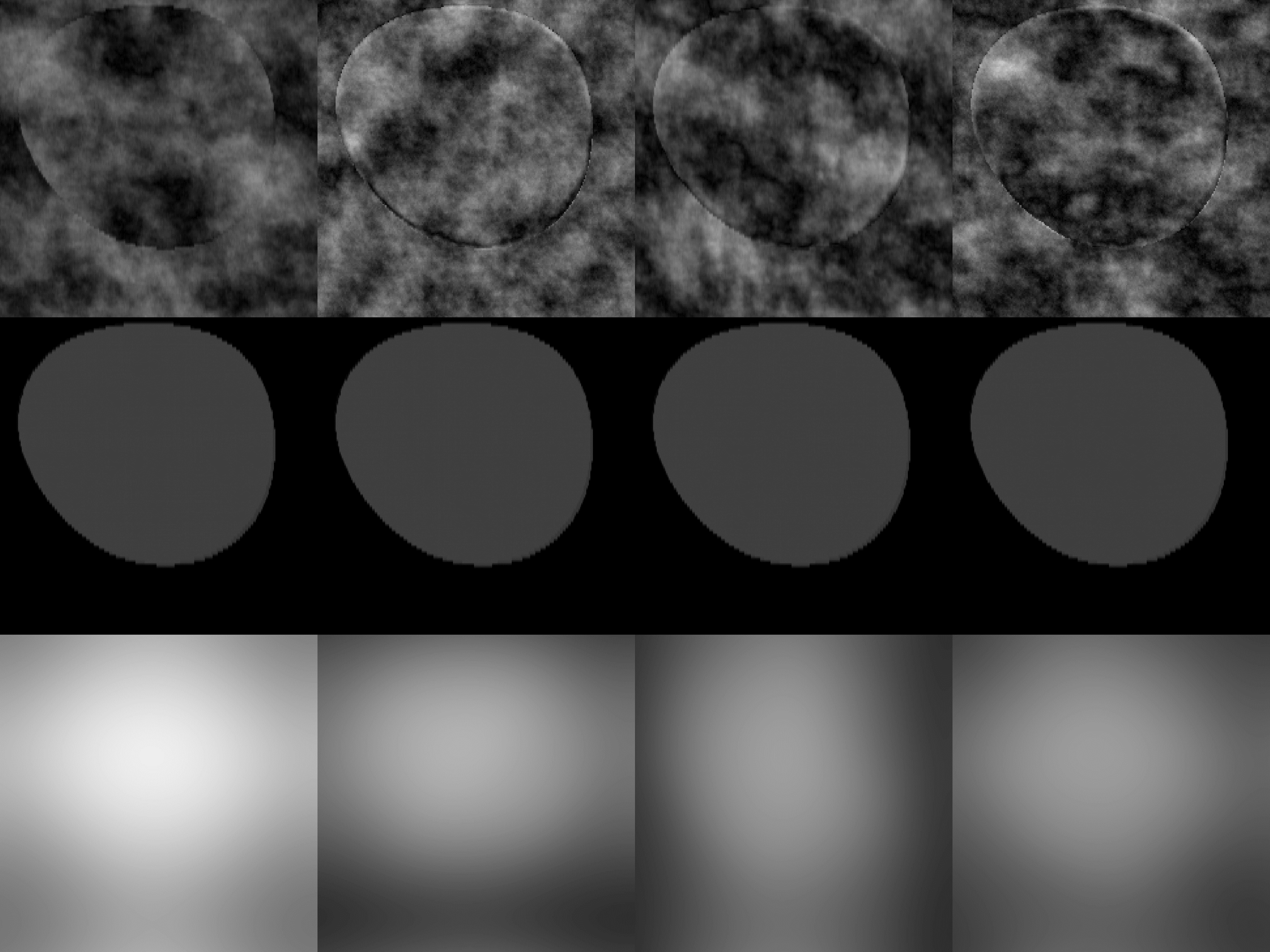}}\hspace{12pt}
    \subfloat[]{\includegraphics[width=0.4\textwidth]{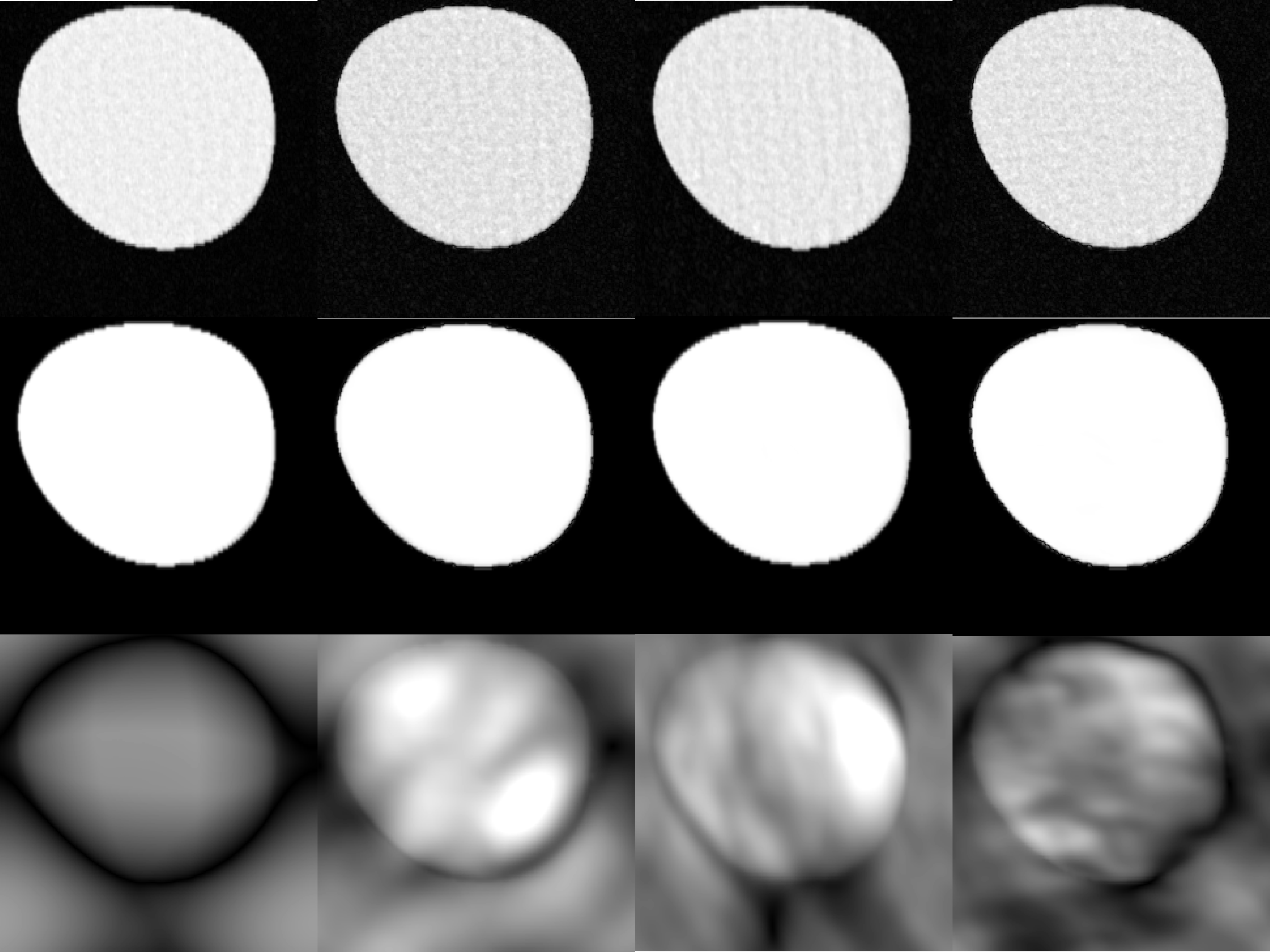}}\\
    \caption{Comparison of least \(\ell_2\)-norm recovery using (a) Bernoulli and (b) Gaussian sampling for 75\% undersampling. The image used corresponds to the simulated aortic flow test case presented in Section~\ref{sec:simAorticFlow}. In each case, the first row shows the magnitude of the expected reconstructed image conditioned on the sampling mask, showing the bias is substantially larger for Bernoulli sampling than for Gaussian sampling. The second row shows the magnitude of the expected reconstruction over both noise and sampling set. As expected, the bias for Bernoulli sampling is larger than for Gaussian sampling. Finally, the third row shows the magnitude of \(K_{\mu, x}\). In this case, the range of values  is much higher for Bernoulli sampling, the range being \([0, 6]\) suggesting that there are spatial correlations that persist in the noiseless limit. In contrast, for Gaussian sampling, the range is \([0, 0.002]\).}
    \label{fig:l2Recon:images}
\end{figure}

From closed-form expressions for \(\whvxO\) we can also determine the expectation and covariance over both \(\Omega\) and \(\vz\). In fact, using the sampling model introduced in Section~\ref{sec:mathematicalModel:undersampling} we deduce that
\[
    \ev_{\Omega,z}\bset{\whx_\Omega(\vc{S}_\Omega(\mF\vx + \sigma\vz))} = \ev_\Omega\bset{\ev_{z}\bset{\mF^*\vc{S}_\Omega^*(\vc{S}_\Omega\mF \vx + \sigma \vc{z})}} = \mF^* \ev_\Omega\bset{\vc{S}_\Omega^*\vc{S}_\Omega}\mF \vx = \mF^* \vc{D}_{\mu} \mF \vx := \vc{C}_{K_\mu} \vx
\]
where \(\vc{D}_\mu := \diag(\vc{\mu})\) and \(\vc{C}_{K_\mu}\) represents circular convolution by the function \(K_\mu\) defined implicitly by the above expression; \(\mu\) is as defined in Section~\ref{sec:mathematicalModel:undersampling}. In contrast to~\eqref{eq:l2NormRecovery:estimate}, on average the estimate \(\whvx\) is not the approximation of \(\vx\) by a trigonometric polynomial, but instead \(\vx\) circularly convolved with \(K_\mu\). Fig.~\ref{fig:l2Recon:images} shows the expectations, where once again the bias is larger for Bernoulli sampling.

The variance over \(\Omega\) is given by the following proposition.
\begin{proposition}
\label{prop:l2NormRecovery:covariance:samplingNoise}
    Define \(w\) as
    \[
        w(\vec{\xi}) = \mu(\vec{\xi}) (1- \mu(\vec{\xi}) |Fx(\vec{\xi})|^2
    \]
    and let \(\vc{K}_{\mu, x}\) denote its inverse Fourier transform. Then
    \[
        \var_{\Omega,z}\bset{\whx_\Omega(\vc{S}_\Omega(\mF\vx + \sigma\vz))} = 2\sigma^2 \vc{C}_{K_\mu} + \vc{C}_{K_{\mu, x}}.
    \]
\end{proposition}

As expected, this is consistent with the leading terms in~\eqref{eq:limitCovariance:samplingNoise}. The first term is independent of the true image, proportional to \(\sigma^2\), and it is due to both random sampling and the measurement noise. The second term depends on the true image, it is independent of \(\sigma^2\), and it is the joint effect of random sampling and the true image. Since both terms represent convolutions, neither is localized.

Fig.~\ref{fig:l2Recon:images} shows the convolution kernels \(K_{\mu, x}\) associated to Bernoulli and Gaussian sampling. Although neither kernel is local, for Gaussian sampling the kernel is significantly smaller in magnitude. This is because both the true image has a substantial low-frequency component, and Gaussian sampling samples low-frequencies more often. In contrast, the convolution kernel for Bernoulli sampling is almost constant, showing long-range spatial correlations across the entire image.

\subsection{Reconstruction by Compressed Sensing}
\label{sec:recoveryCompressedSensing}

Images, including MR images, are {\it compressible} and their representation in a suitable basis or frame only involves a few number of significant terms. For instance, the wavelet decomposition of an image is typically sparse and most of its wavelet coefficients are either zero or negligible. Based on this observation, a sufficient number of measurements to reconstruct an image should be proportional to the number of its non-negligible wavelet coefficients. Compressed Sensing formalizes this idea and provides conditions under which an image can be recovered exactly from an incomplete set of linear measurements by solving a convex optimization problem. 

\subsubsection{Exact recovery}

We consider wavelets as the basis on which images have a sparse representation. We denote the wavelet transform as \(\vc{W}\). Compressed Sensing asserts that the wavelet coefficients of the true image can be recovered by solving
\begin{equation}
    \label{opt:l1NormRecovery}
    \begin{aligned}
        & \underset{\whva}{\text{minimize}} & & \nrm{\whva}_1\\
        & \text{subject to} & & \mFO\mW^{-1}\whva = \mFO\vx
    \end{aligned}
\end{equation}
In other words, we find the image with wavelet coefficients of least \(\ell_1\)-norm that fits the measurements. If we let \(\whvaO\) be the optimal solution to~\eqref{opt:l1NormRecovery} then the recovered image is \(\whvxO = \mW^{-1}\whvaO\).

Exact recovery can be achieved when both the wavelet coefficients \(\va := \mW\vx\) of the true image \(\vx\) are sparse and the measurement matrix \(\mPhO := \mFO\mW^{-1}\) preserves the information contained in \(\va\). To make quantitative statements about these conditions, we say a vector is \(s\)-sparse if it has at most \(s\) non-zero coefficients. The property that ensures the measurement matrix preserves the information on a \(s\)-sparse vector is the Restricted Isometry Property (RIP). For an integer \(s\in \set{1,\ldots, n}\) we define the restricted isometry constant \(\delta_s\) as the smallest quantity such that
\[
    (1-\delta_s)\nrm{\vc{\alpha}}_2^2 \leq \nrm{\mPhO\va}_2^2 \leq (1+ \delta_s)\nrm{\vc{\alpha}}_2^2
\]
holds for all \(s\)-sparse \(\vc{\alpha}\), and the measurement matrix has the RIP for \(s\)-sparse vectors if \(\delta_{2s} < 1\). We have the following proposition~(see~\cite{Candes2005a} and~\cite[Thm.~6.9]{Foucart2013}).

\begin{proposition}
\label{prop:l1ExactRecovery}
    Let \(\va\) be the the wavelet transform of the true image \(\vx\) and let \(s\) be the cardinality of its support. If \(\delta_{2s} < 1\), then \(\va\) is the unique \(s\)-sparse solution to \(\mPhO\va = \mFO\vx\). If \(\delta_{2s} < 1/3\), then \(\va\) is the unique minimizer to~\eqref{opt:l1NormRecovery}.
\end{proposition}
The main difficulty in practice is to determine when a given measurement matrix satisfies these conditions; it can be NP-Hard to compute the restricted isometry constants~\cite{Tillmann2014}. In practice, random constructions satisfy these conditions with high probability. Although the main focus has been of Gaussian matrices, in our case the randomness enters only by choosing the subset of measured Fourier coefficients at random. When the wavelet transform is orthogonal only \(m \sim s\log n\) Fourier coefficients sampled uniformly at random are necessary~\cite{Candes2006}. In this case, we cannot characterize explicitly the behavior of the solution to~\eqref{opt:l1NormRecovery} when the measurements are corrupted by noise as there is no closed-form expression for \(\whvxO\). However, the theory of {\em partial smoothness}~\cite{Lewis2002, Drusvyatskiy2013} can be applied in this case to show it is differentiable outside of an exceptional set. In particular, in the real case the resulting reconstruction method is piecewise linear; since we are working with complex images, this is no longer the case. However, our results show the covariance behaves as if it were piecewise linear. Therefore, Theorem~\ref{thm:noiselessLimit:momentsNoise} and~\ref{thm:noiselessLimit:momentsSamplingNoise} can be applied to \(\whvxO\) outside this exceptional set as the variance vanishes.

\subsubsection{Robust \(\ell_1\)-recovery}
\label{sec:recoveryCompressedSensing:l1Recovery}

When measurements are corrupted by noise the equality constraint in~\eqref{opt:l1NormRecovery} is typically replaced by a constraint on the norm of the residual, and
\begin{equation}
    \label{opt:l1NormRecoveryNoisy}
    \begin{aligned}
        & \underset{\whva}{\text{minimize}} & & \nrm{\whva}_1\\
        & \text{subject to} & & \nrm{\vy -\mPhO\whva}_2 \leq \eta
    \end{aligned}
\end{equation}
is solved instead. The parameter \(\eta\) adjusts the allowed deviations and is usually proportional to \(\sigma\). Let \(\whvaO\) be the optimal solution to~\eqref{opt:l1NormRecoveryNoisy} and let \(\whvxO = \mW\whvaO\). We have the following error guarantee~\cite[Thm.~6.1]{Foucart2013}.

\begin{proposition}
\label{prop:l1RobustRecovery}
    Suppose the assumptions of Proposition~\ref{prop:l1ExactRecovery} hold. Furthermore, suppose that \(\sigma \nrm{\vz}_2 \leq \eta\) and that \(\delta_{2s} < 4/\sqrt{41}\). Then there is a constant \(C > 0\) depending on \(\delta_{2s}\) such that
    \[
        \nrm{\whvaO -\va}_2 \leq C\eta.
    \]
\end{proposition}

From Proposition~\ref{prop:l1RobustRecovery} we deduce the error estimate for \(\whvxO\)
\begin{equation}
\label{eq:robustnessForX}
    \nrm{\whvxO - \vx}_2 \leq C_\Omega\eta \nrm{\mW^{-1}}_2
\end{equation}
In the above, we have made explicit the dependence of the constant on the sampling set \(\Omega\). In particular, when the wavelet transform used is orthogonal, the same bound holds for \(\widehat{\vc{\alpha}}_k\) and \(\whvx_k\). As for~\eqref{opt:l1NormRecovery}, there is no closed-form for the solution to~\eqref{opt:l1NormRecoveryNoisy}. However, the theory of partial smoothness also holds for the solution to this problem, and Theorem~\ref{thm:noiselessLimit:momentsNoise} and Theorem~\ref{thm:noiselessLimit:momentsSamplingNoise} may be applied at the points of differentiability of \(\whvxO\). 

\subsubsection{Shrinkage effect and debiasing}
\label{sec:recoveryCompressedSensing:shrinkageAndDebias}

Although Proposition~\ref{prop:l1RobustRecovery} provides error bounds for \(\whvaO\), its entries tend to underestimate those of \(\va\). Similarly to the case of \(\ell_2\)-methods, this amounts to a bias that underestimates the effect of the estimated variables. In the statistics literature, this {\em shrinkage effect} has been studied for the LASSO~\cite{Tibshirani1996}.

To mitigate this effect, a popular approach is to {\it debias} the estimate by replacing \(\whvaO\) for an ordinary least-squares (OLS) estimate \(\whvaOdb\) that has the same support as \(\whvaO\) (see, e.g.~\cite[Ch.~6]{Hastie2015}). We can interpret \(\widehat{T}_\Omega := \supp(\whvaO)\) as an estimate for \(\supp(\va)\). Then, we can debias the reconstruction obtained from~\eqref{opt:l1NormRecoveryNoisy} by solving
\[
    \whvaOdb = \lim\nolimits_{\lambda\to 0}\Argmin\set{\nrm{\vy - \mPhO\whva}_2^2 + \lambda\nrm{\whva}_2^2:\, \supp(\whva) \subset \widehat{T}_\Omega}
\]
and defining the unbiased reconstruction \(\whvxOdb = \mW^{-1}\whvaOdb\). The limit in the above expression ensures the uniqueness of the solution, i.e., the single-valuedness of \(\Argmin\). Furthermore, it coincides either with the minimum \(\ell_2\)-norm solution to \(\vy = \mPhO\whva\) among all those supported on \(\widehat{T}_\Omega\) or with the OLS estimate. We expect \(\whvaOdb\) to have a smaller bias, or no bias at all, as long as \(\widehat{T}_\Omega\) is a good estimate for the support of \(\supp(\va)\).

The debiased estimate can be written as
\[
    \whvxOdb = \vc{B}_{\widehat{T}_\Omega} \vy\quad\mbox{for}\quad \vc{B}_{\widehat{T}_\Omega}:=\mW \vc{S}_{\widehat{T}_\Omega}^* (\mFO\mW \vc{S}_{\widehat{T}_\Omega}^*)^+
\]
where \(^+\) denotes the pseudo-inverse. This is {\it not} linear in \(\vy\) as \(\widehat{T}_\Omega\) itself depends on \(\vy\). However, its behavior in the noiseless limit is similar to that of a linear reconstruction method. We provide the following heuristic. Suppose for any sufficiently small perturbation \(\vc{h}\) the supports of \(\whaO(\mFO\vx + \vc{h})\) and \(\whaO(\mFO\vx)\) are the same. Assuming differentiability, in the noiseless limit we have
\[
    \ev_\vz\bset{\whvxOdb} = \vc{B}_{\widehat{T}_{\Omega}^0}\mFO\vx + o(1)\quad\mbox{and}\quad \var_\vz\bset{\whvxOdb} = 2\sigma^2 \vc{B}_{\widehat{T}_{\Omega}^0}\vc{B}_{\widehat{T}_{\Omega}^0}^* + o(\sigma^2)
\]
where \(\widehat{T}_\Omega^{0}\) is the support of \(\whvaO(\mFO\vx)\). There is a clear similarity between the expressions obtained for least \(\ell_2\)-norm recovery and the debiased estimate. When \(\widehat{T}_\Omega^0\) contains the true support \(\supp(\va)\) and \(\mFO\mW\vc{S}_{\widehat{T}_\Omega^0}\) has a left-inverse, then
\[
    \vc{B}_{\widehat{T}_{\Omega}^0}\mFO\vx = \mW \vc{S}_{\widehat{T}_\Omega^0}^* (\mFO\mW \vc{S}_{\widehat{T}_\Omega^0}^*)^+ \mFO\mW\vc{S}_{\widehat{T}_\Omega^0}^*\vc{S}_{\widehat{T}_\Omega^0}\va = \mW\vc{S}_{\widehat{T}_\Omega^0}^*\vc{S}_{\widehat{T}_\Omega^0}\va = \vx
\]
and in the noiseless limit, the debiased estimate \(\whvxOdb\) is unbiased. We do not provide a theoretical justification for this heuristic, as this would go beyond the scope of this work. However, we will provide numerical results in Section~\ref{sec:numExperiments} to justify its soundness.

\subsection{Reconstruction by Stagewise Orthogonal Matching Pursuit}\label{sec:omp}

Under suitable conditions, the solution to~\eqref{opt:l1NormRecovery} is the sparsest solution to the linear system \(\mPhO\va = \mFO\vx\). However, solving~\eqref{opt:l1NormRecovery} or~\eqref{opt:l1NormRecoveryNoisy} could be computationally expensive in some applications. For this reason, a number of {\it heuristic algorithms} have been proposed in the literature to find or approximate such a solution. These algorithms do not necessarily solve an optimization problem, and are thus best understood as implementing implicitly an estimator \(\whvxO\) which may be substantially different than that resulting from ~\eqref{opt:l1NormRecoveryNoisy}. 
In particular, it may be harder to characterize their regularity or asymptotic behavior. In this work, we consider the Stagewise Orthogonal Matching Pursuit (stOMP~\cite{donoho2012sparse}), a variant of the Orthogonal Matching Pursuit (OMP) algorithm, but other choices are also possible including CoSaMP~\cite{needell2009cosamp}, ROMP~\cite{needell2009uniform}, and others.

The stOMP algorithm is a \emph{greedy} recovery heuristic for sparse signals designed to find the minimum \(\ell^{0}\)-norm $\va$ so that $\nrm{\vy -\mPhO\whva}_2 \leq \eta$, for a given tolerance $\eta$. It constructs a sequence of iterates \(\set{\va^{(k)}}\) aiming to reduce, at each iteration, the \(\ell^2\)-norm of the residual \(\vr^{(k)} = \vy - \mPhO\va^{(k)}\) as much as possible~\cite{tropp2007signal}.
This is accomplished by alternating the discovery of new components in the support of \(\va^{(k)}\) (referred to as \emph{sweep} stage or \emph{matching filter}) with the computation of the least-square projection of \(\vy\) onto the space generated by an increasing number of \emph{atoms} from \(\mPhO\).
Starting with \(\va^{(0)} = 0\), new non-zero components \(\boldsymbol{\ell}^{(k)} = (\ell_{1}^{(k)}, \ell_{2}^{(k)},\dots)\) are added to the support of \(\va^{(k-1)}\) at the \(k\)-th iteration based on 
\begin{equation}\label{ompSweep}
  \Big|\left[\mPhO^{*}\,\vr^{(k-1)}\right]_{\boldsymbol{\ell}^{(k)}}\Big|\ge t_{k}\,\sigma_{k} = 2\cdot \Vert \vr^{(k-1)}\Vert_{2}/\sqrt{m_{k}}.
\end{equation}
where $m_{k}$ is the cardinality of $\Omega_{k}$.
We then evaluate \(\va^{(k)}\) as a least-squares approximation for \(\vy\), having \(\supp(\va^{(k)}) = \supp(\va^{(k-1)})\cup\set{\boldsymbol{\ell}^{(k)}}\), computed using LSQR~\cite{paige1982lsqr,paige1982algorithm}. The total number of iterations is fixed and equal to $10$.

Note how the inner products on the left-hand-side of~\eqref{ompSweep} are just an efficient \emph{proxy} to determine the columns of \(\mPhO\) maximally correlated with \(\vr^{(k-1)}\). Unlike correlations, inner products are affected by column norms and the effectiveness of this proxy is maximized for cases where all columns have the same norms. Since the Fourier transform is unitary and we consider orthogonal wavelet basis, i.e., Haar or Daubechies, all columns have equal norm when \(\Omega\) comprises all frequencies. 
However, when the frequency information is undersampled, the norm decays selectively, based on the frequency content of each wavelet atom.
For variable density Gaussian undersampling, the high-frequencies are rarely sampled, reducing significantly the norm of the atoms at small scales.
As a result, the stOMP reconstructions will have a tendency towards promoting coefficient representations associated with coarse scale wavelets, thus oversmoothing the reconstructed image.
For large undersampling ratio and smooth wavelets, this reduction in norm can extend to coarse scale atoms, as discussed in Section~\ref{sec:simAorticFlow}.

\section{Numerical experiments}\label{sec:results}

\subsection{Experimental setup}\label{sec:numExperiments}

Our numerical experiments include four test cases, three with synthetic velocity fields and one acquired through 4D flow MRI.
Our first experiment considers Hagen-Poiseuille flow in an ideal cylindrical vessel. Two-dimensional images are obtained by slicing both the fluid field and the binary cylindrical mask in two directions, i.e., orthogonal and parallel to the cylinder axis, respectively.
The second dataset consists of a 2D slice from a stabilized finite element simulation of blood flow in a patient-specific thoracic aorta, performed with the SimVascular flow solver~\cite{updegrove2017simvascular}, and projected onto a three-dimensional regular lattice.
The last dataset is obtained from MRI on a healthy 70-year old volunteer. The field-of-view is 340$\times$233$\times$77 mm, resulting in a grid with 192$\times$132$\times$32 voxels of size equal to 1.77$\times$1.77$\times$2.4 mm.
Twenty phase-averaged velocity fields were acquired over one cardiac cycle, with temporal spacing of 38.4 ms. The SNR for the phase-averaged in-vivo measurements varied between 3.5 and 3.7. This resulted in a large velocity standard deviation of v = 0.19 m/s, equal to a Gaussian noise of 31\% of the maximum velocity in the ascending aorta during the systolic phase of the cycle.
Segmentation of the ascending aorta, descending aorta, aortic arch, left subclavian artery, left common carotid artery, and brachiocephalic artery was performed in SimVascular~\cite{updegrove2017simvascular}, using image data containing a combination of signal intensities and systolic velocity module, allowing for better identification of the blood volume.

Numerical tests were performed at four different noise levels, setting the noise standard deviation as a percent of the average $k$-space amplitudes, with values equal to 1\%, 5\%, 10\% and 30\%, with corresponding signal to noise ratio SNR$_{\vy}$ $1\times 10^{4}$, $400$, $100$ and approximately $11$, respectively. 
For the MRI test case, we applied an additional 10\% $k$-space noise.

\begin{figure}[ht!]
\centering
\includegraphics[width=0.8\textwidth]{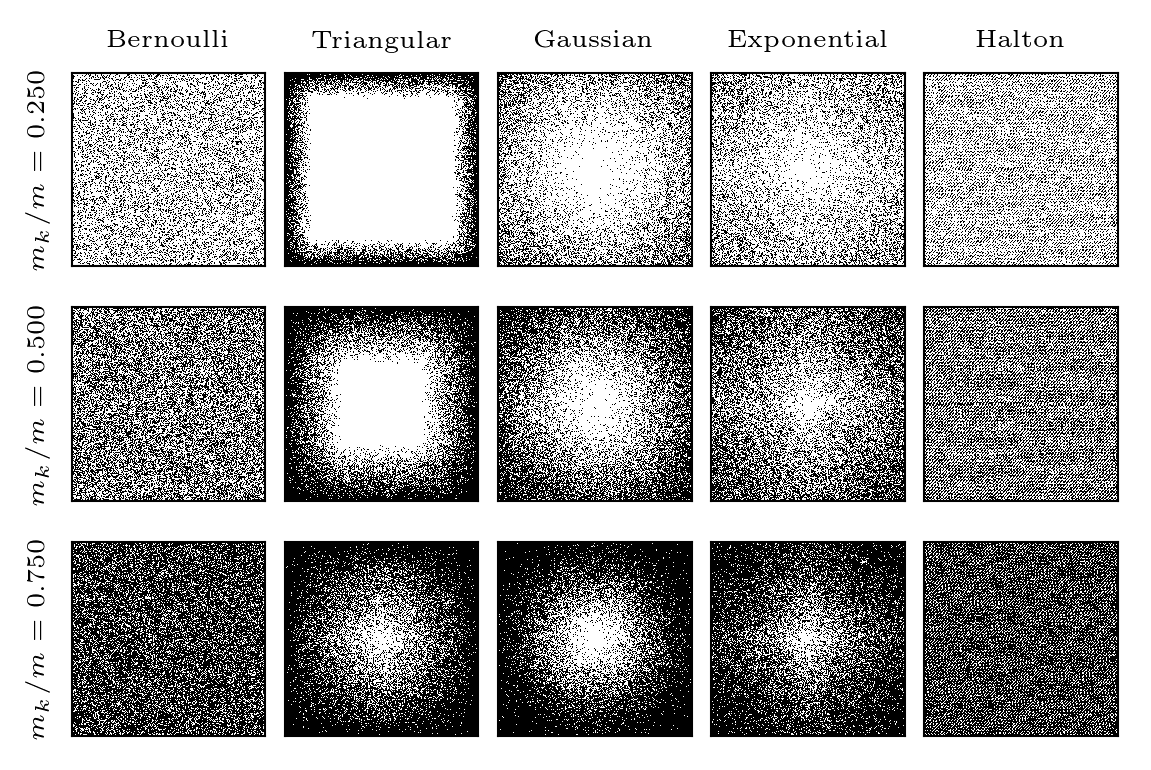}
\caption{$k$-space masks for various undersampling ratios.}\label{fig:samplingMasks}
\end{figure}

Masks were generated with variable \emph{undersampling ratio} $\mu(\vec{\xi})\in\{0.25, 0.5, 0.75\}$ using the patterns shown in Figure~\ref{fig:samplingMasks}, i.e., Bernoulli, variable density triangular, Gaussian, exponential and Halton. While preliminary results were generated using all the masks in Figure~\ref{fig:samplingMasks}, the result sections only contain images generated through the Bernoulli and variable density Gaussian masks, as these were found to produced marked differences in the noise correlations, consistent with the discussion in Section~\ref{sec:recoveryOrthogonalProjection}. 
Additionally, results for larger undersampling ratios, i.e. $\mu(\vec{\xi})\in\{0.85, 0.90, 0.95\}$ were generated for the simulated aortic flow test case.
For each of these tests, we generated 100 sets of synthetic $k$-space measurements each containing a different noise realization but with fixed undersampling mask, and performed image reconstruction leveraging shared memory parallelism. For the simulated aortic flow, $k$-space measurements were also generated by considering randomness both in the noise and in the sampling set $\Omega_{k} = \Omega,\,k=\{0,1,2,3\}$.
Image magnitude and velocities were reconstructed using Fourier-Haar and Fourier-Db8 measurement operators $\mPhO$, with the approaches discussed in Sections~\ref{sec:recoveryCompressedSensing:l1Recovery}, ~\ref{sec:recoveryCompressedSensing:shrinkageAndDebias}, and~\ref{sec:omp}, abbreviated in what follows using CS, CSDEB and stOMP, respectively. 

The source code as well as all the scripts used to generate the pictures in the next sections are based on the \emph{MRI noise analysis} library (mrina) available through a public GitHub repository at \url{https://github.com/desResLab/mrina}.

\subsection{Result quantities and notation}

We considered \emph{baseline} conditions characterized by 10\% k-space noise, a variable density Gaussian mask with 75\%  undersampling (only 1 every four frequencies are sampled, i.e., 4x compression), and image recovery through the \emph{compressed sensing} biased estimator approach discussed in Section~\ref{sec:recoveryCompressedSensing:l1Recovery}.
This case was selected as it contains enough noise to sensibly affect the recovered images, and possibly enough undersampling to challenge the reconstruction algorithms. Experiments were conducted by changing the parameters with respect to this baseline case and observing the resulting changes according to a number of metrics defined below.

Let $b_{k},\,k=0,\dots,3$ represent a reconstruction containing either four complex images ($b_{k}=x_{k}$), or image density plus velocities (i.e., $\{b_{0},b_{1},b_{2},b_{3}\} = \{\rho,v_{1},v_{2},v_{3}\}$).
Additionally, $b^{t}_{k}$ is the true image (assumed known) while $b^{a}_{k}$ is the average reconstruction over 100 different noise and undersampling mask realizations.
Reconstruction artifacts are identified through the quantity 
\begin{equation}\label{equ:rec_artifacts}
e_{k,s}(\vec{r}_{j}) = 2\,\frac{b_{k}(\vec{r}_{j}) - b^{s}_{k}(\vec{r}_{j})}{\vert b_{k}(\vec{r}_{j})\vert + \vert b^{s}_{k}(\vec{r}_{j})\vert},\,\,\text{for}\,\,k=0,\dots,3,\,s\in\{t,a\},
\end{equation}
with $e_{k,s}(\vec{r}_{j})\in [-2,2]$ for $s\in\{t,a\}$. The errors in \eqref{equ:rec_artifacts} are plotted with respect to both the average and true images, where the first represents the variance generated between recoveries, while the second incorporates both recovery bias and variance. 

Consider now $\widehat{v}_{k}^{(i)}(\vec{r}_{j})$ to be the $k$-th velocity component at the $\vec{r}_{j}$ pixel from the $i$-th reconstructed image.
Velocity correlations were determined by randomly selecting 50 pairs of locations $\{\vec{r}_{j},\vec{r}_{k}\}$ distant $d$ voxels apart and computing the sample Pearson correlation across the 100 reconstructed images as
\begin{equation}\label{equ:res_corr}
r_{d,k} = \frac{\sum_{i=1}^{100} [\widehat{v}_{k}^{(i)}(\vec{r}_{j})-\bar{v}_{k}(\vec{r}_{j})]\cdot[\widehat{v}_{k}^{(i)}(\vec{r}_{k})-\bar{v}_{k}(\vec{r}_{k})]}{\sqrt{\sum_{i=1}^{100} [\widehat{v}_{k}^{(i)}(\vec{r}_{j})-\bar{v}_{k}(\vec{r}_{j})]^2}\,\sqrt{\sum_{i=1}^{100} [\widehat{v}_{k}^{(i)}(\vec{r}_{k})-\bar{v}_{k}(\vec{r}_{k})]^2}},\,\,\text{for}\,k=1,2,3,
\end{equation}
where $\bar{v}_{k}(\vec{r}_{j})$ and $\bar{v}_{k}(\vec{r}_{k})$, $k=1,2,3$ denote the pixel velocities at $\vec{r}_{j}$ and $\vec{r}_{k}$ averaged over 100 reconstructions.
To make sure the correlations are representative of the true signal, the pairs $\{\vec{r}_{j},\vec{r}_{k}\}$ are selected only within the image region occupied by fluid. 

Three metrics of \emph{relative} mean squared error are used, one for the signal magnitude
\begin{equation}\label{opt:mse_mag}
\text{MSE}_{k,\text{mag},s} = \frac{\sum_{i=1}^{n}\left(\vert b_{k}(\vec{r}_{i})\vert - \vert  b^{s}_{k}(\vec{r}_{i})\vert\right)^{2}}{\sum_{j=1}^{n}\vert b^{s}_{k}(\vec{r}_{j})\vert^{2}},\,\,\text{for}\,\,k=0,\dots,3,\,\,s\in\{t,a\},
\end{equation}
one for the argument
\begin{equation}\label{opt:mse_ang}
\text{MSE}_{k,\text{ang},s} = \frac{1}{n}\cdot\sum_{i=1}^{n}\left\{\arg\left[b_{k}(\vec{r}_{i})\right] - \arg\left[b^{s}_{k}(\vec{r}_{i})\right]\right\}^{2},\,\,k=0,\dots,3,\,\,s\in\{t,a\},
\end{equation}
and one for complex $b_{k}=x_{k},\,k=0,\dots,3$ ($u=\text{cmx}$) or for $b_{0}=\rho$ and $b_{k}=v_{k},\,k=1,\dots,3$ ($u=\text{vel}$)
\begin{equation}\label{opt:mse_cmx}
\text{MSE}_{k,u,s} = \frac{\sum_{i=1}^{n}\vert b_{k}(\vec{r}_{i}) - b^{s}_{k}(\vec{r}_{i})\vert^{2}}{\sum_{j=1}^{n}\vert  b^{s}_{k}(\vec{r}_{j})\vert^{2}},\,\,k=0,\dots,3,\,u\in\{\text{cmx},\text{vel}\},\,s\in\{t,a\}.
\end{equation}
Percent errors for image density and velocity components are defined as
\begin{equation}\label{opt:mse_den}
\text{PE}_{k,s} = m_{i=1}^{n}\left[ \dfrac{\vert b_{k}(\vec{r}_{i}) - b^{s}_{k}(\vec{r}_{i})\vert}{\vert b^{s}_{k}(\vec{r}_{i})\vert}\cdot 100\right],\,\,k=0,\dots,3,\,\,s\in\{t,a\},
\end{equation}
where $m_{i=1}^{n}[\cdot]$ is the median computed over all the pixels of a single image. In the next sections, we report the mean value of $\text{PE}_{k,s}$ error over 100 reconstructions. Finally, all MSE and PE are computed within the fluid region, and the use of the median avoids the percent errors to be affected by negligibly small true or average image signals at the boundary of such region.

\subsection{Hagen-Poiseuille flow}\label{sec:poiseuilleFlowResult}

We consider a velocity field from a diffusion dominated, steady state, Hagen-Poiseuille flow in an ideal cylindrical domain. Two orthogonal slices are considered, the first perpendicular to the longitudinal axis of the cylinder, the second containing it (referred to as the \emph{orthogonal} and \emph{longitudinal} slice, respectively). The image density appears either as a solid disc (orthogonal slice) or a rectangle (longitudinal slice) and only one velocity component is non zero, varying smoothly following a parabolic velocity profile.

\begin{figure}[ht!]
\newcommand\tw{0.065}
\centering
\setlength{\tabcolsep}{0pt}
\renewcommand{\arraystretch}{0}
%
%
\subfloat[]{
\begin{tabular}[b]{c | c | c | c}%
\includegraphics[width=\tw\linewidth]{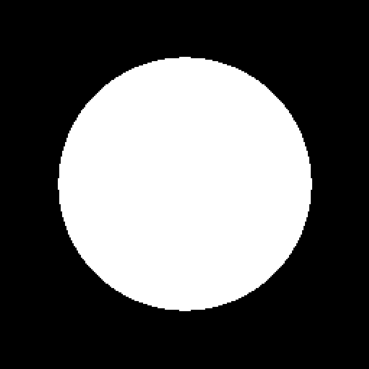} & 
\includegraphics[width=\tw\linewidth]{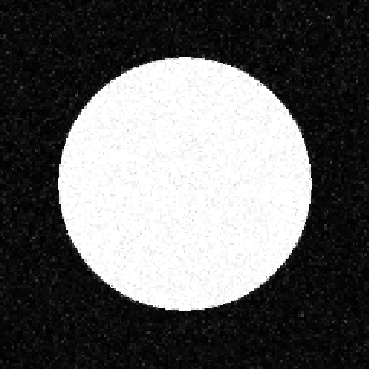} & 
\includegraphics[width=\tw\linewidth]{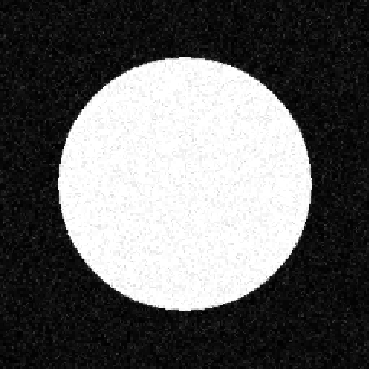} & 
\includegraphics[width=\tw\linewidth]{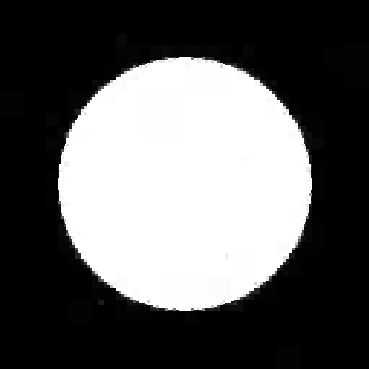}\\
\hline
\includegraphics[width=\tw\linewidth]{imgs/p1/true_k0.png} & 
\includegraphics[width=\tw\linewidth]{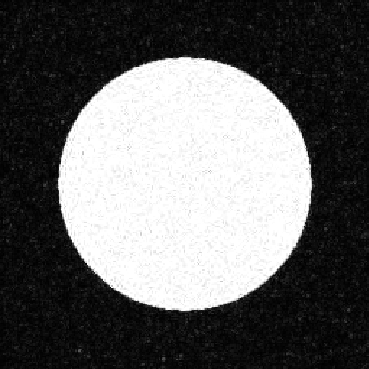} & 
\includegraphics[width=\tw\linewidth]{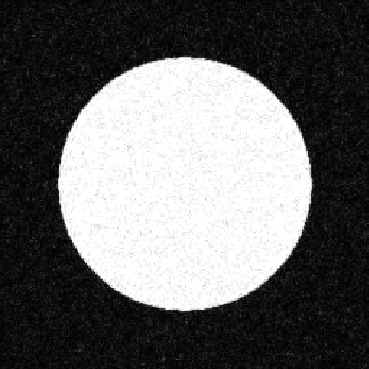} & 
\includegraphics[width=\tw\linewidth]{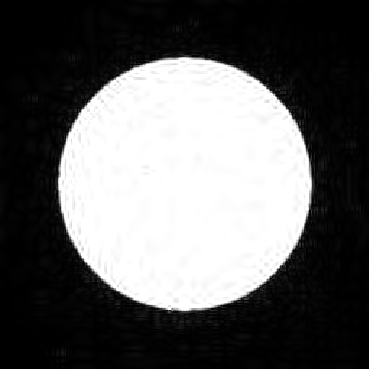}\\
\hline
\includegraphics[width=\tw\linewidth]{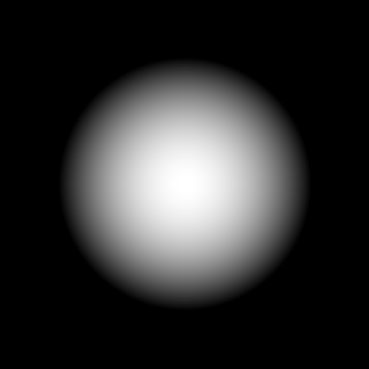} & 
\includegraphics[width=\tw\linewidth]{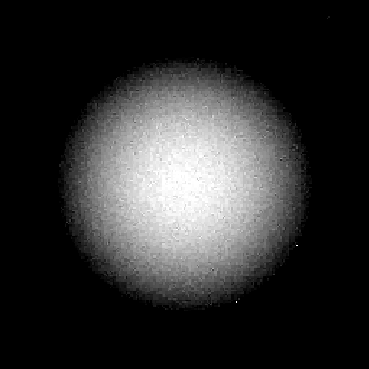} & 
\includegraphics[width=\tw\linewidth]{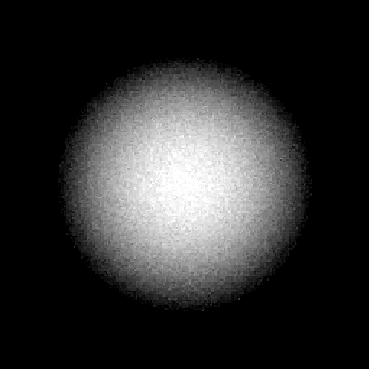} & 
\includegraphics[width=\tw\linewidth]{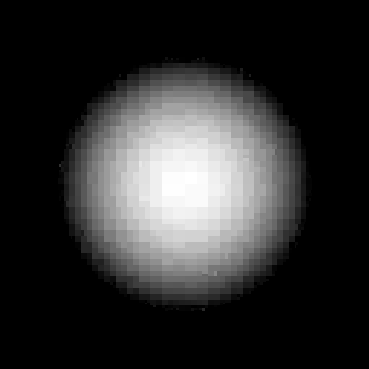}\\
\hline
\includegraphics[width=\tw\linewidth]{imgs/p1/true_k3.png} & 
\includegraphics[width=\tw\linewidth]{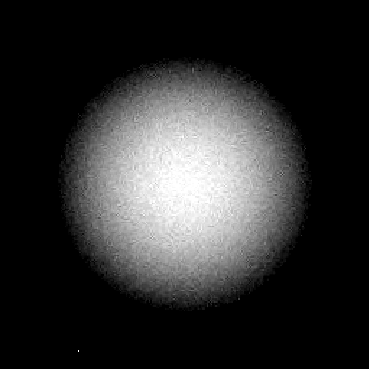} & 
\includegraphics[width=\tw\linewidth]{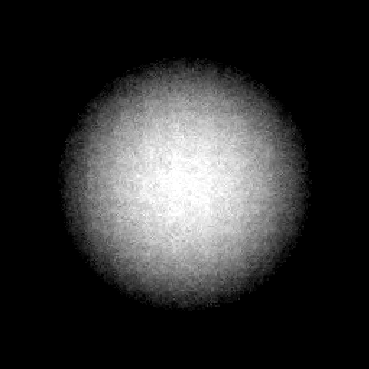} & 
\includegraphics[width=\tw\linewidth]{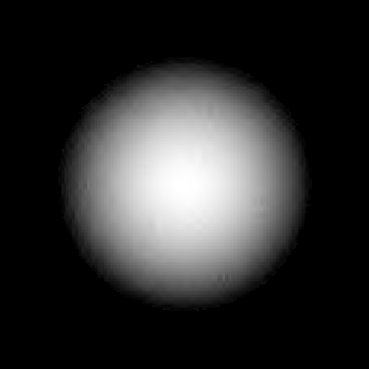}\\
\end{tabular}\label{fig:pA1_img_A}
}
\subfloat[]{
\begin{tabular}[b]{c | c | c | c}%
\includegraphics[width=\tw\linewidth]{imgs/p1/true_k0.png} & 
\includegraphics[width=\tw\linewidth]{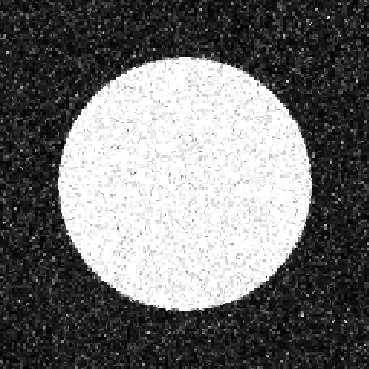} & 
\includegraphics[width=\tw\linewidth]{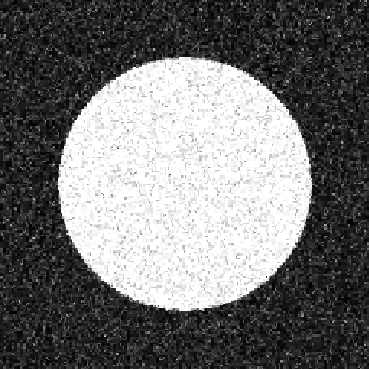} & 
\includegraphics[width=\tw\linewidth]{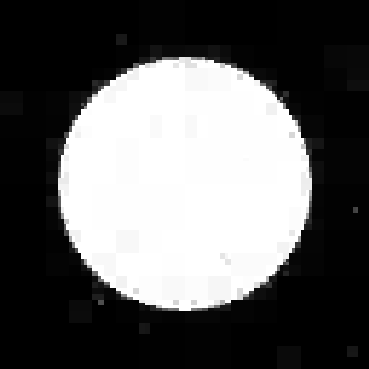}\\
\hline
\includegraphics[width=\tw\linewidth]{imgs/p1/true_k0.png} & 
\includegraphics[width=\tw\linewidth]{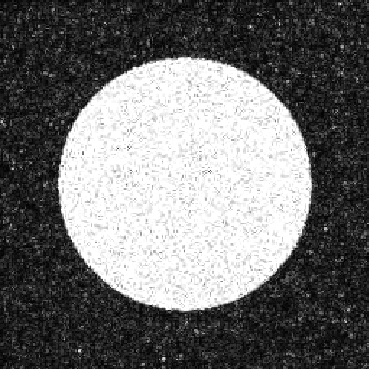} & 
\includegraphics[width=\tw\linewidth]{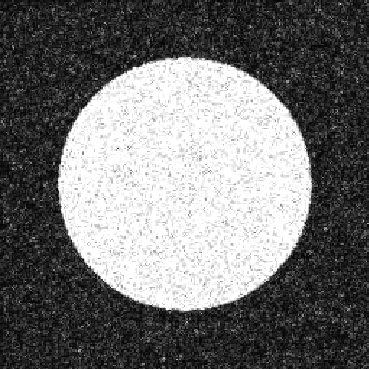} & 
\includegraphics[width=\tw\linewidth]{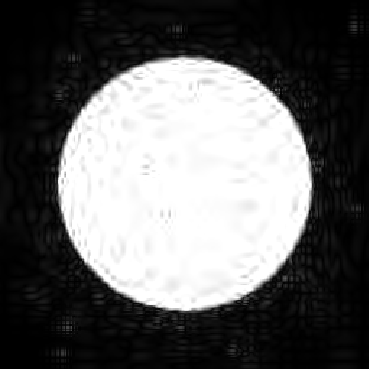}\\
\hline
\includegraphics[width=\tw\linewidth]{imgs/p1/true_k3.png} & 
\includegraphics[width=\tw\linewidth]{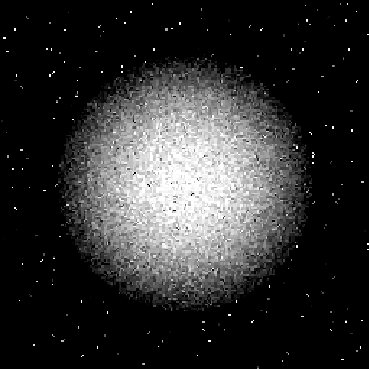} & 
\includegraphics[width=\tw\linewidth]{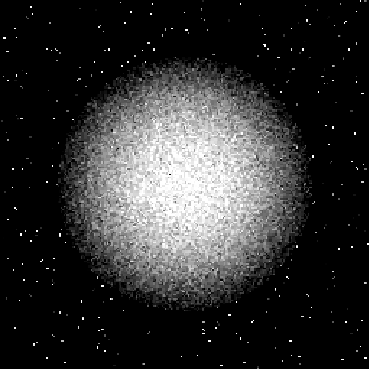} & 
\includegraphics[width=\tw\linewidth]{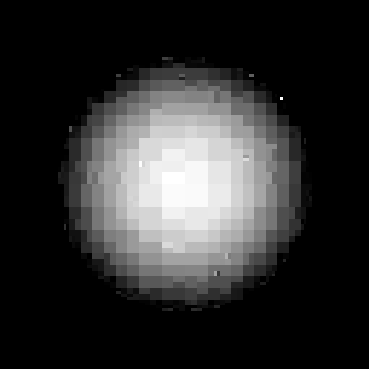}\\
\hline
\includegraphics[width=\tw\linewidth]{imgs/p1/true_k3.png} & 
\includegraphics[width=\tw\linewidth]{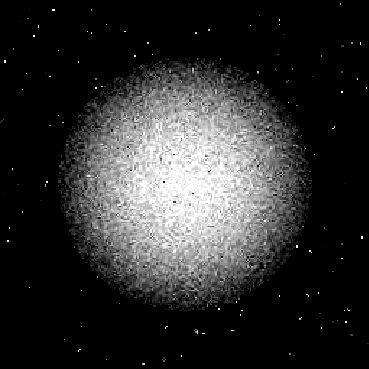} & 
\includegraphics[width=\tw\linewidth]{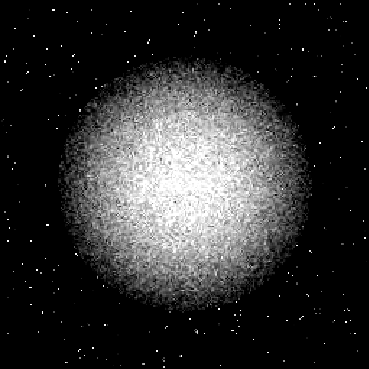} & 
\includegraphics[width=\tw\linewidth]{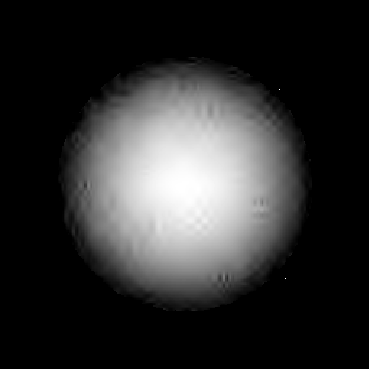}\\
\end{tabular}\label{fig:pA1_img_B}
}
\subfloat[]{
\begin{tabular}[b]{c | c | c}%
\includegraphics[width=\tw\linewidth]{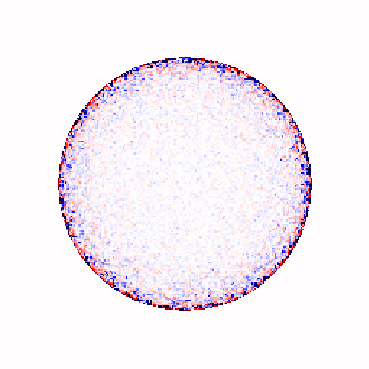} & 
\includegraphics[width=\tw\linewidth]{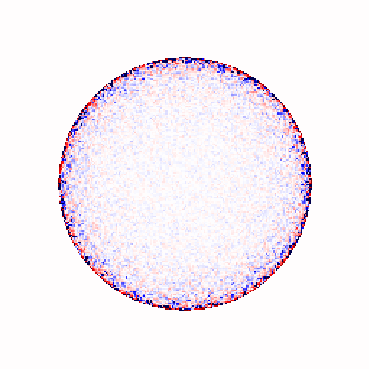} & 
\includegraphics[width=\tw\linewidth]{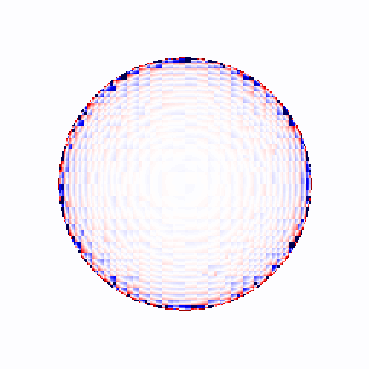}\\
\hline
\includegraphics[width=\tw\linewidth]{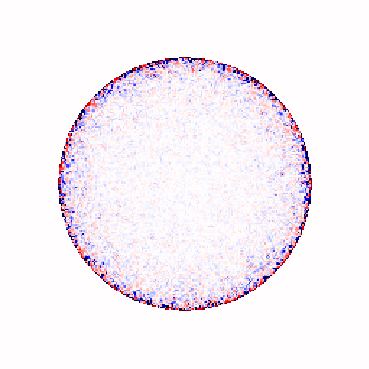} & 
\includegraphics[width=\tw\linewidth]{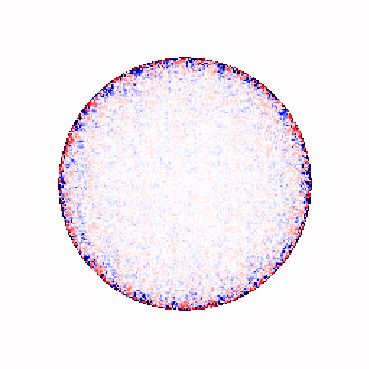} & 
\includegraphics[width=\tw\linewidth]{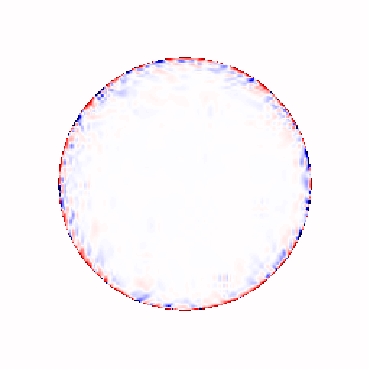}\\
\hline
\includegraphics[width=\tw\linewidth]{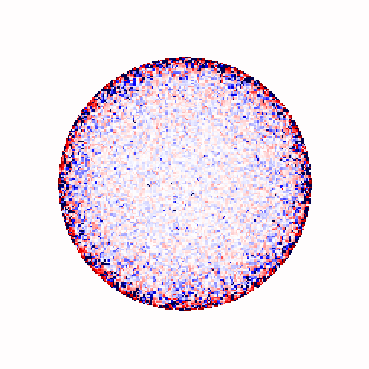} & 
\includegraphics[width=\tw\linewidth]{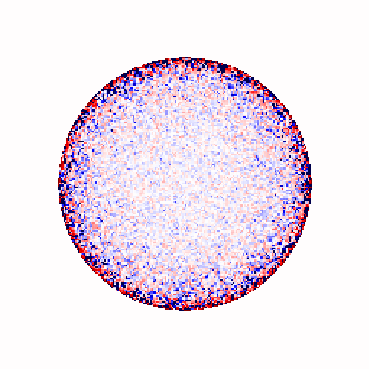} & 
\includegraphics[width=\tw\linewidth]{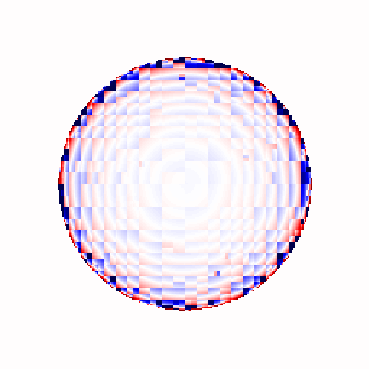}\\
\hline
\includegraphics[width=\tw\linewidth]{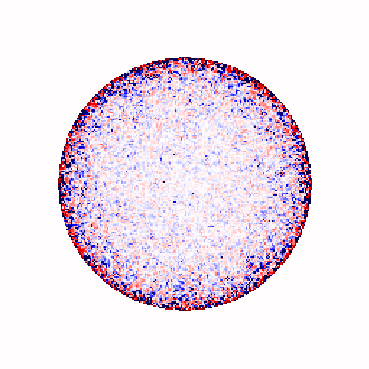} & 
\includegraphics[width=\tw\linewidth]{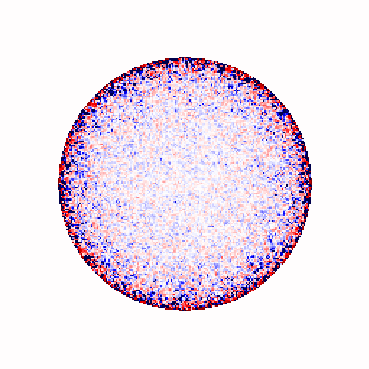} & 
\includegraphics[width=\tw\linewidth]{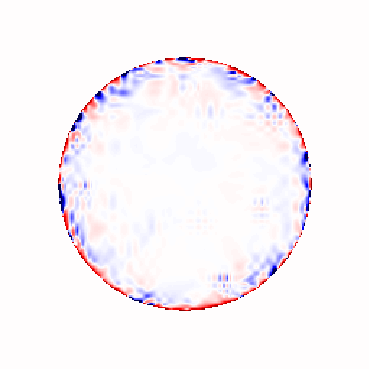}\\
\end{tabular}\label{fig:pA1_img_C}
}
\subfloat[]{
\begin{tabular}[b]{c | c | c}%
\includegraphics[width=\tw\linewidth]{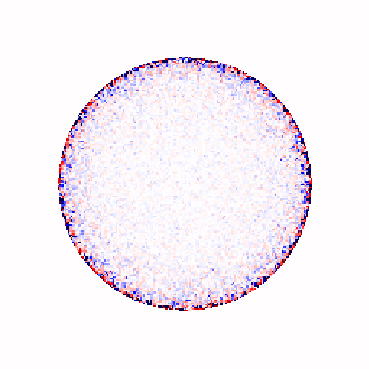} & 
\includegraphics[width=\tw\linewidth]{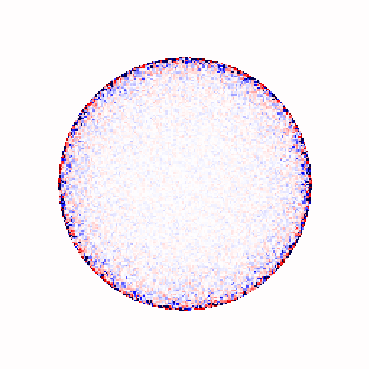} & 
\includegraphics[width=\tw\linewidth]{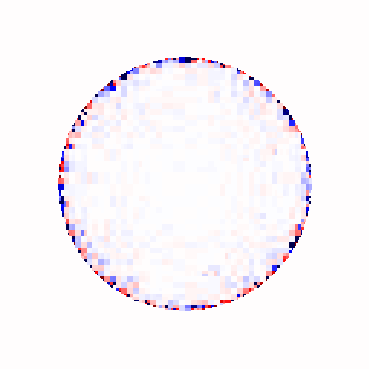}\\
\hline
\includegraphics[width=\tw\linewidth]{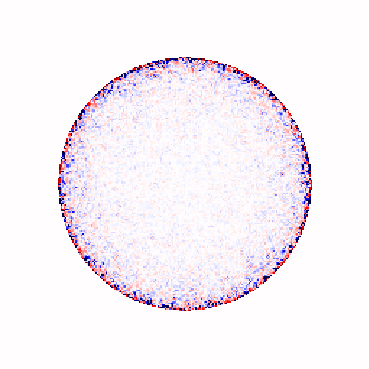} & 
\includegraphics[width=\tw\linewidth]{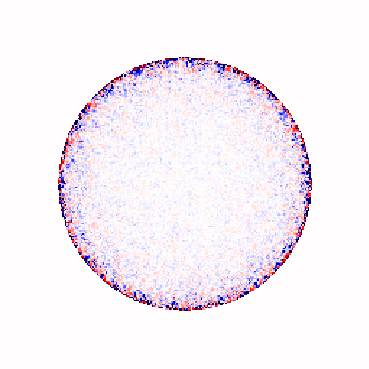} & 
\includegraphics[width=\tw\linewidth]{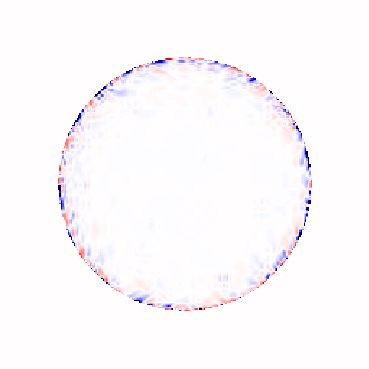}\\
\hline
\includegraphics[width=\tw\linewidth]{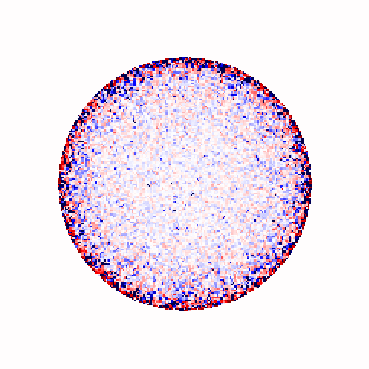} & 
\includegraphics[width=\tw\linewidth]{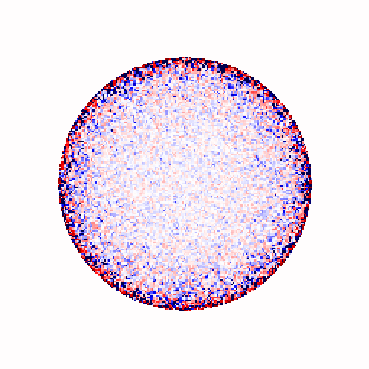} & 
\includegraphics[width=\tw\linewidth]{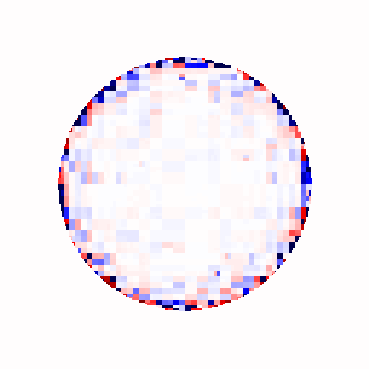}\\
\hline
\includegraphics[width=\tw\linewidth]{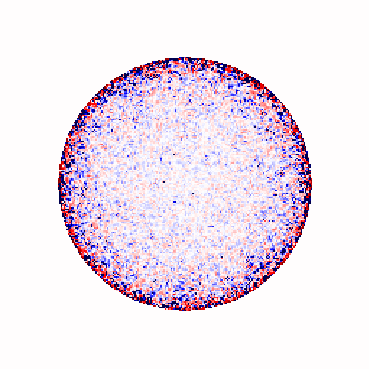} & 
\includegraphics[width=\tw\linewidth]{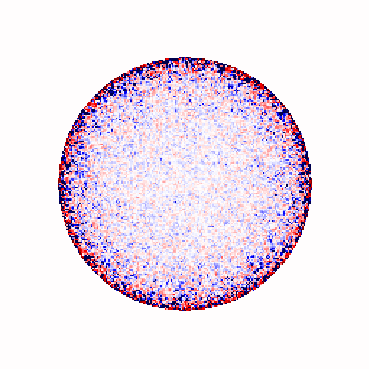} & 
\includegraphics[width=\tw\linewidth]{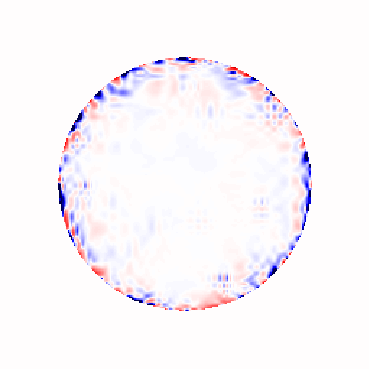}\\
\end{tabular}\label{fig:pA1_img_D}
}\\
%
%
\subfloat[]{
\begin{tabular}[b]{c | c | c | c}%
\includegraphics[width=\tw\linewidth]{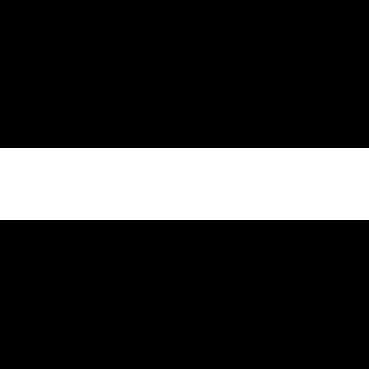} & 
\includegraphics[width=\tw\linewidth]{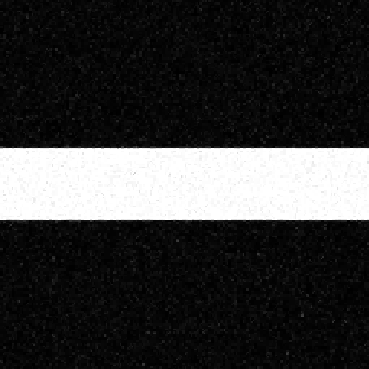} & 
\includegraphics[width=\tw\linewidth]{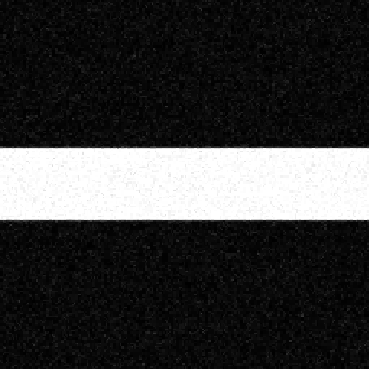} & 
\includegraphics[width=\tw\linewidth]{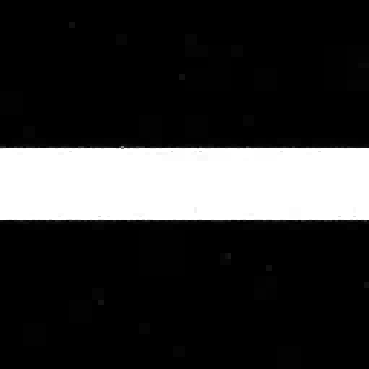}\\
\hline
\includegraphics[width=\tw\linewidth]{imgs/p2/true_k0.png} & 
\includegraphics[width=\tw\linewidth]{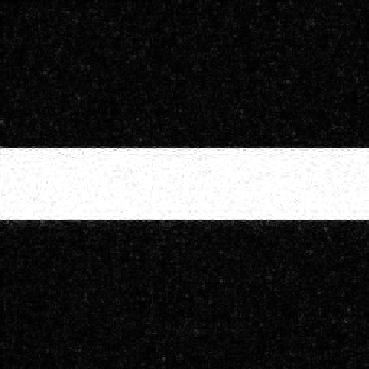} & 
\includegraphics[width=\tw\linewidth]{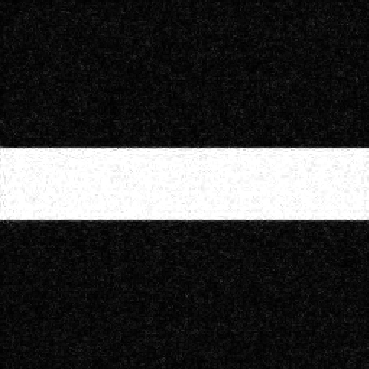} & 
\includegraphics[width=\tw\linewidth]{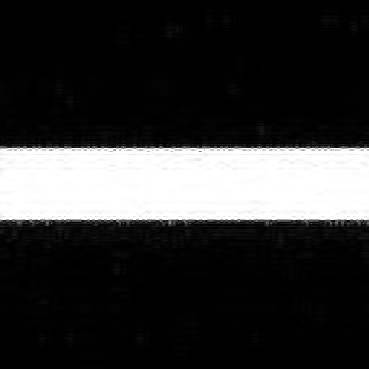}\\
\hline
\includegraphics[width=\tw\linewidth]{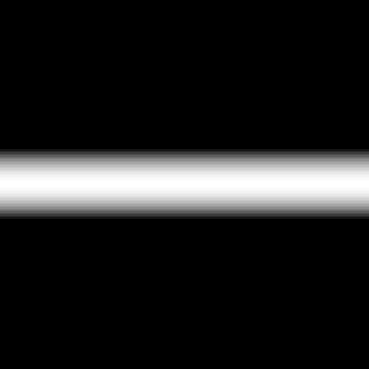} & 
\includegraphics[width=\tw\linewidth]{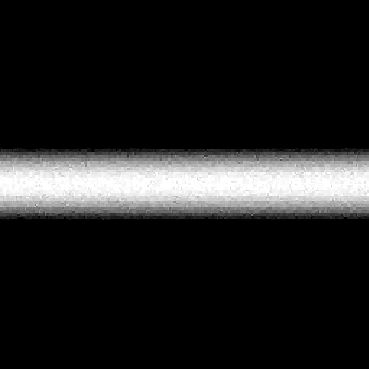} & 
\includegraphics[width=\tw\linewidth]{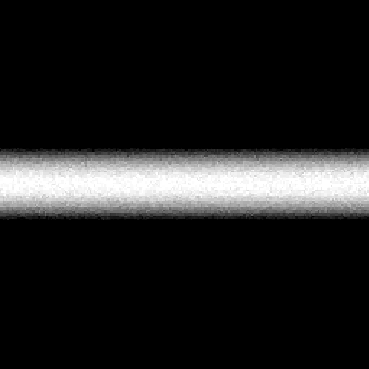} & 
\includegraphics[width=\tw\linewidth]{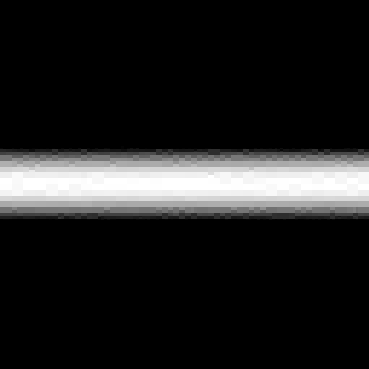}\\
\hline
\includegraphics[width=\tw\linewidth]{imgs/p2/true_k3.png} & 
\includegraphics[width=\tw\linewidth]{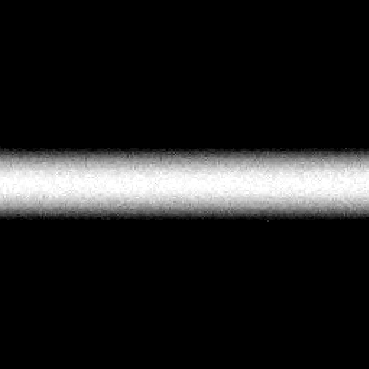} & 
\includegraphics[width=\tw\linewidth]{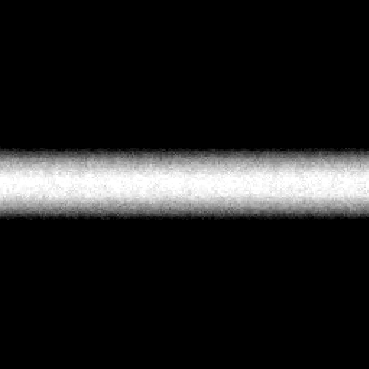} & 
\includegraphics[width=\tw\linewidth]{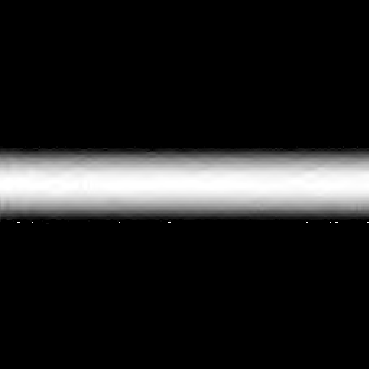}\\
\end{tabular}\label{fig:pA1_img_E}
}
\subfloat[]{
\begin{tabular}[b]{c | c | c | c}%
\includegraphics[width=\tw\linewidth]{imgs/p2/true_k0.png} & 
\includegraphics[width=\tw\linewidth]{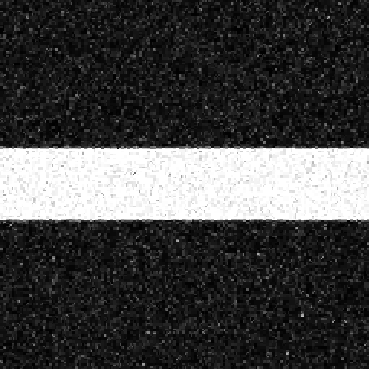} & 
\includegraphics[width=\tw\linewidth]{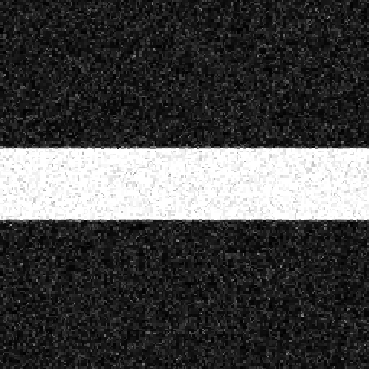} & 
\includegraphics[width=\tw\linewidth]{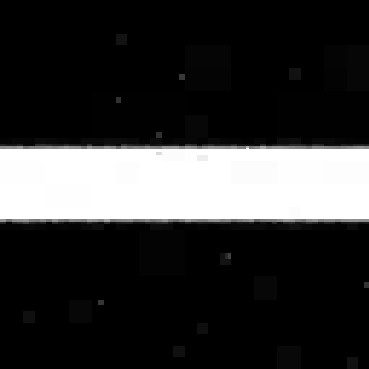}\\
\hline
\includegraphics[width=\tw\linewidth]{imgs/p2/true_k0.png} & 
\includegraphics[width=\tw\linewidth]{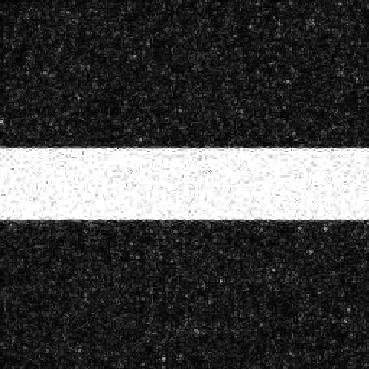} & 
\includegraphics[width=\tw\linewidth]{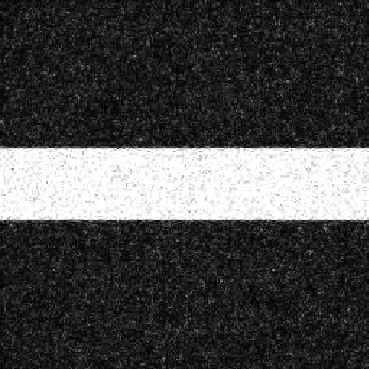} & 
\includegraphics[width=\tw\linewidth]{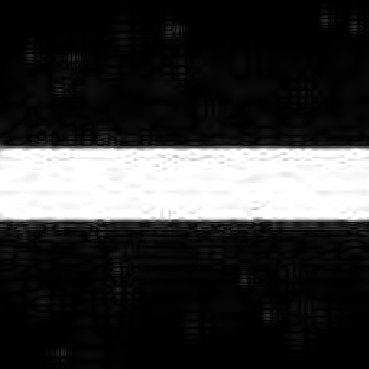}\\
\hline
\includegraphics[width=\tw\linewidth]{imgs/p2/true_k3.png} & 
\includegraphics[width=\tw\linewidth]{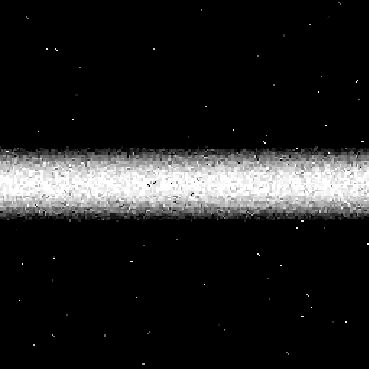} & 
\includegraphics[width=\tw\linewidth]{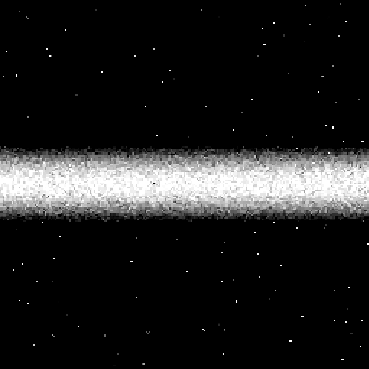} & 
\includegraphics[width=\tw\linewidth]{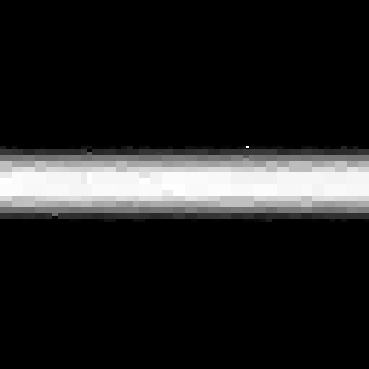}\\
\hline
\includegraphics[width=\tw\linewidth]{imgs/p2/true_k3.png} & 
\includegraphics[width=\tw\linewidth]{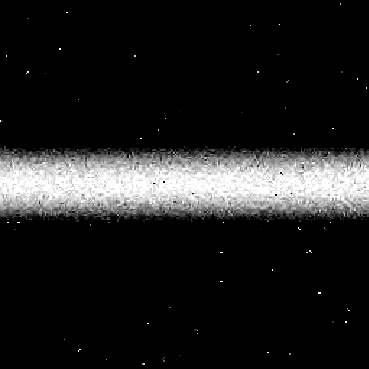} & 
\includegraphics[width=\tw\linewidth]{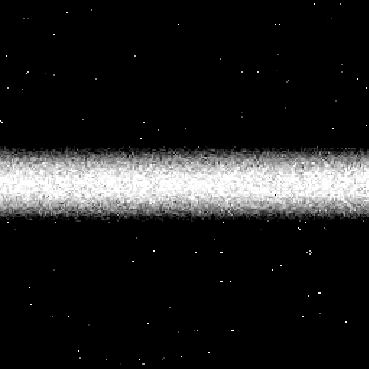} & 
\includegraphics[width=\tw\linewidth]{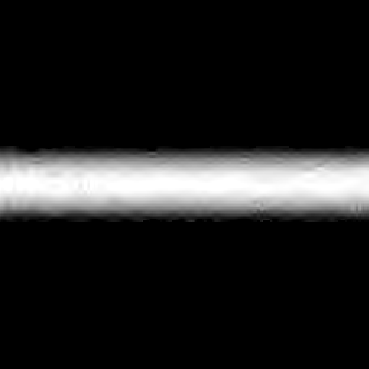}\\
\end{tabular}\label{fig:pA1_img_F}
}
\subfloat[]{
\begin{tabular}[b]{c | c | c}%
\includegraphics[width=\tw\linewidth]{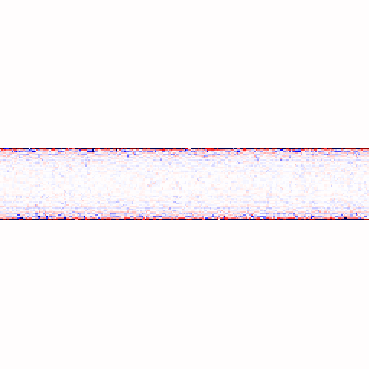} & 
\includegraphics[width=\tw\linewidth]{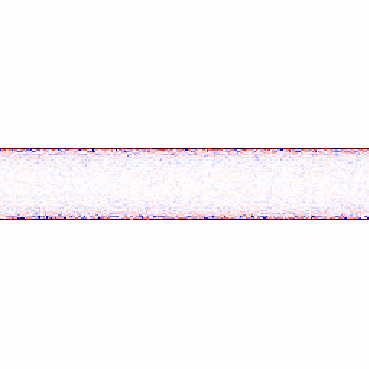} & 
\includegraphics[width=\tw\linewidth]{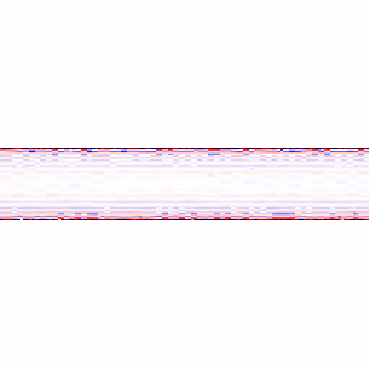}\\
\hline
\includegraphics[width=\tw\linewidth]{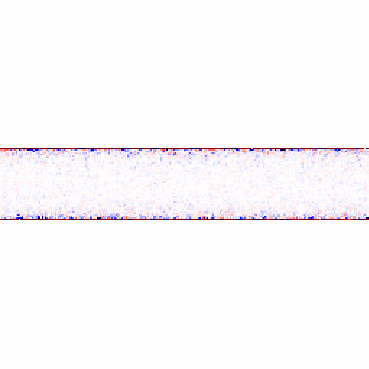} & 
\includegraphics[width=\tw\linewidth]{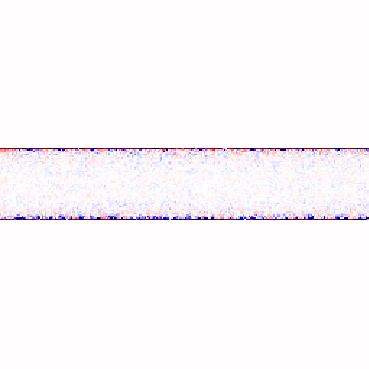} & 
\includegraphics[width=\tw\linewidth]{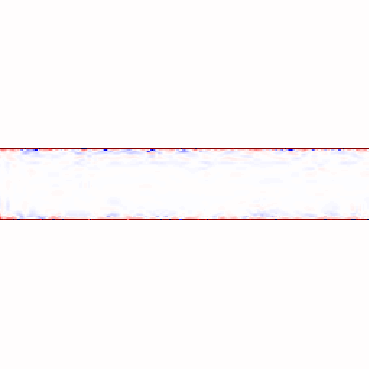}\\
\hline
\includegraphics[width=\tw\linewidth]{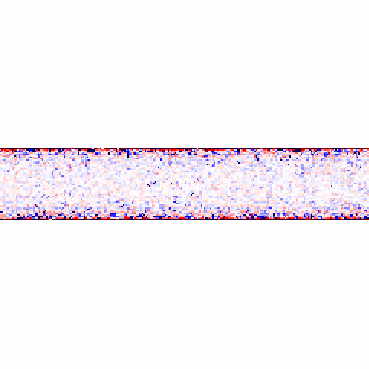} & 
\includegraphics[width=\tw\linewidth]{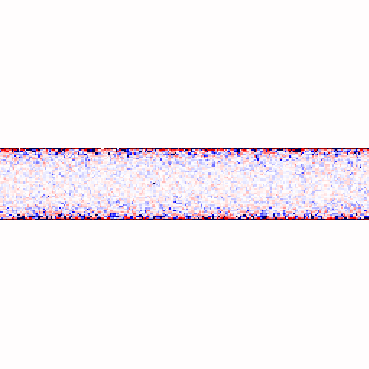} & 
\includegraphics[width=\tw\linewidth]{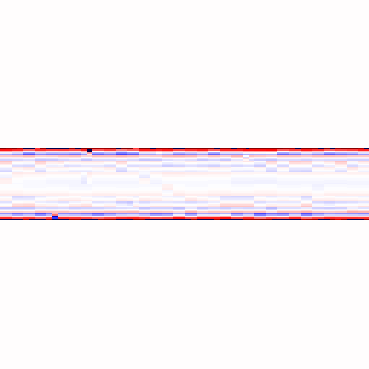}\\
\hline
\includegraphics[width=\tw\linewidth]{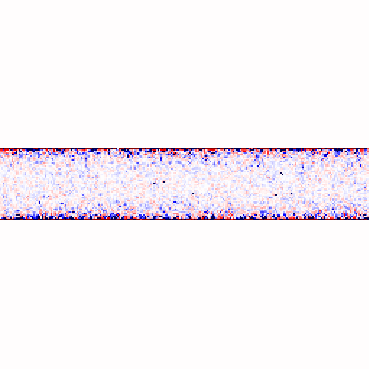} & 
\includegraphics[width=\tw\linewidth]{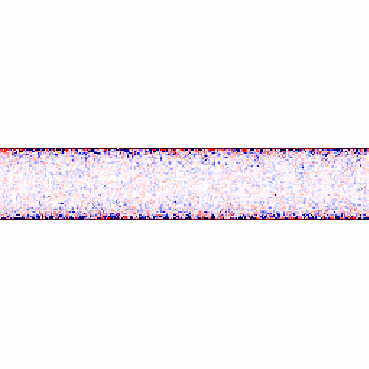} & 
\includegraphics[width=\tw\linewidth]{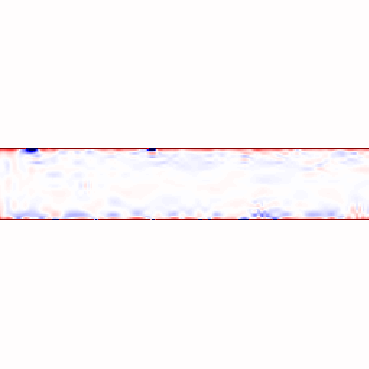}\\
\end{tabular}\label{fig:pA1_img_G}
}
\subfloat[]{
\begin{tabular}[b]{c | c | c}%
\includegraphics[width=\tw\linewidth]{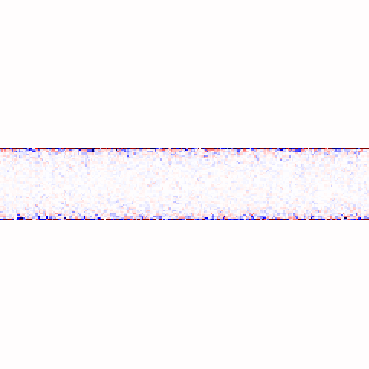} & 
\includegraphics[width=\tw\linewidth]{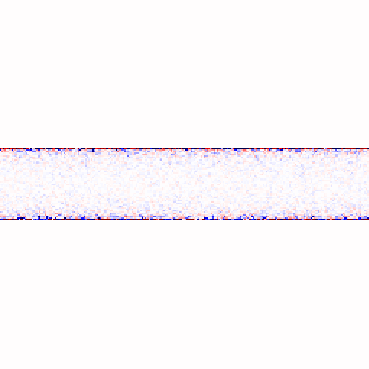} & 
\includegraphics[width=\tw\linewidth]{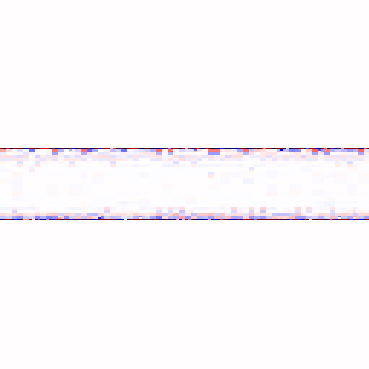}\\
\hline
\includegraphics[width=\tw\linewidth]{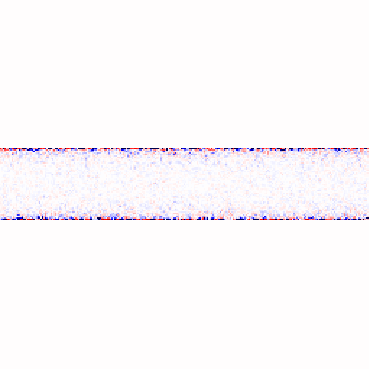} & 
\includegraphics[width=\tw\linewidth]{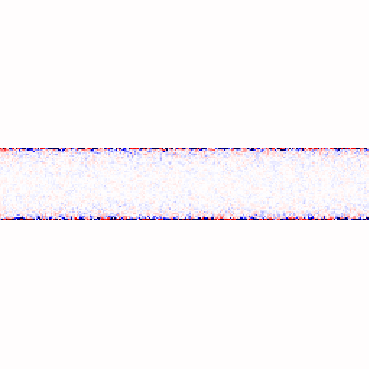} & 
\includegraphics[width=\tw\linewidth]{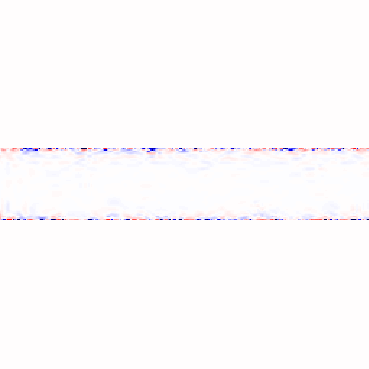}\\
\hline
\includegraphics[width=\tw\linewidth]{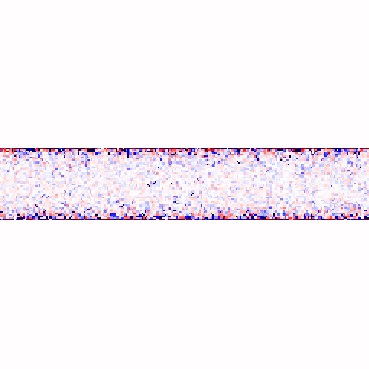} & 
\includegraphics[width=\tw\linewidth]{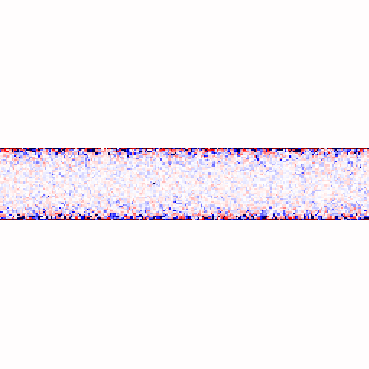} & 
\includegraphics[width=\tw\linewidth]{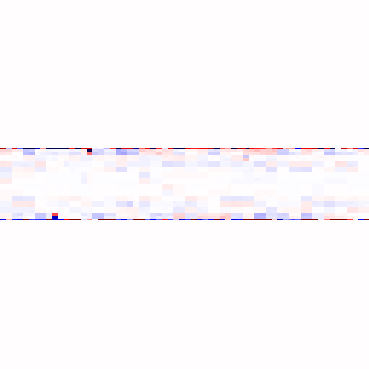}\\
\hline
\includegraphics[width=\tw\linewidth]{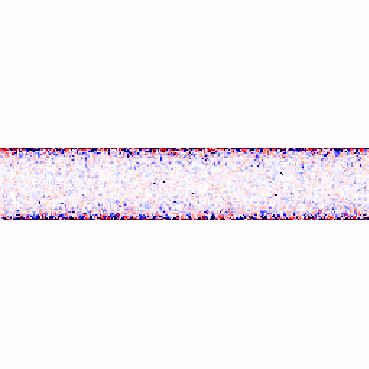} & 
\includegraphics[width=\tw\linewidth]{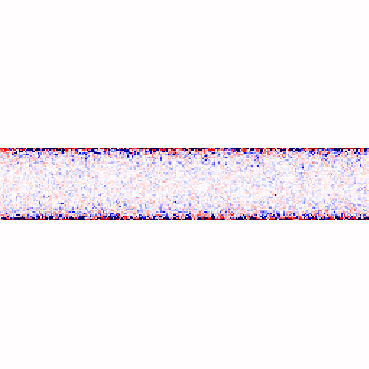} & 
\includegraphics[width=\tw\linewidth]{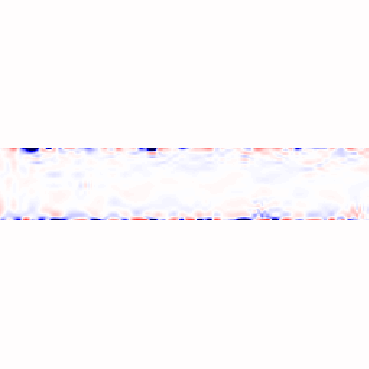}\\
\end{tabular}\label{fig:pA1_img_H}
}
\caption{Recovered images from orthogonal (a,b,c,d) and longitudinal (e,f,g,h) slices of Hagen-Poiseuille flow.
The columns in (a,b,e,f) represent the true image followed by the CS, CSDEB and stOMP reconstruction. The true image is instead omitted in (c,d,g,h). 
The reconstructions in (a,e) and (b,f) are obtained with 10\% and 30\% $k$-space noise, respectively. Images in the first and third rows are produced using Haar wavelets, those in the second and forth row using Db8 wavelets.  
The reconstruction artifacts~\eqref{equ:rec_artifacts} with respect to the true underlying image and average reconstruction are shown in (c,g) and (d,h), respectively. The first two rows in (c,d,g,h) have 10\% noise and the second two rows have 30\% noise. The first and third row have Haar wavelets, while the second and forth row use Db8 wavelets.}\label{fig:pA1A2_img}
\end{figure}

Figure~\ref{fig:pA1A2_img} shows the ability of all recovery algorithms to accurately reconstruct both the image density and velocity profiles under baseline conditions. Reconstructions appear to be robust (small difference with average reconstruction) with errors concentrated at the edge of the cylinder.

\begin{figure}[ht!]
\centering
\subfloat[]{\includegraphics[width=0.18\textwidth]{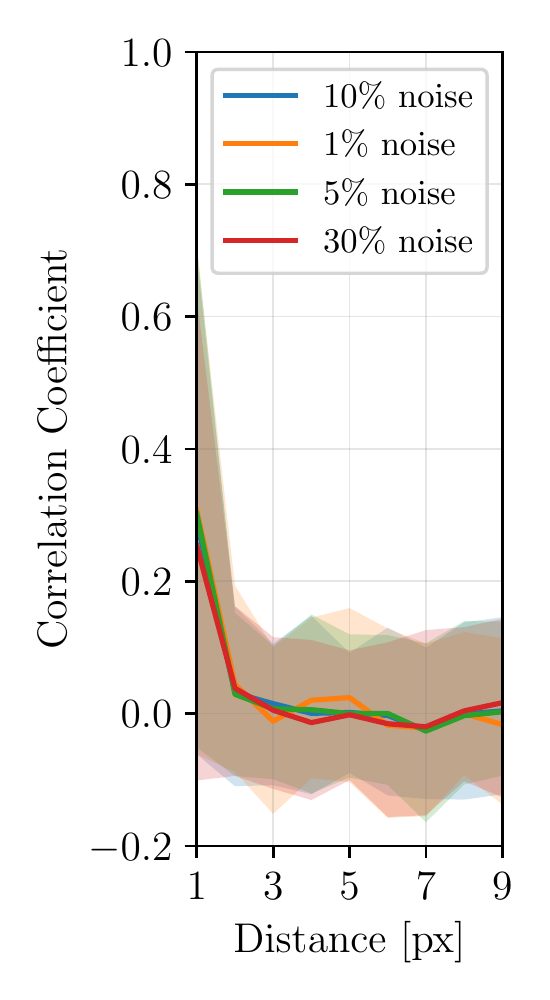}}$\,$
\subfloat[]{\includegraphics[width=0.18\textwidth]{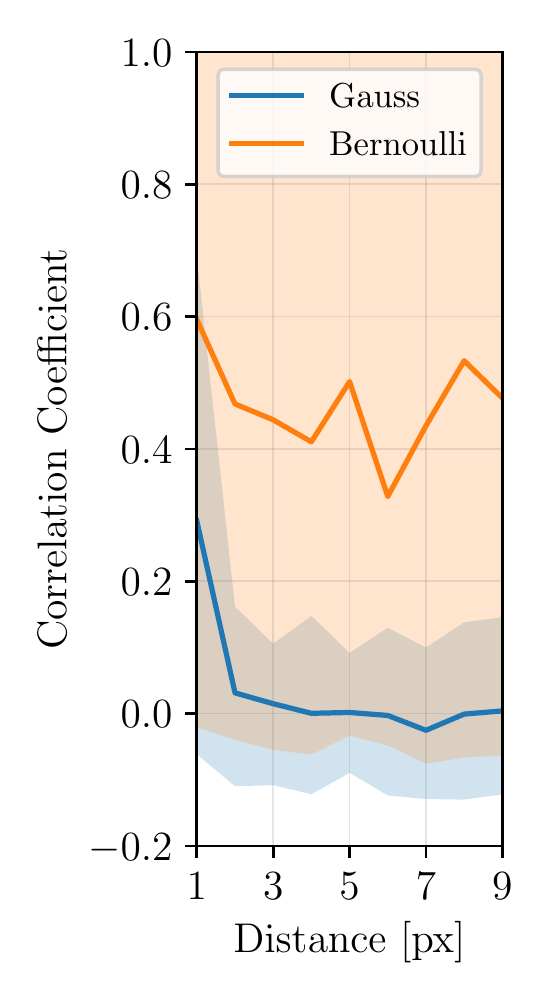}}$\,$
\subfloat[]{\includegraphics[width=0.18\textwidth]{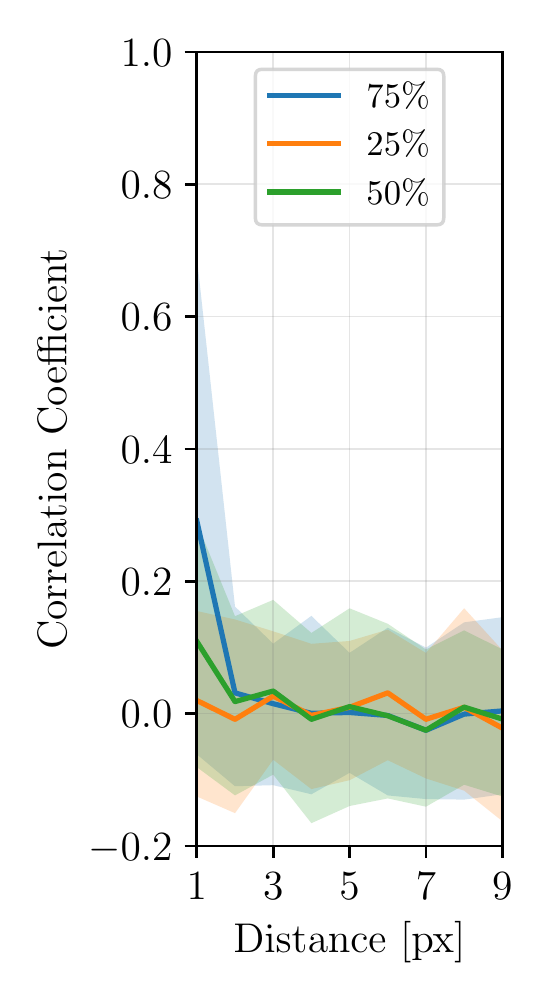}}$\,$
\subfloat[]{\includegraphics[width=0.18\textwidth]{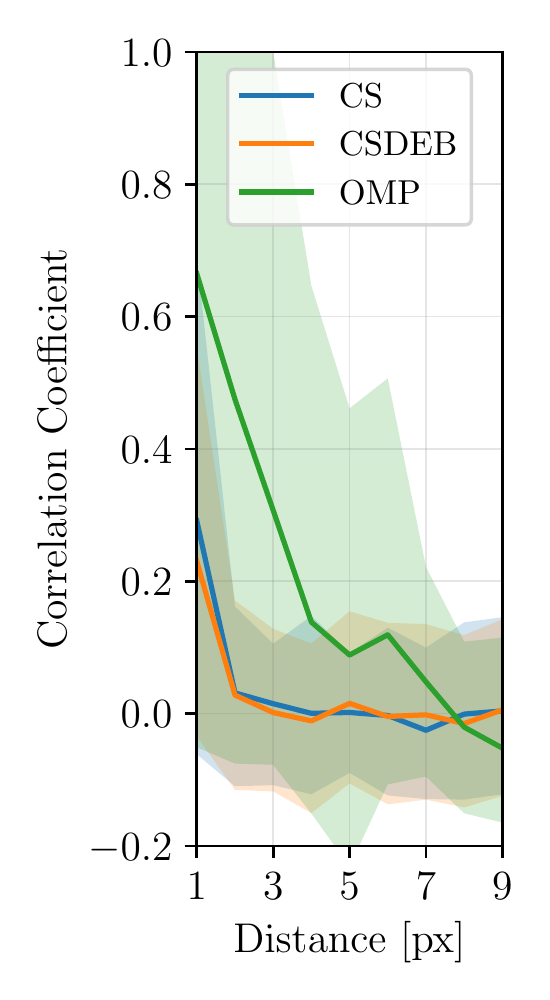}}$\,$
\subfloat[]{\includegraphics[width=0.18\textwidth]{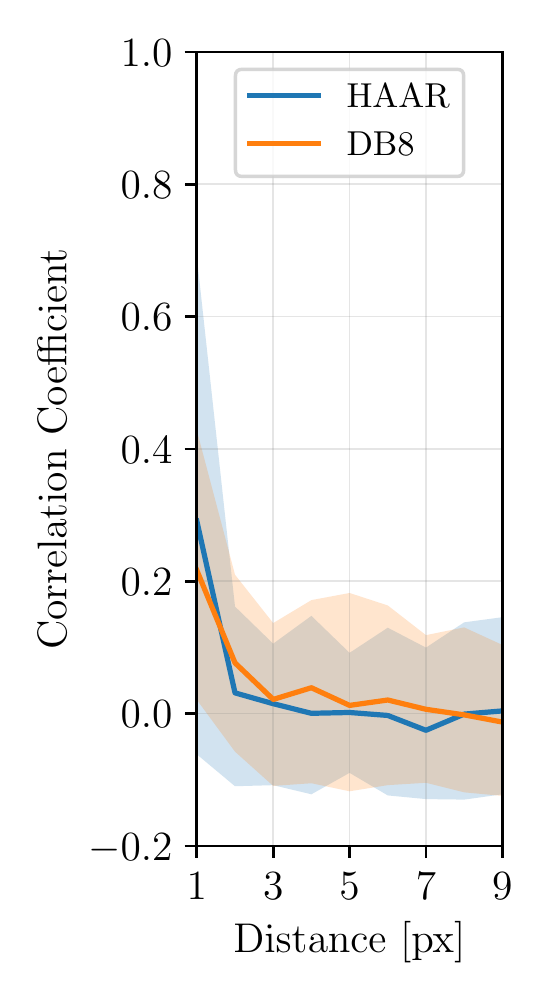}}
\caption{Hagen-Poiseuille flow test case. Correlation coefficients obtained by perturbing recovery parameters from baseline conditions. The correlations are only shown for the out-of-plane velocity ($k=3$), and are very similar to those obtained for the image density ($k=0$).}\label{fig:pA1A2_corr}
\end{figure}

\begin{figure}[ht!]
\centering
\includegraphics{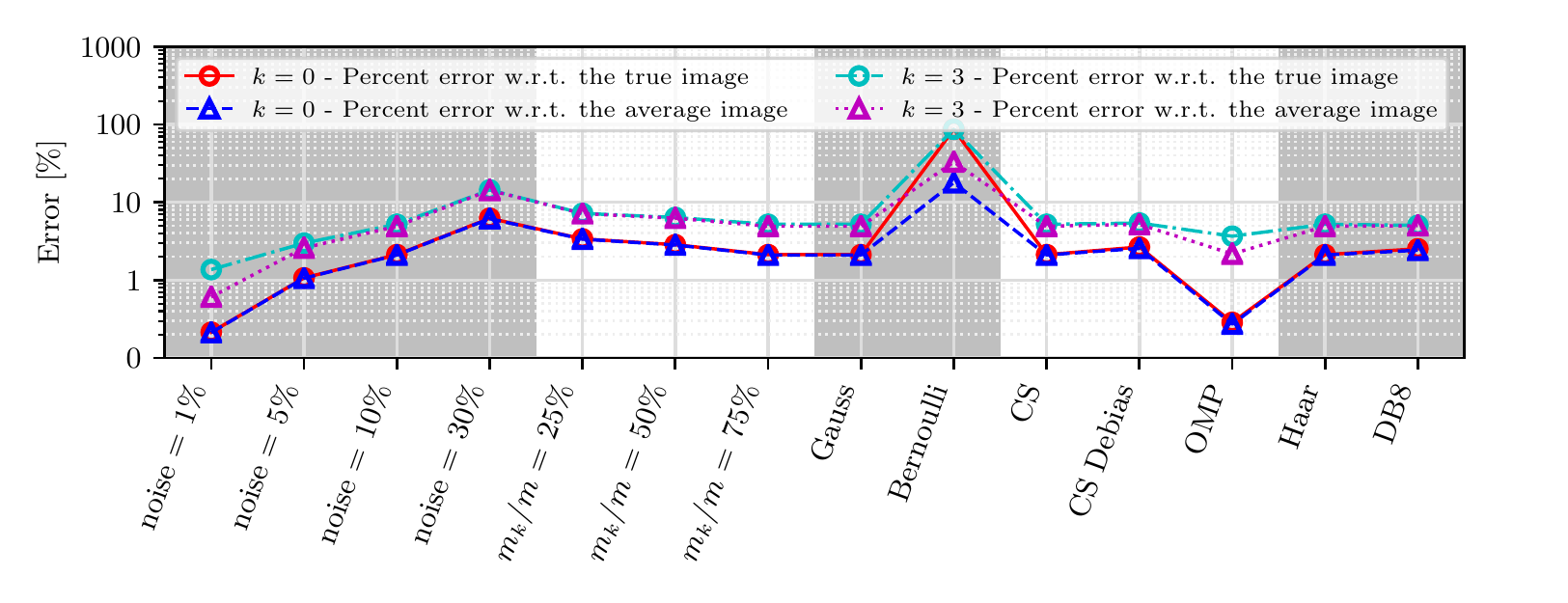}
\caption{Summary of percent errors $\text{PE}_{k,s},\,k\in\{0,3\},\,s\in\{t,a\}$ for Hagen-Poiseuille flow.}\label{fig:pA1A2_perc_error}
\end{figure}

The correlation decreases rapidly with the distance between pixels, and is approximately zero, on average, for distances greater than two-pixels, as shown in Figure~\ref{fig:pA1A2_corr}.
The selection of 50 different pairs of pixels to compute the signed correlation, induces a variability whose amplitude remains practically constant for a distance greater than two pixels. This variability is attributed to the finite sample size.
The only exception is observed for the stOMP recovery algorithm, where correlations were found higher than for the other methods, despite the lower PE in the velocity components and image density (Figure~\ref{fig:pA1A2_perc_error}).
Changes in the undersampling ratio affect the correlations at one pixel distance, but correlations at larger distances remains unaltered, as shown in Figure~\ref{fig:pA1A2_corr}.
Bernoulli undersampling does not seem to affect the correlations for the image intensity, but significantly affects the correlation of the noise in the velocity reconstructions. This occurs when Bernoulli undersampling discards some of the low frequencies that are important to represent the smooth velocity profile. 

A summary of the percent errors in the Hagen-Poiseuille flow test case is shown in Figure~\ref{fig:pA1A2_perc_error}, showing smaller errors in the image intensity ($k=0$) than the axial velocity component ($k=3$).
The reconstruction errors increase with the $k$-space noise, but reduce with an increasing undersampling ratio, as a variable density Gaussian mask preserves the relevant frequency information even for 75\% undersampling ratio. Conversely, Bernoulli undersampling removes relevant signal content at low frequencies, inducing large reconstruction variance and bias.
All recovery methods seem to perform equally well, with stOMP producing smaller percent errors, particularly when reconstructing a uniform image density.
Finally, no significant difference is observed when using Fourier-Haar or Fourier-Db8 operators.


%
%

\subsection{Simulated aortic flow}\label{sec:simAorticFlow}

We analyze a simulated velocity field in the thoracic aorta. The anatomic model has been created in SimVascular~\cite{updegrove2017simvascular} from the 4D flow MRI scan of a 70-years-old volunteer.
A Dirichlet boundary condition is applied at the aortic inlet, matching the flow rate at this location measured from 4D flow MRI images. A resistance boundary condition is applied to the model outlets, evaluated from the ratio between a cardiac output equal to 3.73 L/min and an average aortic pressure of 93 mmHg.
The total resistance is distributed in parallel between the outlets using Murray's law~\cite{sherman1981connecting} with an exponent of 2.0, considering an uniform resistance across branch outlets.
A finite element discretization with few hundred thousand tetrahedral elements and an edge size of roughly 0.7 mm was used to simulate ten heart cycles using a time step of 7.296$\times 10^{-4}$ s, until the resulting pressure, flow rate and volume time histories were found to be periodic.
After identifying the systolic flow peak in the last simulated heart cycle, the descending aorta was sliced orthogonal to its centerline path and projected onto a 256$\times$256 acquisition lattice.

\begin{figure}[ht!]
\centering
\setlength{\tabcolsep}{0pt}
\renewcommand{\arraystretch}{0}
\subfloat[]{
\begin{tabular}[b]{c | c | c | c}%
\includegraphics[width=0.06\linewidth]{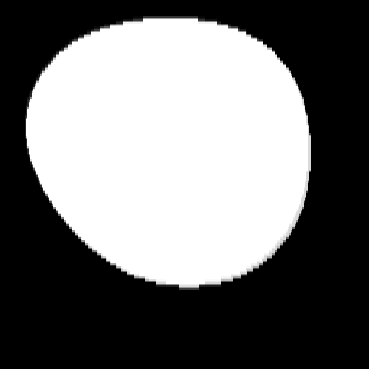} & 
\includegraphics[width=0.06\linewidth]{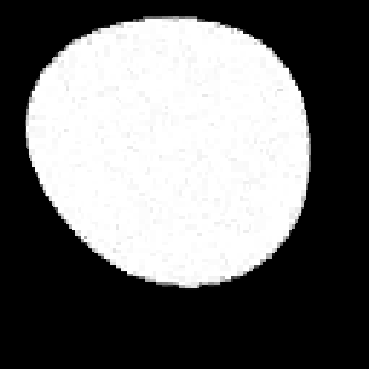} & 
\includegraphics[width=0.06\linewidth]{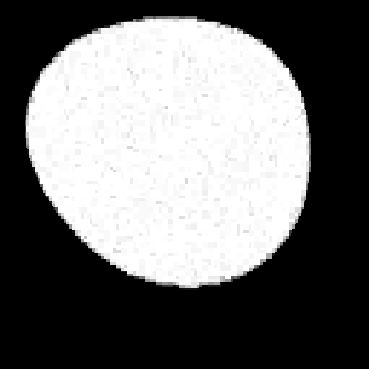} & 
\includegraphics[width=0.06\linewidth]{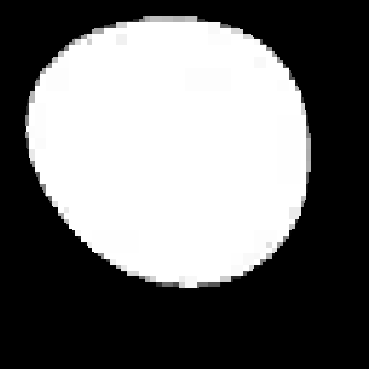}\\
\hline
\includegraphics[width=0.06\linewidth]{imgs/ao_id/true_k0.png} & 
\includegraphics[width=0.06\linewidth]{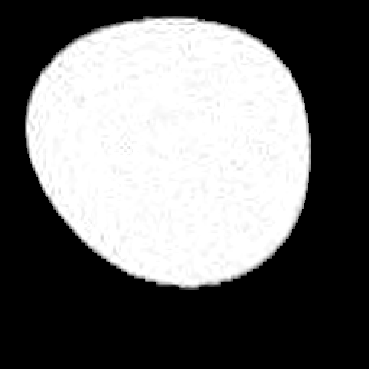} & 
\includegraphics[width=0.06\linewidth]{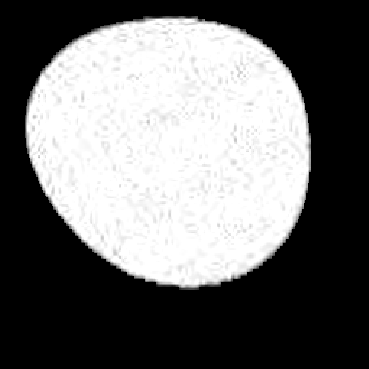} & 
\includegraphics[width=0.06\linewidth]{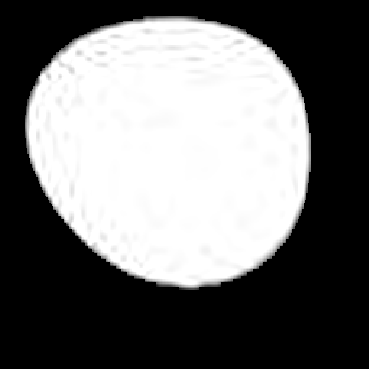}\\
\hline
\includegraphics[width=0.06\linewidth]{imgs/ao_id/true_k0.png} & 
\includegraphics[width=0.06\linewidth]{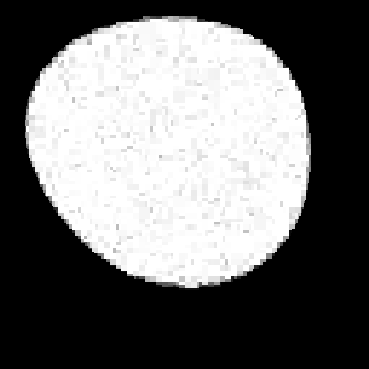} & 
\includegraphics[width=0.06\linewidth]{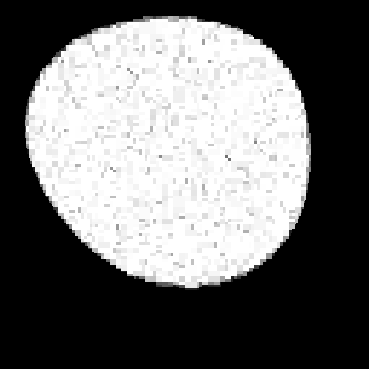} & 
\includegraphics[width=0.06\linewidth]{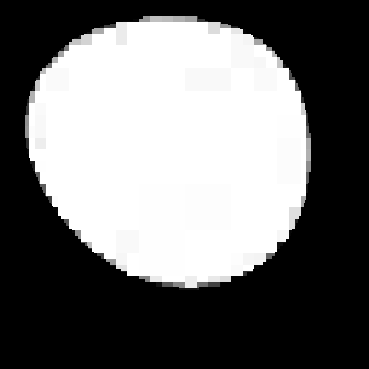}\\
\hline
\includegraphics[width=0.06\linewidth]{imgs/ao_id/true_k0.png} & 
\includegraphics[width=0.06\linewidth]{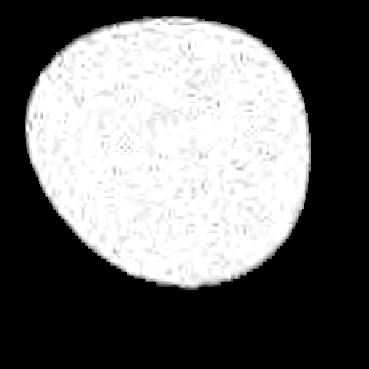} & 
\includegraphics[width=0.06\linewidth]{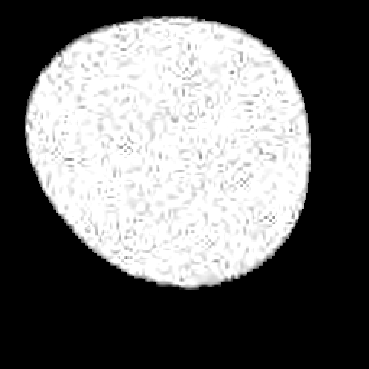} & 
\includegraphics[width=0.06\linewidth]{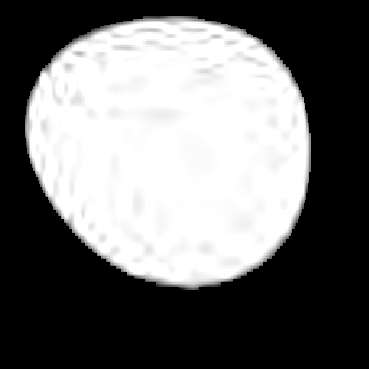}\\
\end{tabular}\label{fig:aortaideal_img_A}
}
\subfloat[]{
\begin{tabular}[b]{c | c | c | c}%
\includegraphics[width=0.06\linewidth]{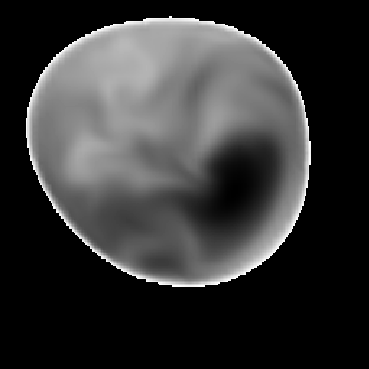} & 
\includegraphics[width=0.06\linewidth]{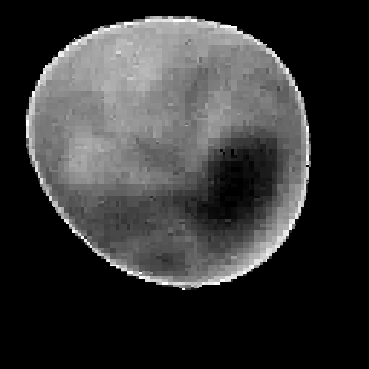} & 
\includegraphics[width=0.06\linewidth]{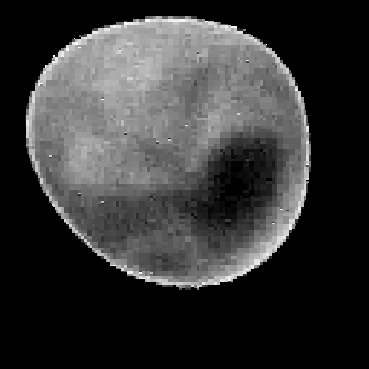} & 
\includegraphics[width=0.06\linewidth]{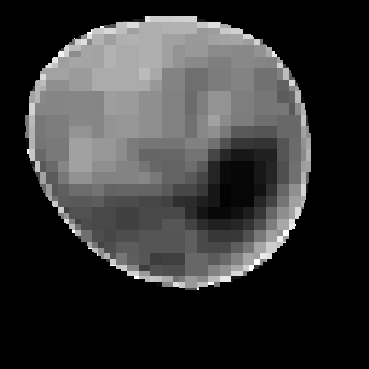}\\
\hline
\includegraphics[width=0.06\linewidth]{imgs/ao_id/true_k1.png} & 
\includegraphics[width=0.06\linewidth]{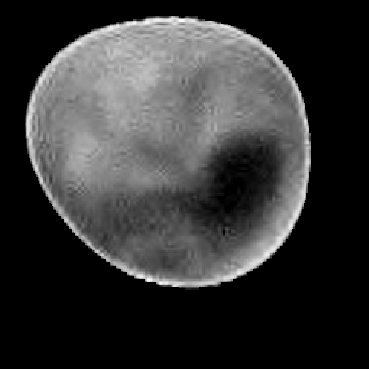} & 
\includegraphics[width=0.06\linewidth]{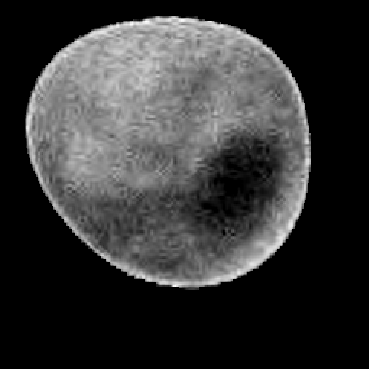} & 
\includegraphics[width=0.06\linewidth]{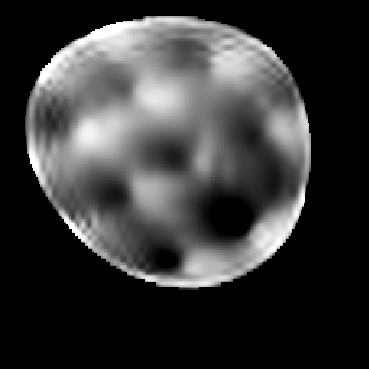}\\
\hline
\includegraphics[width=0.06\linewidth]{imgs/ao_id/true_k1.png} & 
\includegraphics[width=0.06\linewidth]{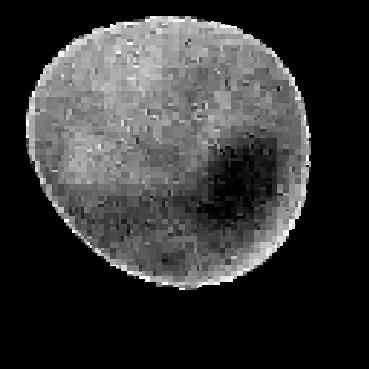} & 
\includegraphics[width=0.06\linewidth]{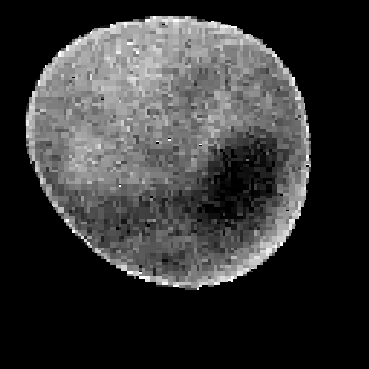} & 
\includegraphics[width=0.06\linewidth]{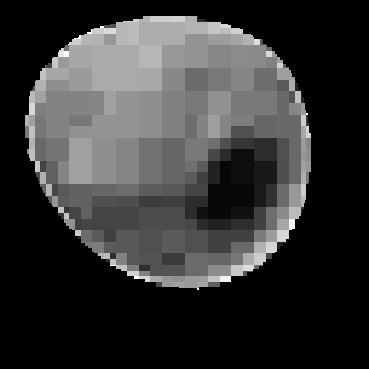}\\
\hline
\includegraphics[width=0.06\linewidth]{imgs/ao_id/true_k1.png} & 
\includegraphics[width=0.06\linewidth]{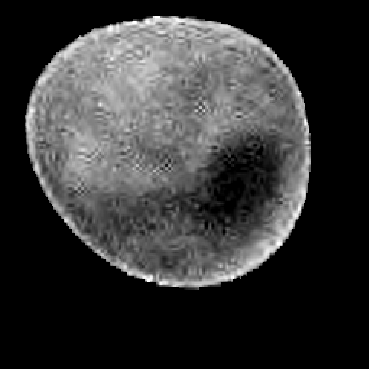} & 
\includegraphics[width=0.06\linewidth]{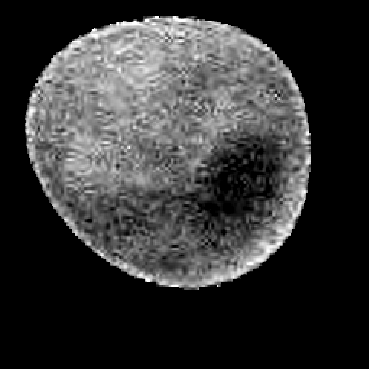} & 
\includegraphics[width=0.06\linewidth]{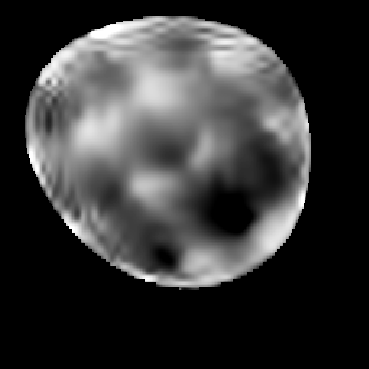}\\
\end{tabular}\label{fig:aortaideal_img_B}
}
\subfloat[]{
\begin{tabular}[b]{c | c | c | c}%
\includegraphics[width=0.06\linewidth]{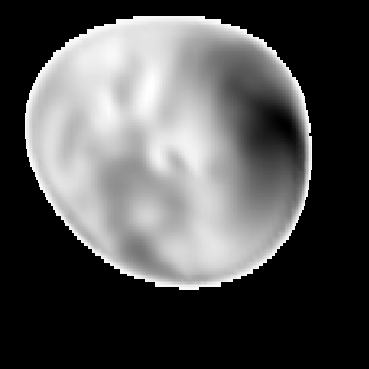} & 
\includegraphics[width=0.06\linewidth]{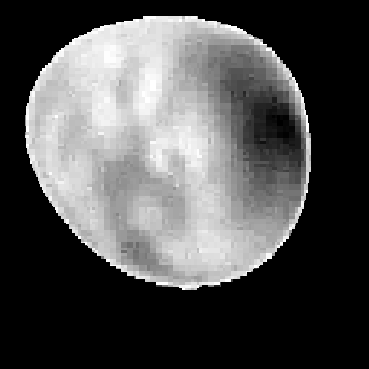} & 
\includegraphics[width=0.06\linewidth]{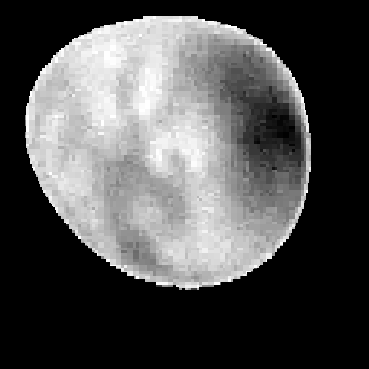} & 
\includegraphics[width=0.06\linewidth]{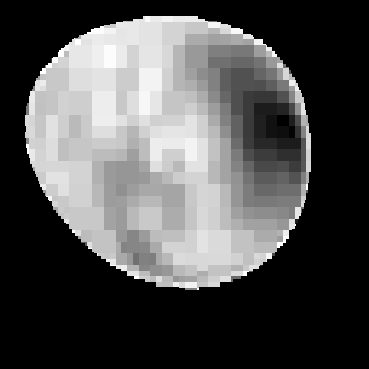}\\
\hline
\includegraphics[width=0.06\linewidth]{imgs/ao_id/true_k2.png} & 
\includegraphics[width=0.06\linewidth]{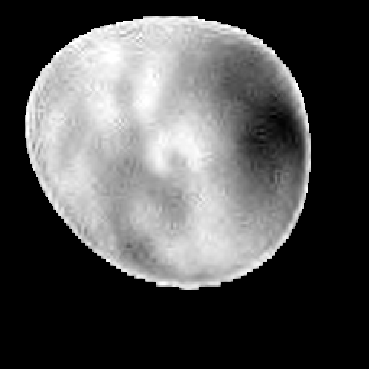} & 
\includegraphics[width=0.06\linewidth]{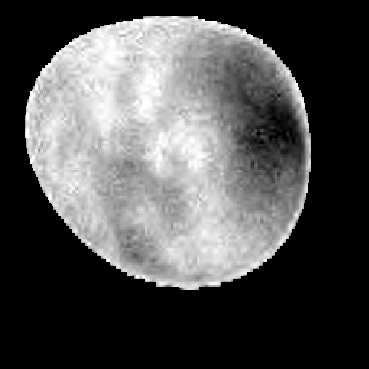} & 
\includegraphics[width=0.06\linewidth]{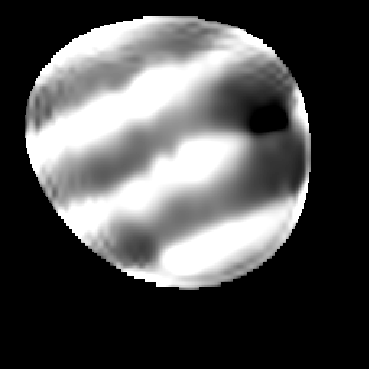}\\
\hline
\includegraphics[width=0.06\linewidth]{imgs/ao_id/true_k2.png} & 
\includegraphics[width=0.06\linewidth]{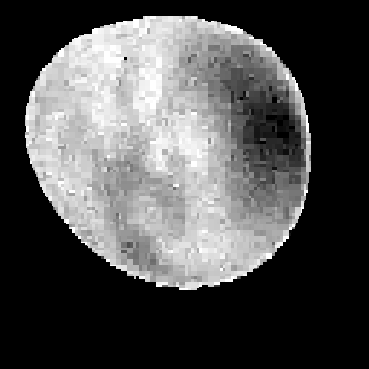} & 
\includegraphics[width=0.06\linewidth]{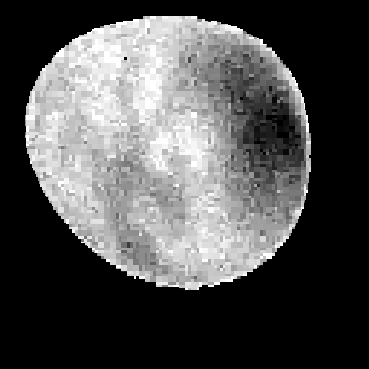} & 
\includegraphics[width=0.06\linewidth]{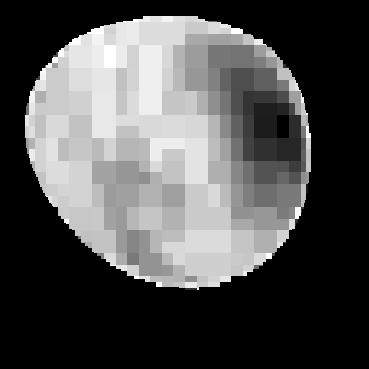}\\
\hline
\includegraphics[width=0.06\linewidth]{imgs/ao_id/true_k2.png} & 
\includegraphics[width=0.06\linewidth]{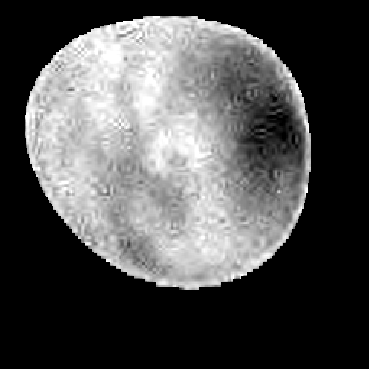} & 
\includegraphics[width=0.06\linewidth]{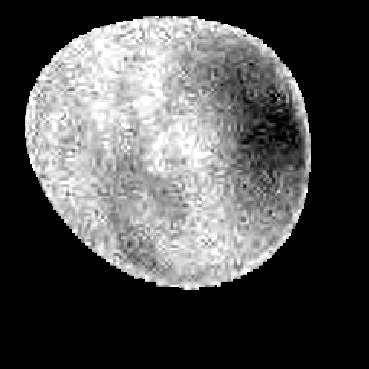} & 
\includegraphics[width=0.06\linewidth]{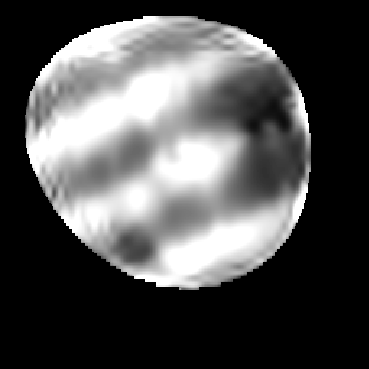}\\
\end{tabular}\label{fig:aortaideal_img_C}
}
\subfloat[]{
\begin{tabular}[b]{c | c | c | c}%
\includegraphics[width=0.06\linewidth]{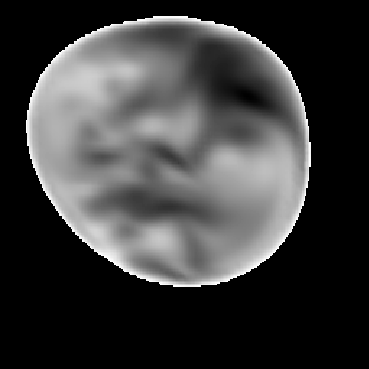} & 
\includegraphics[width=0.06\linewidth]{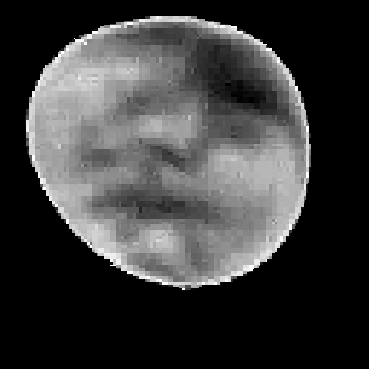} & 
\includegraphics[width=0.06\linewidth]{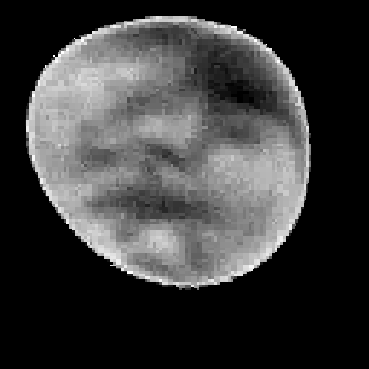} & 
\includegraphics[width=0.06\linewidth]{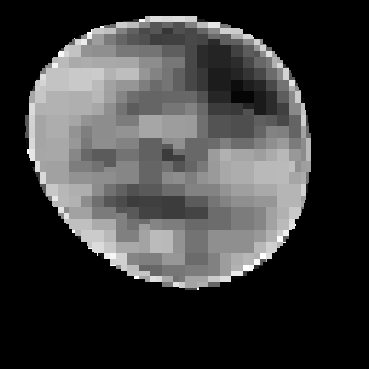}\\
\hline
\includegraphics[width=0.06\linewidth]{imgs/ao_id/true_k3.png} & 
\includegraphics[width=0.06\linewidth]{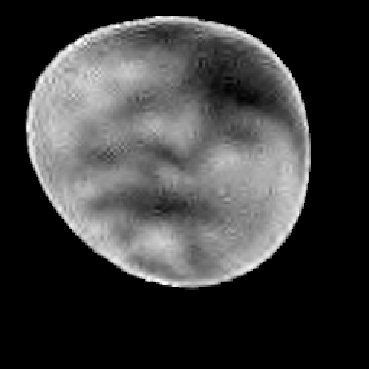} & 
\includegraphics[width=0.06\linewidth]{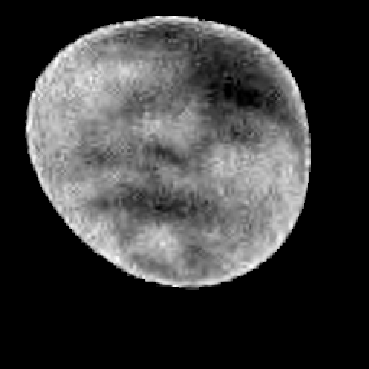} & 
\includegraphics[width=0.06\linewidth]{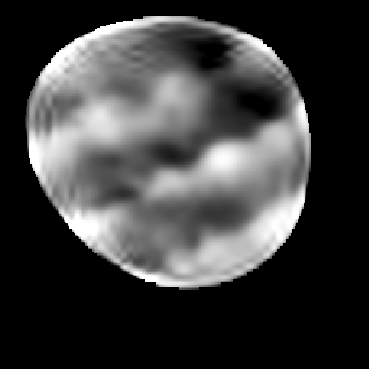}\\
\hline
\includegraphics[width=0.06\linewidth]{imgs/ao_id/true_k3.png} & 
\includegraphics[width=0.06\linewidth]{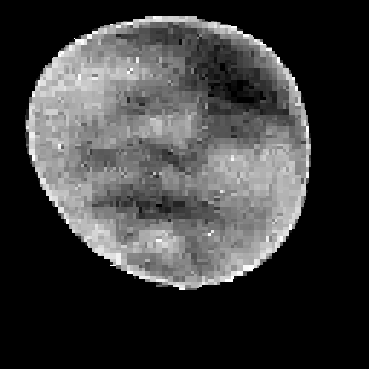} & 
\includegraphics[width=0.06\linewidth]{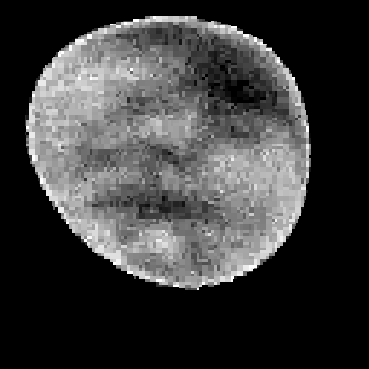} & 
\includegraphics[width=0.06\linewidth]{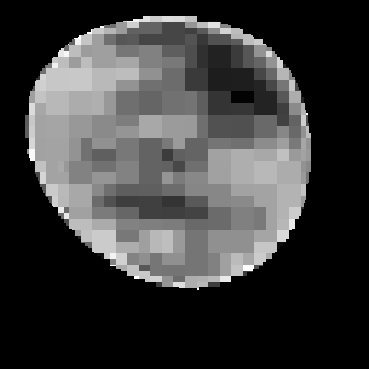}\\
\hline
\includegraphics[width=0.06\linewidth]{imgs/ao_id/true_k3.png} & 
\includegraphics[width=0.06\linewidth]{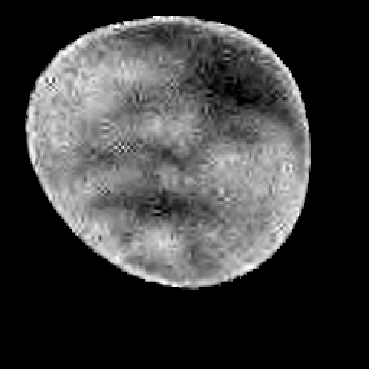} & 
\includegraphics[width=0.06\linewidth]{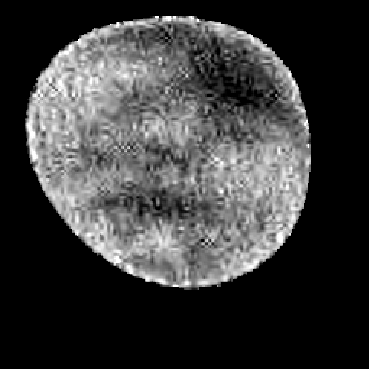} & 
\includegraphics[width=0.06\linewidth]{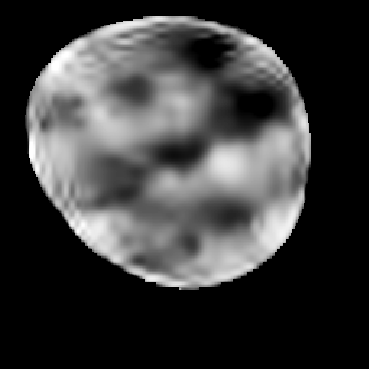}\\
\end{tabular}\label{fig:aortaideal_img_D}
}

\subfloat[]{
\begin{tabular}[b]{c | c | c}%
\includegraphics[width=0.06\linewidth]{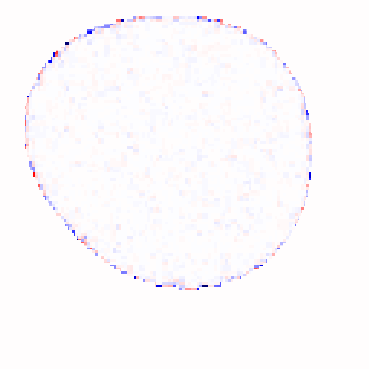} & 
\includegraphics[width=0.06\linewidth]{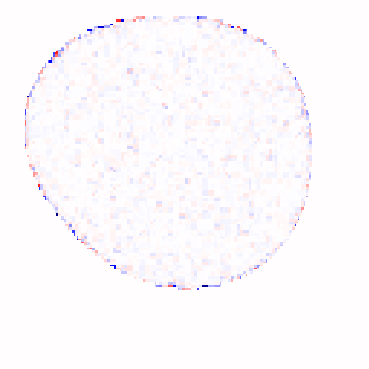} & 
\includegraphics[width=0.06\linewidth]{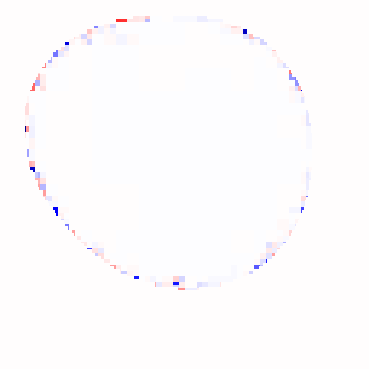}\\
\hline
\includegraphics[width=0.06\linewidth]{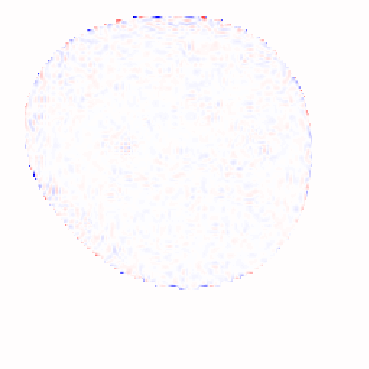} & 
\includegraphics[width=0.06\linewidth]{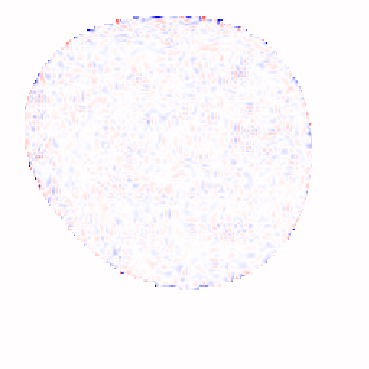} & 
\includegraphics[width=0.06\linewidth]{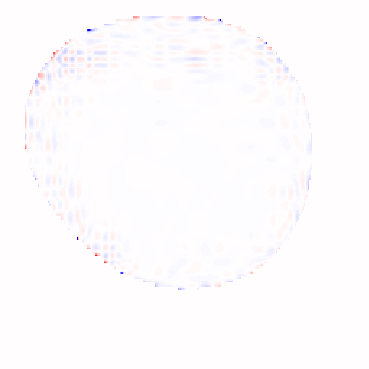}\\
\hline
\includegraphics[width=0.06\linewidth]{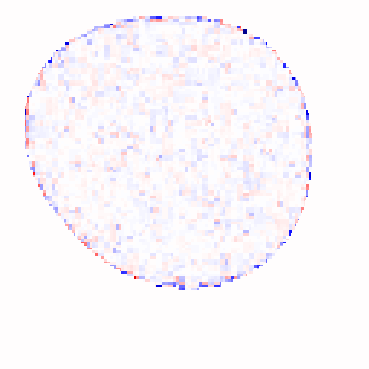} & 
\includegraphics[width=0.06\linewidth]{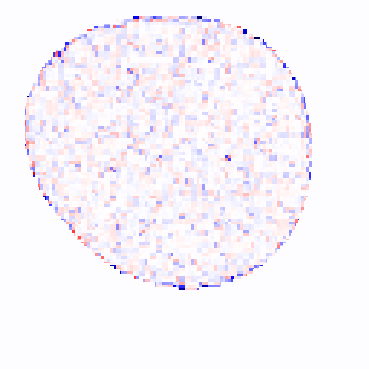} & 
\includegraphics[width=0.06\linewidth]{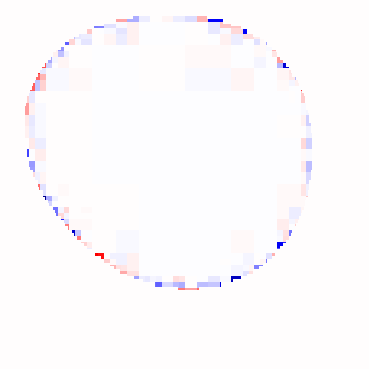}\\
\hline
\includegraphics[width=0.06\linewidth]{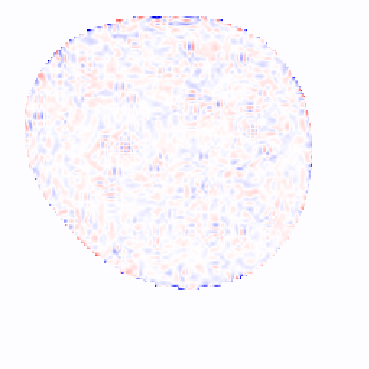} & 
\includegraphics[width=0.06\linewidth]{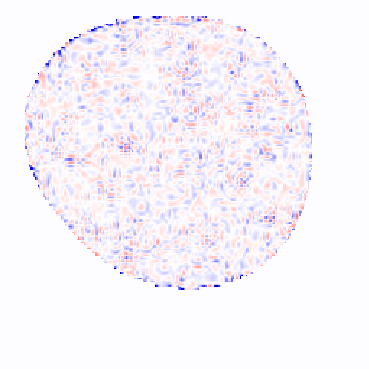} & 
\includegraphics[width=0.06\linewidth]{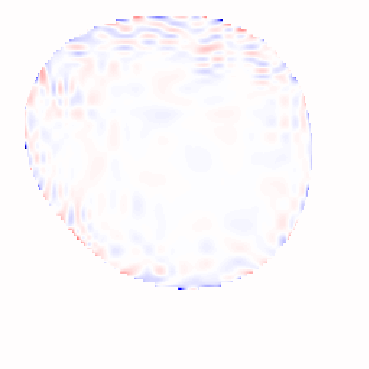}\\
\end{tabular}\label{fig:aortaideal_img_E}
}
\subfloat[]{
\begin{tabular}[b]{c | c | c}%
\includegraphics[width=0.06\linewidth]{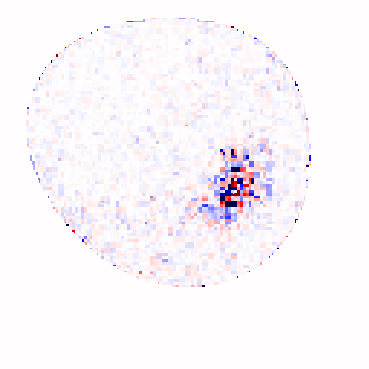} & 
\includegraphics[width=0.06\linewidth]{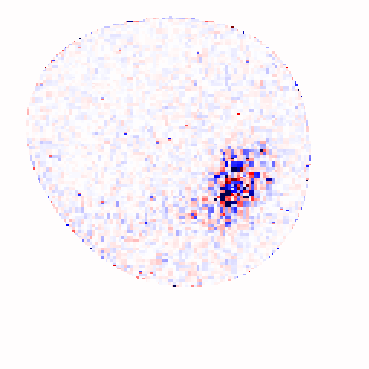} & 
\includegraphics[width=0.06\linewidth]{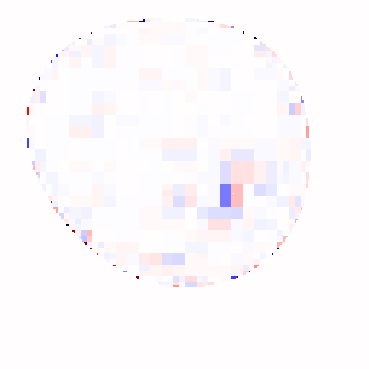}\\
\hline
\includegraphics[width=0.06\linewidth]{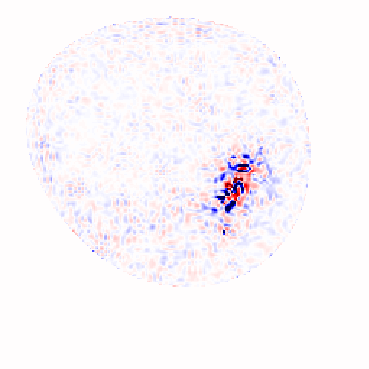} & 
\includegraphics[width=0.06\linewidth]{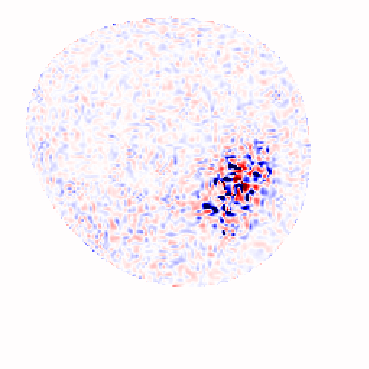} & 
\includegraphics[width=0.06\linewidth]{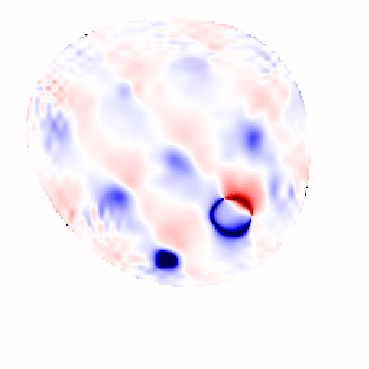}\\
\hline
\includegraphics[width=0.06\linewidth]{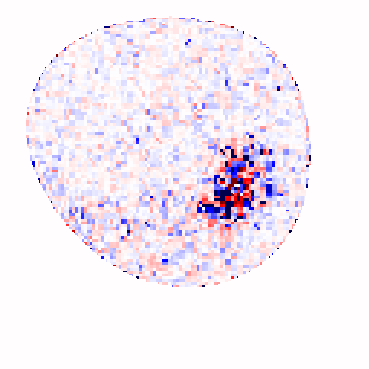} & 
\includegraphics[width=0.06\linewidth]{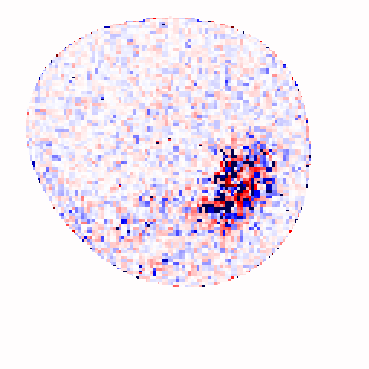} & 
\includegraphics[width=0.06\linewidth]{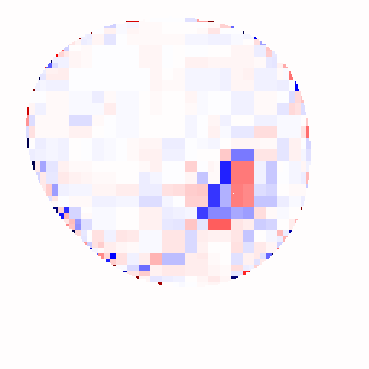}\\
\hline
\includegraphics[width=0.06\linewidth]{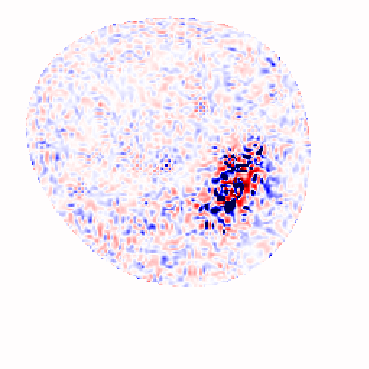} & 
\includegraphics[width=0.06\linewidth]{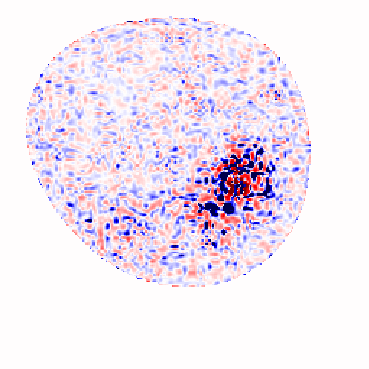} & 
\includegraphics[width=0.06\linewidth]{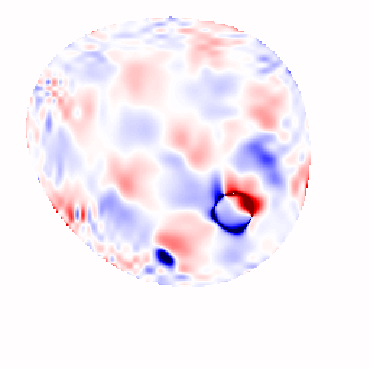}\\
\end{tabular}\label{fig:aortaideal_img_F}
}
\subfloat[]{
\begin{tabular}[b]{c | c | c}%
\includegraphics[width=0.06\linewidth]{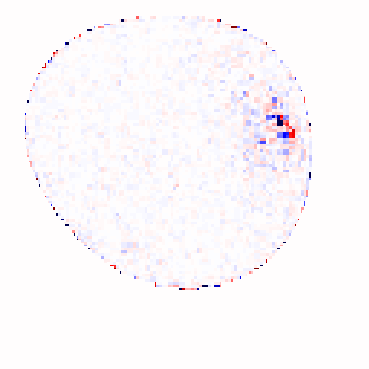} & 
\includegraphics[width=0.06\linewidth]{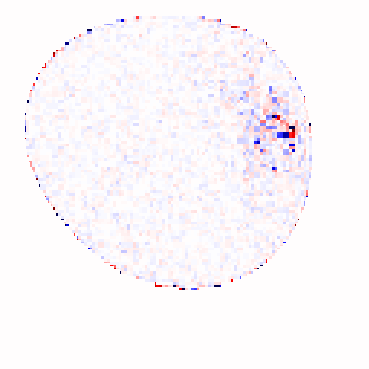} & 
\includegraphics[width=0.06\linewidth]{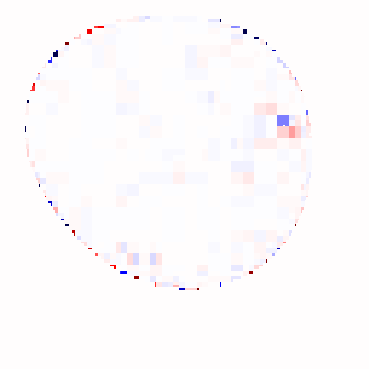}\\
\hline
\includegraphics[width=0.06\linewidth]{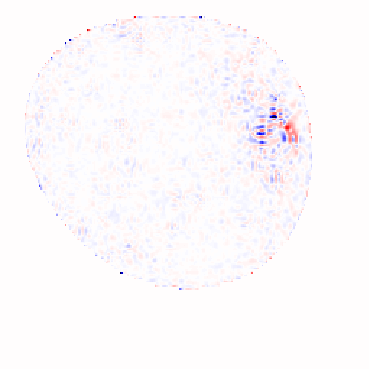} & 
\includegraphics[width=0.06\linewidth]{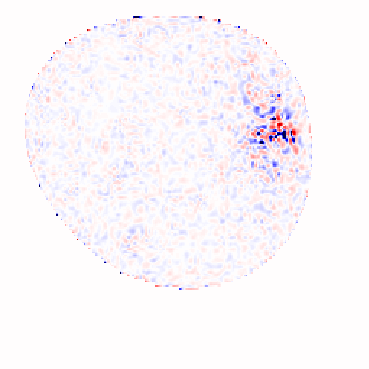} & 
\includegraphics[width=0.06\linewidth]{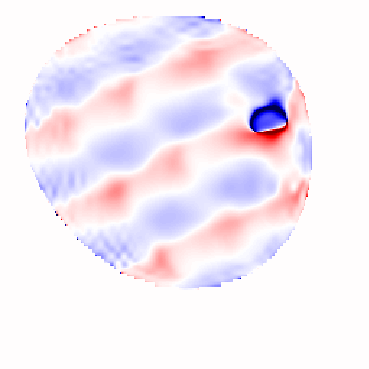}\\
\hline
\includegraphics[width=0.06\linewidth]{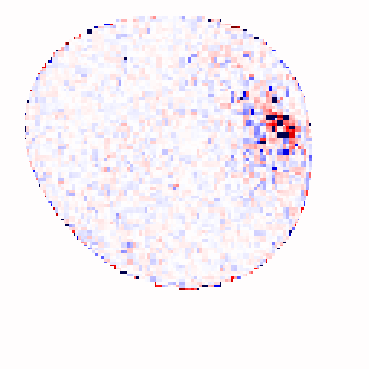} & 
\includegraphics[width=0.06\linewidth]{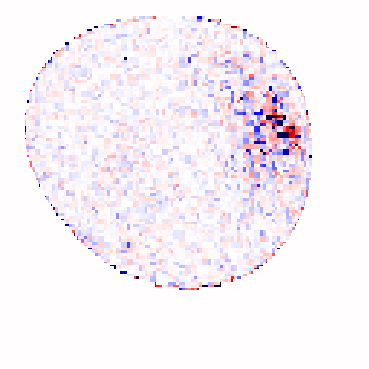} & 
\includegraphics[width=0.06\linewidth]{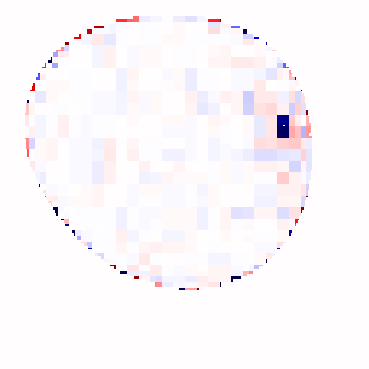}\\
\hline
\includegraphics[width=0.06\linewidth]{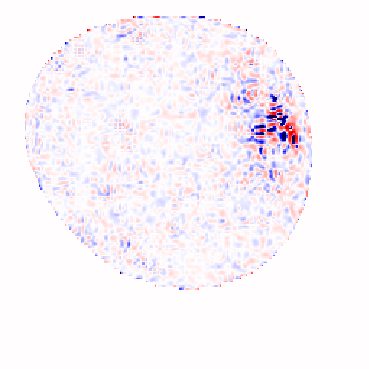} & 
\includegraphics[width=0.06\linewidth]{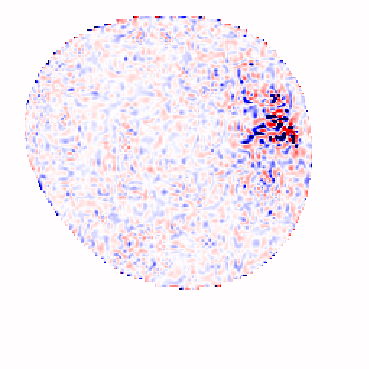} & 
\includegraphics[width=0.06\linewidth]{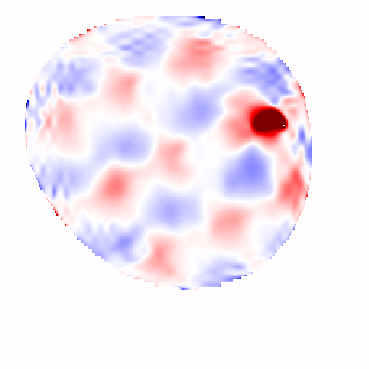}\\
\end{tabular}\label{fig:aortaideal_img_G}
}
\subfloat[]{
\begin{tabular}[b]{c | c | c}%
\includegraphics[width=0.06\linewidth]{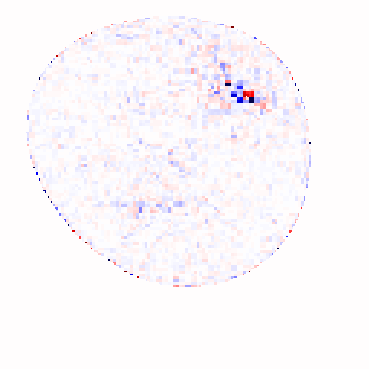} & 
\includegraphics[width=0.06\linewidth]{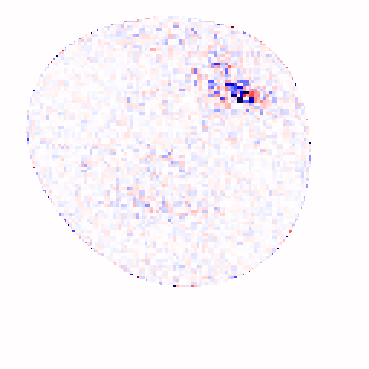} & 
\includegraphics[width=0.06\linewidth]{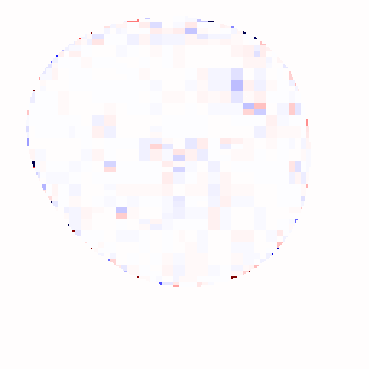}\\
\hline
\includegraphics[width=0.06\linewidth]{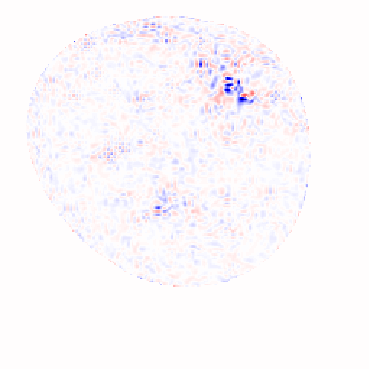} & 
\includegraphics[width=0.06\linewidth]{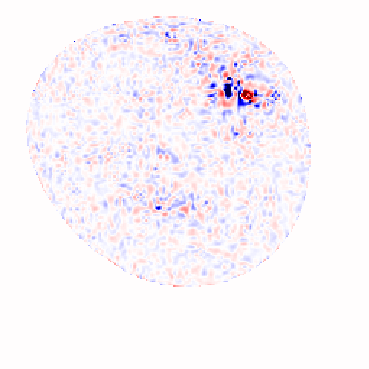} & 
\includegraphics[width=0.06\linewidth]{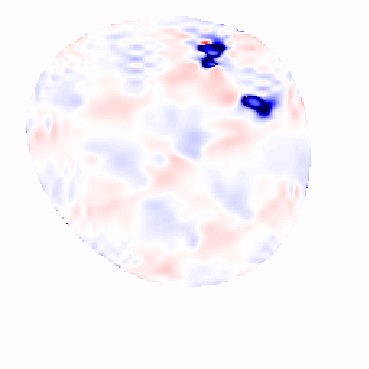}\\
\hline
\includegraphics[width=0.06\linewidth]{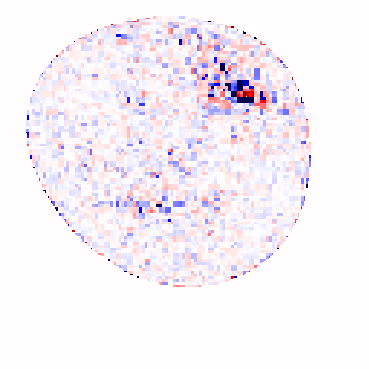} & 
\includegraphics[width=0.06\linewidth]{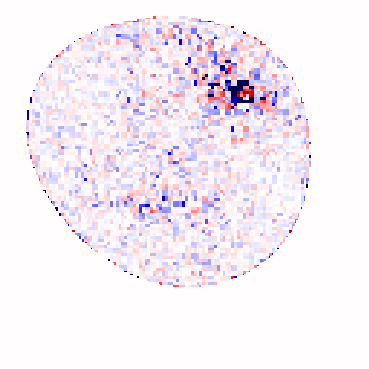} & 
\includegraphics[width=0.06\linewidth]{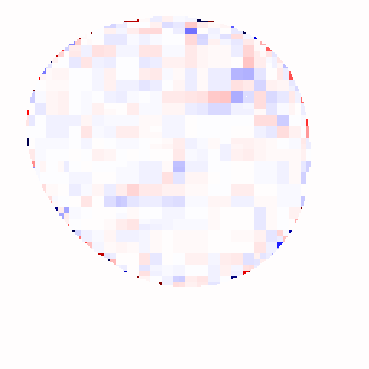}\\
\hline
\includegraphics[width=0.06\linewidth]{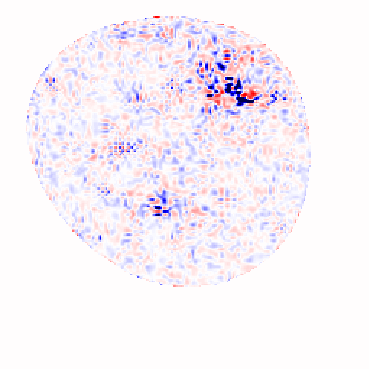} & 
\includegraphics[width=0.06\linewidth]{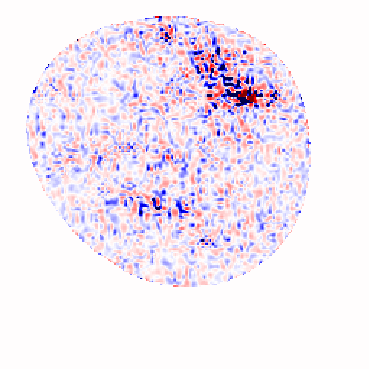} & 
\includegraphics[width=0.06\linewidth]{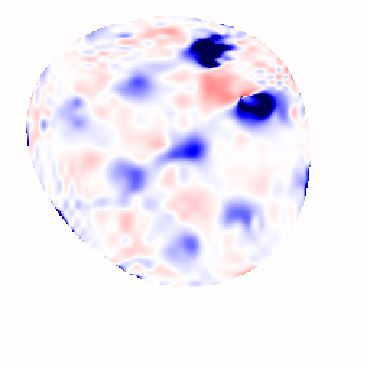}\\
\end{tabular}\label{fig:aortaideal_img_H}
}
\caption{Recovered images from simulated aortic flow with 95\% undersampling. Images in (a,b,c,d) refer to $k\in\{0,1,2,3\}$, respectively. The columns in (a,b,c,d) represent the true image followed by results generated using CS, CSDEB and stOMP. The true image is instead omitted in (e,f,g,h), showing the reconstruction errors~\eqref{equ:rec_artifacts}. In all images, the first two rows are obtained with 10\% $k$-space noise, whereas the noise is increased to 30\% in row 3 and 4. In addition, Haar wavelets were used in the first and third rows, Db8 wavelets in the second and forth rows.
Errors in (e,f,g,h) are computed with respect to the average reconstruction (i.e., $s=a$ in~\eqref{equ:rec_artifacts}), but identical error patterns were obtained using the true image instead.}\label{fig:aortaideal_img}
\end{figure}

For this case, three groups of benchmarks were performed including the same \emph{baseline} conditions (i.e. up to 75\% undersampling) analyzed for the Hagen-Poiseuille test case, \emph{large undersampling} with 85\%, 90\% and 95\% undersampling ratios and \emph{variable undersampling} considering randomness in the undersampling mask (i.e., a different mask for each of the 100 reconstructions).
Baseline and variable undersampling results were found to be practically the same as those discussed in the previous section, therefore Figure~\ref{fig:aortaideal_img} focuses on the \emph{large undersampling} regime, with 95\% undersampling (more than 16x $k$-space compression). Even though an increasingly coarse granularity is evident in the reconstructions, particularly for 30\% noise, the low-frequency information is still sufficient to capture relevant flow features.

\begin{figure}[ht!]
\centering
\setlength{\tabcolsep}{0pt}
\renewcommand{\arraystretch}{0}
\subfloat[]{
\includegraphics[]{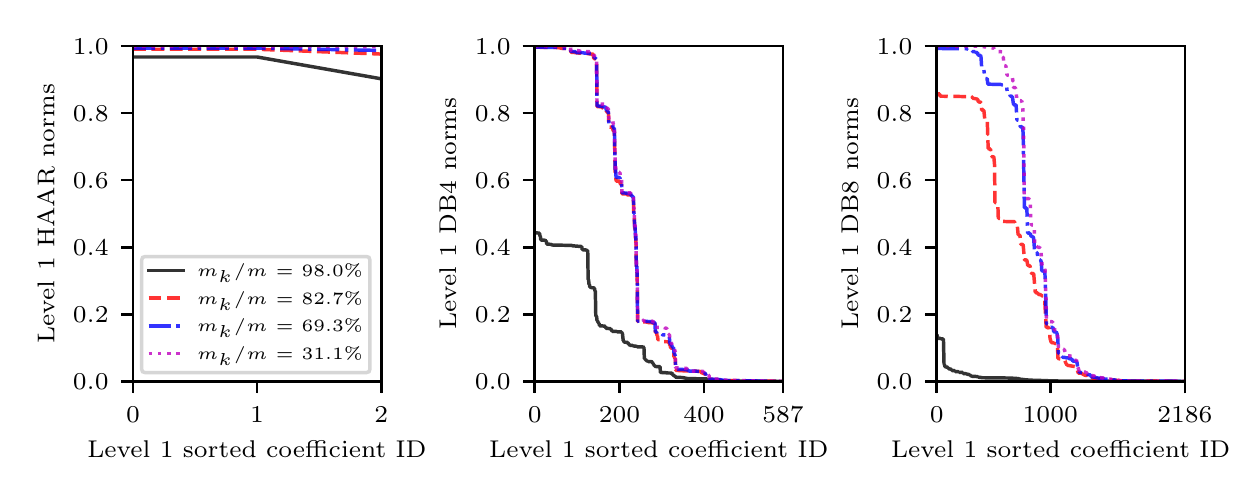}
\label{fig:stomp_problem_plot}
}
\subfloat[]{
\begin{tabular}[b]{c | c | c | c}%
\includegraphics[width=0.1\linewidth]{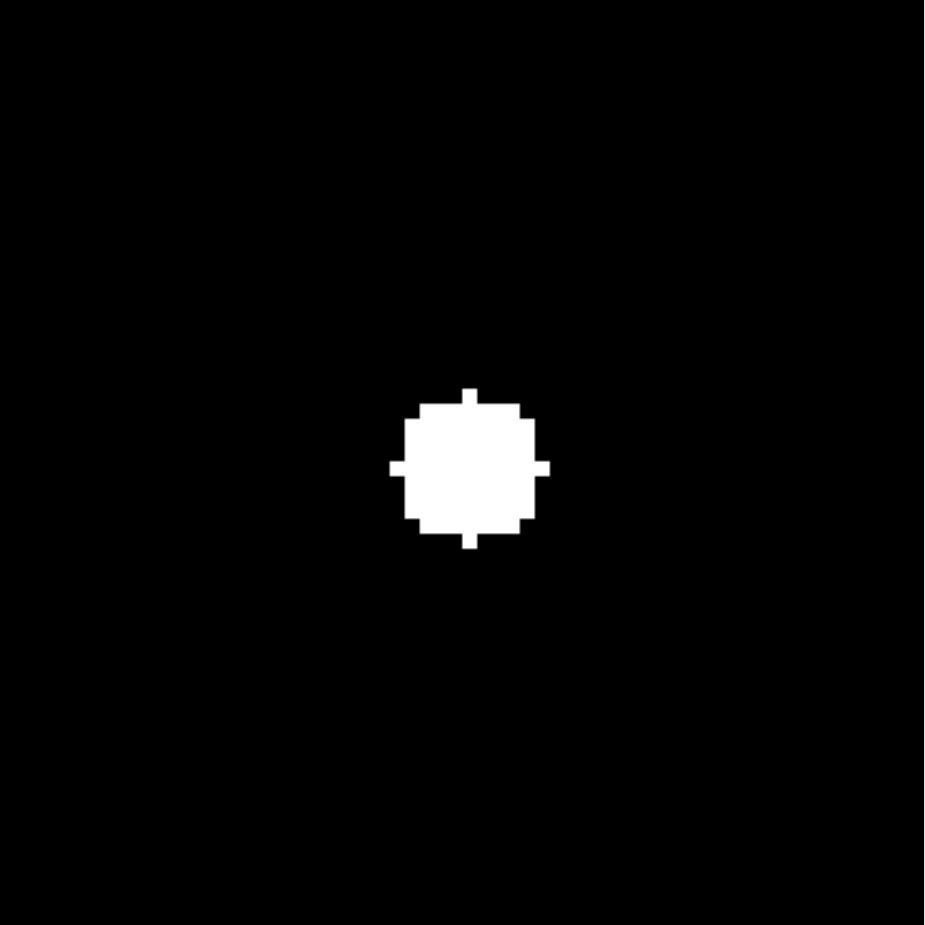} & 
\includegraphics[width=0.1\linewidth]{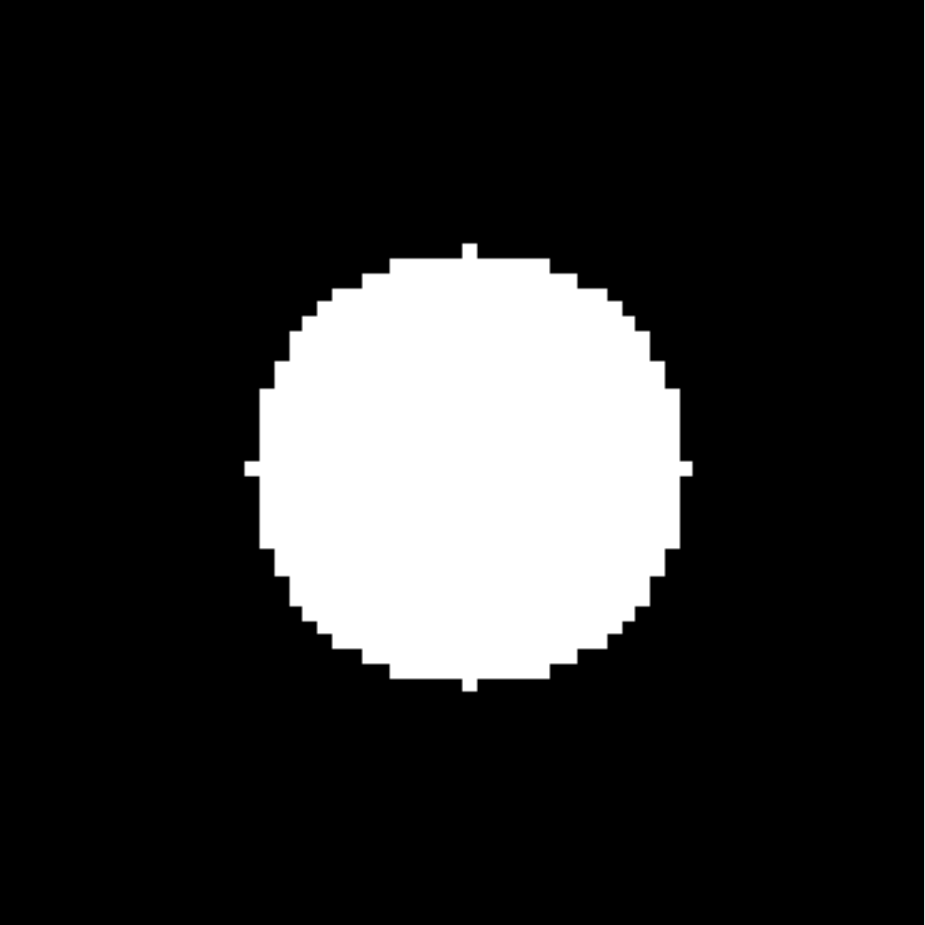}\\ 
\hline
\includegraphics[width=0.1\linewidth]{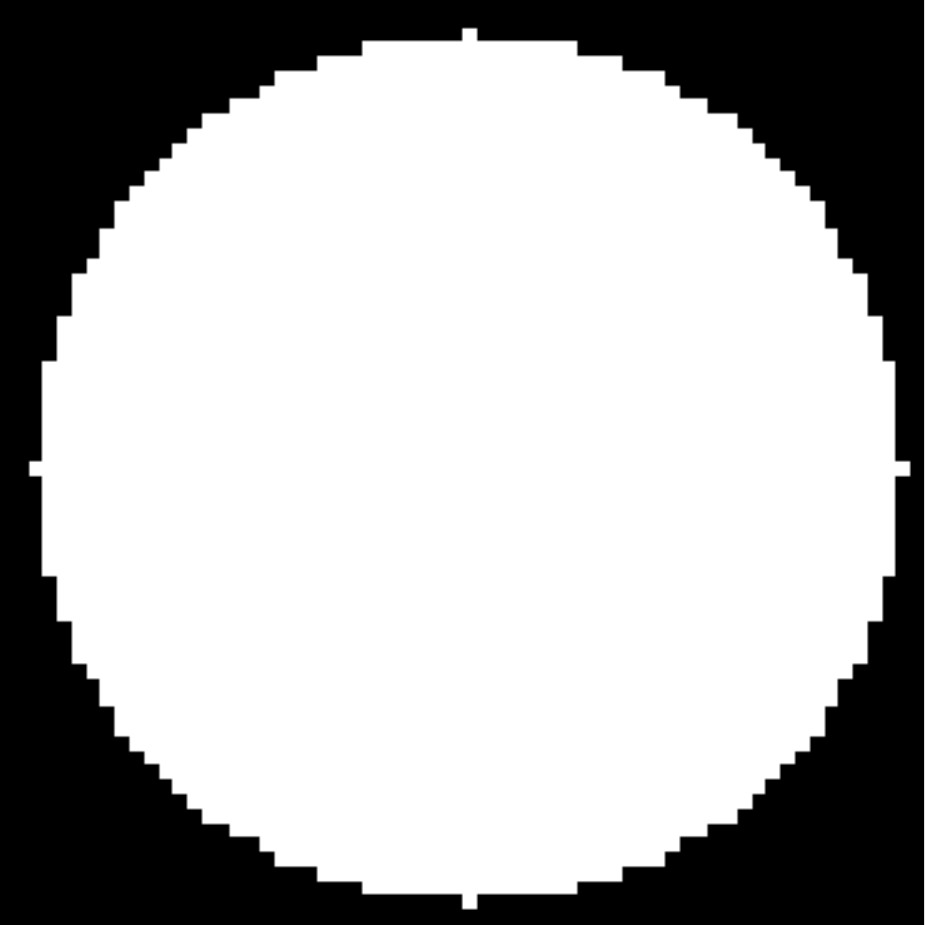} & 
\includegraphics[width=0.1\linewidth]{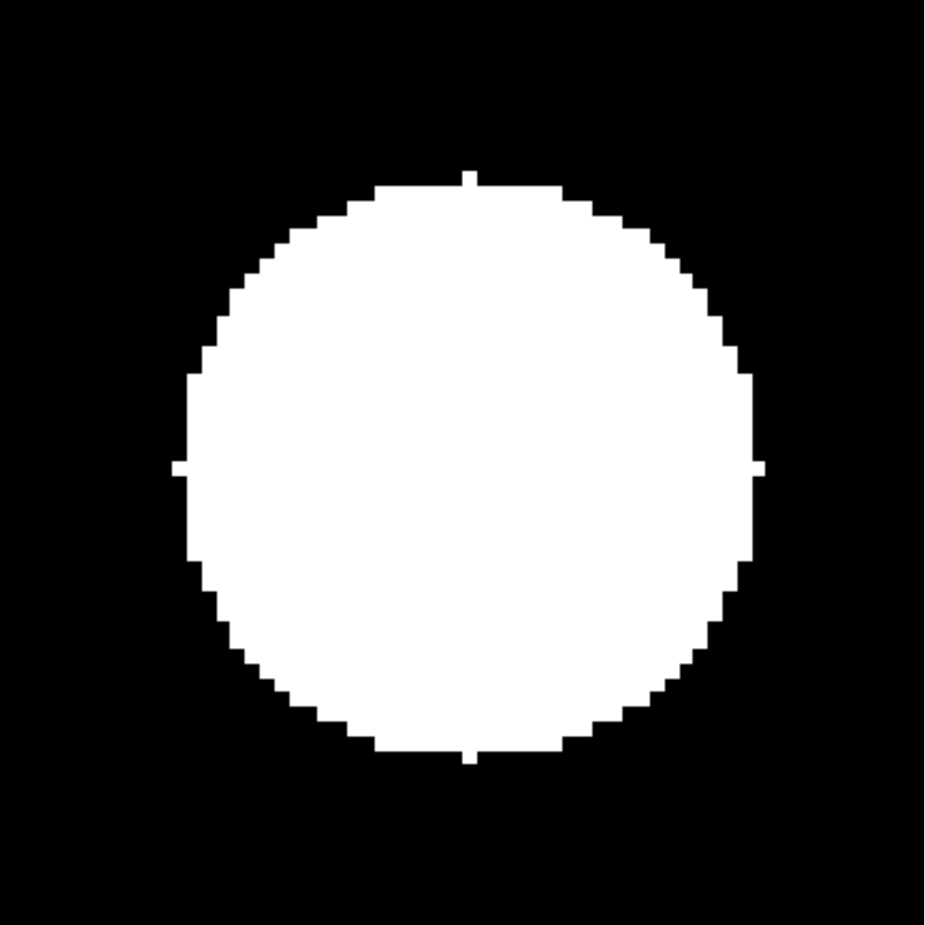}\\ 
\vspace{35pt}
\end{tabular}\label{fig:stomp_problem_table}
}
\caption{Sorted norms of level 1 atoms for Haar, Db4 and Db8 wavelets used in the multilevel decomposition of an image with 64$\times$64 pixels (a). Deterministic patterns with undersampling ratios of 98.0\%, 82.7\%, 69.3\% and 31.1\% (b, from top left in clockwise order), with included frequencies in white.}\label{fig:stomp_problem}
\end{figure}

The only exception is represented by stOMP reconstructions computed using the Db8 wavelet frame, where the image quality is significantly degraded for all velocity components. 
The reason for this is illustrated in Figure~\ref{fig:stomp_problem}, where the norms of the level 1 (coarser level) atoms of \(\mPhO\) are plotted in decreasing order for the Haar, Db4 and Db8 wavelets, following the multilevel decomposition of an image with 64$\times$64 pixels. Various norm profiles are shown, obtained by selectively including a deterministic collection of low $k$-space frequencies (see Figure~\ref{fig:stomp_problem_table}) in \(\Omega\).
The Db8 atom norms decay rapidly for undersampling ratios greater then $\sim$80\%, negatively affecting the ability of stOMP to recognize that such atoms carry relevant image information.

\begin{figure}[ht!]
\centering
\subfloat[]{\includegraphics[width=0.18\textwidth]{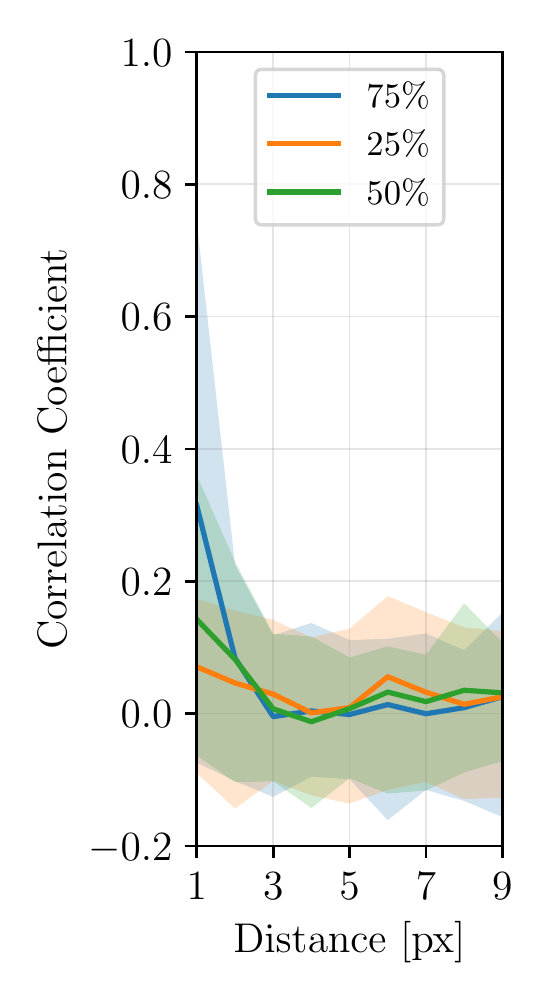}}$\,$
\subfloat[]{\includegraphics[width=0.18\textwidth]{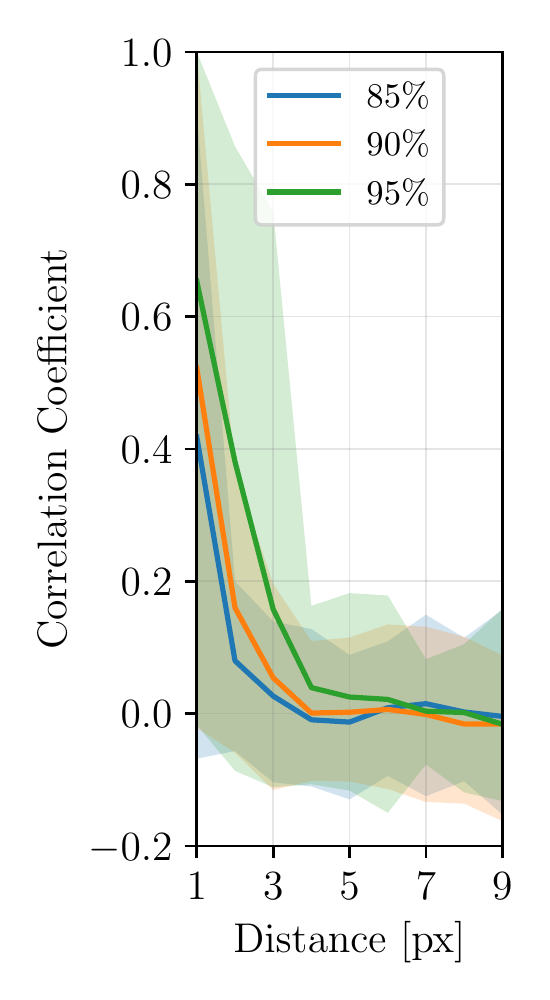}}$\,$
\subfloat[]{\includegraphics[width=0.18\textwidth]{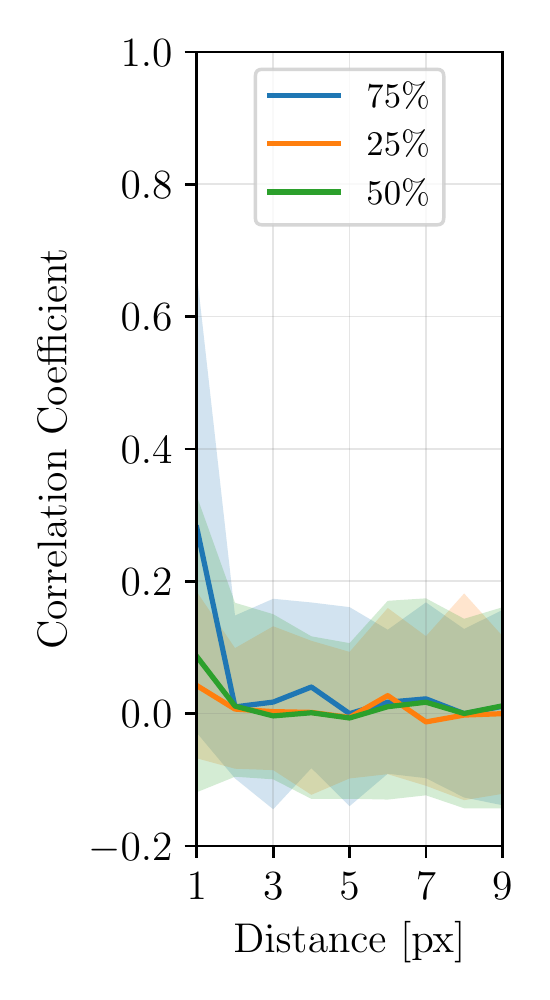}}$\,$
\caption{Reconstruction noise correlations for simulated aortic flow obtained from baseline conditions (a,c) and with large undersampling ratios (b). The correlations in (a) and (b) are generated by keeping the same undersampling mask for all the 100 reconstructions. A different mask was instead generated for each realization in (c).}\label{fig:aortaideal_corr}
\end{figure}

\begin{figure}[ht!]
\centering
\includegraphics[width=0.8\textwidth]{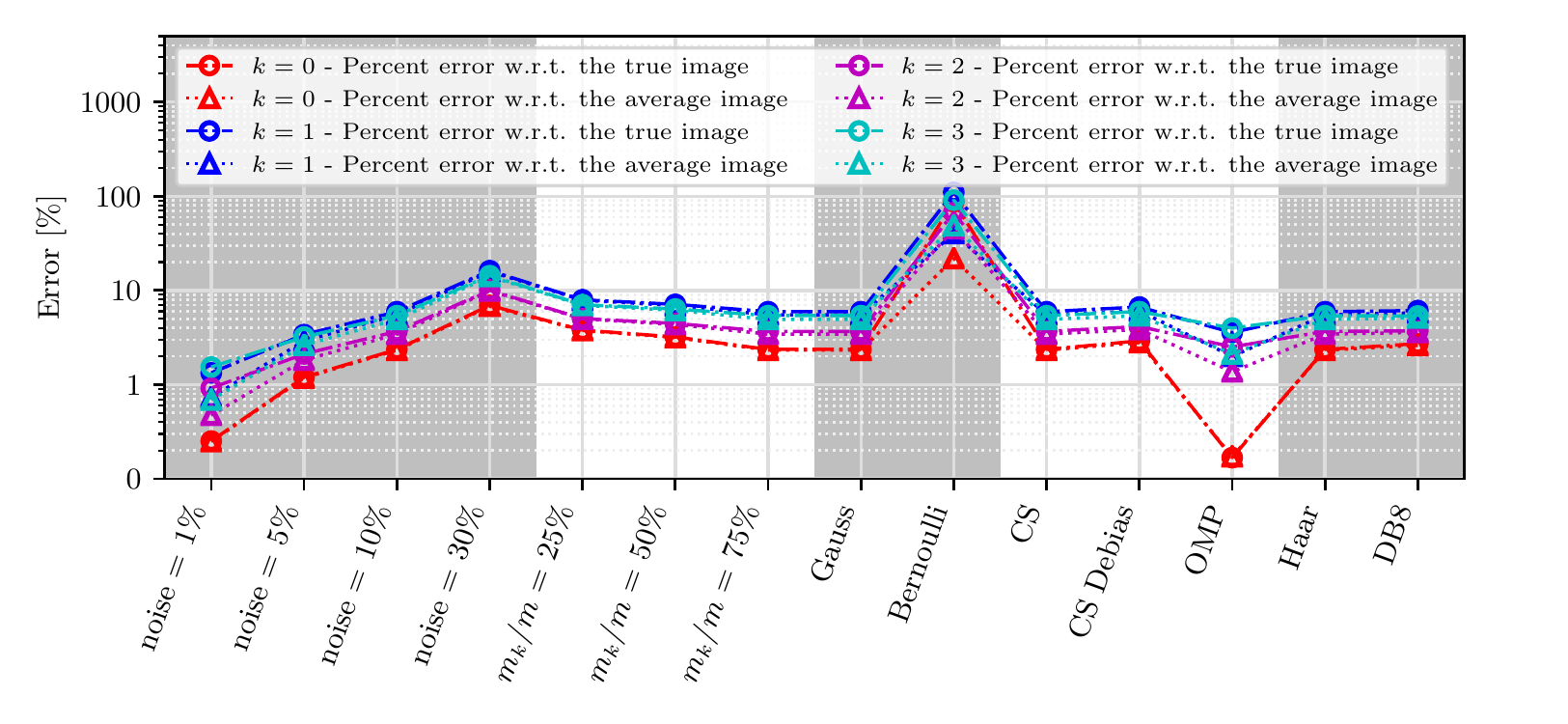}
\caption{Summary of percent errors $\text{PE}_{k,s},\,k\in\{0,1,2,3\},\,s\in\{t,a\}$ for simulated aortic flow. Percent errors remain the same also when introducing randomness in the undersampling pattern.}\label{fig:aortaideal_pe}
\end{figure}

\begin{figure}[ht!]
\centering
\includegraphics[width=0.8\textwidth]{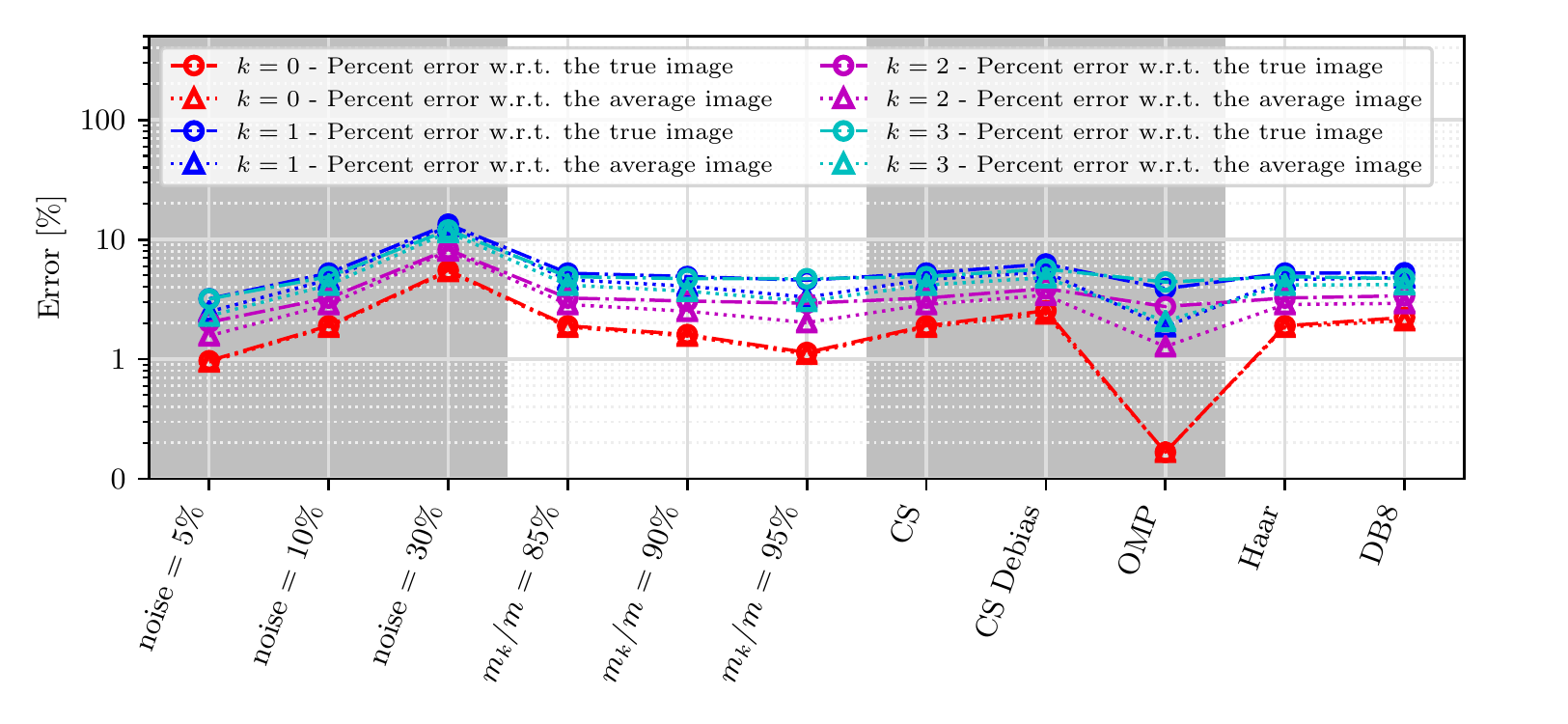}
\caption{Summary of percent errors $\text{PE}_{k,s},\,k\in\{0,1,2,3\},\,s\in\{t,a\}$ for simulated aortic flow from reconstructions with 85\%, 90\% and 95\% undersampling ratios. Where not specified, 85\% undesampling is used to generate the results shown in this plot.}\label{fig:aortaideal_pe_lu}
\end{figure}

The noise correlations shown in Figure~\ref{fig:aortaideal_corr} are not affected by randomness in the undersampling mask, but exhibit a sensible increase for large undersampling ratios, with correlation lengths up to 4-5 pixels for 95\% undersampling and recovery with CS or CSDEB.
The percent errors in Figure~\ref{fig:aortaideal_pe} and~\ref{fig:aortaideal_pe_lu} increase with noise and undersampling. Even with a 16-fold reduction in the acquired $k$-space frequencies, reasonable flow patterns are still observed. 
%
%
%
%
%
%

\subsection{Aortic flow from MRI scan}\label{sec:mriAorticFlow}

The 4D flow MR test case contains a realistic scan of the thoracic aorta, however we has access to the reconstructed images and not the original $k$-space scans. 
Therefore we simulated the effect of noise using an additional 10\% noise intensity and observed the quality of the resulting reconstructions by perturbing the undersampling ratio and mask, the recovery algorithm and the wavelet frame. 

\begin{figure}[ht!]
\centering
\setlength{\tabcolsep}{0pt}
\renewcommand{\arraystretch}{0}
\subfloat[]{
\begin{tabular}[b]{c | c | c | c}
\includegraphics[width=0.06\linewidth]{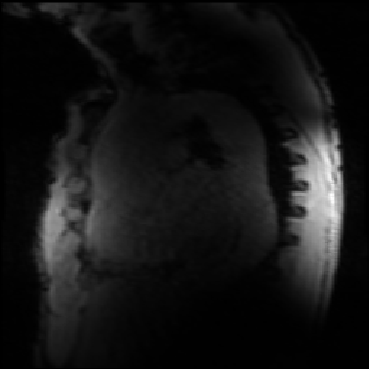} & 
\includegraphics[width=0.06\linewidth]{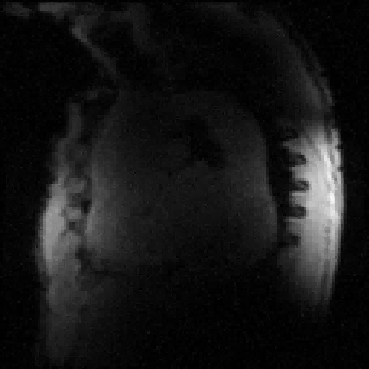} & 
\includegraphics[width=0.06\linewidth]{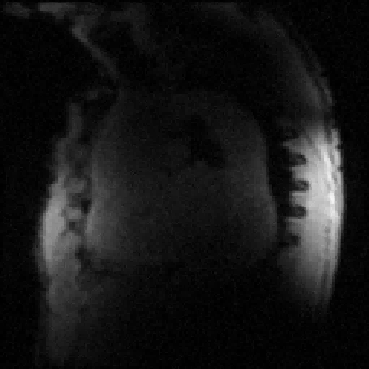} & 
\includegraphics[width=0.06\linewidth]{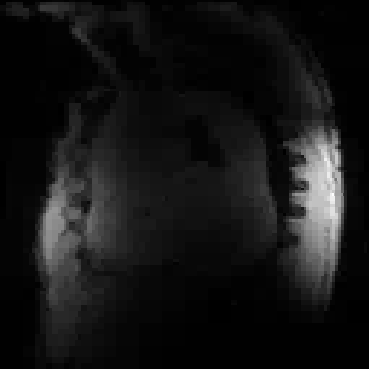}\\
\hline
\includegraphics[width=0.06\linewidth]{imgs/ao_mri/true_k0.png} & 
\includegraphics[width=0.06\linewidth]{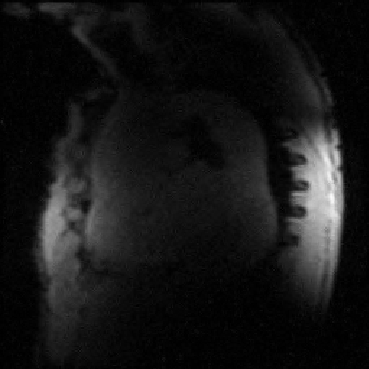} & 
\includegraphics[width=0.06\linewidth]{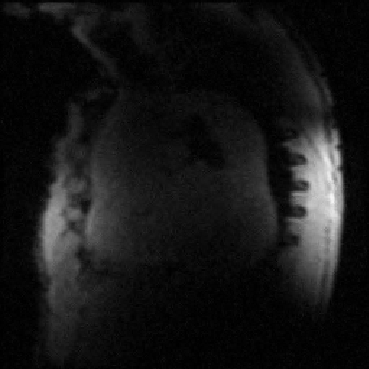} & 
\includegraphics[width=0.06\linewidth]{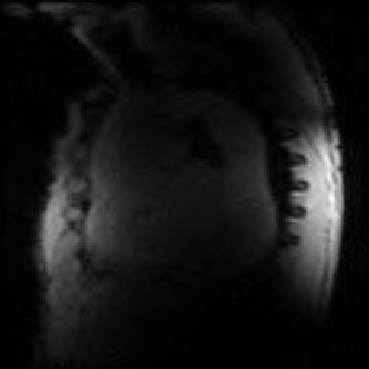}\\
\hline
\includegraphics[width=0.06\linewidth]{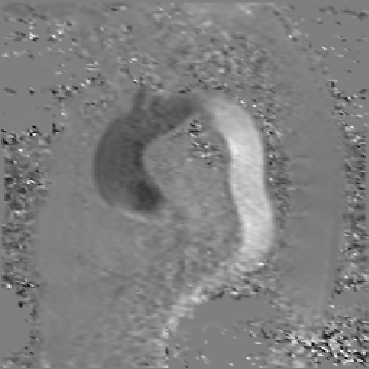} & 
\includegraphics[width=0.06\linewidth]{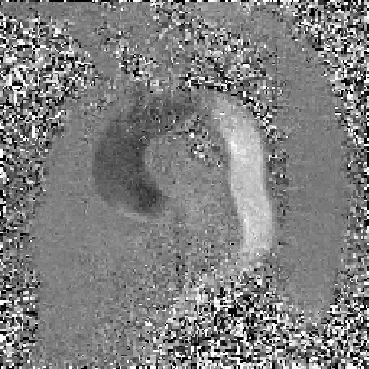} & 
\includegraphics[width=0.06\linewidth]{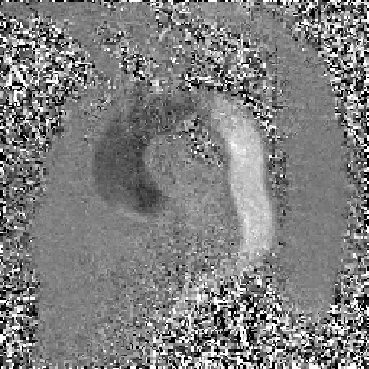} & 
\includegraphics[width=0.06\linewidth]{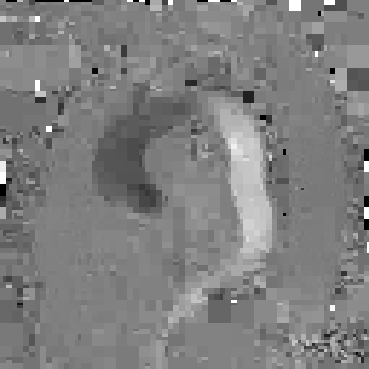}\\
\hline
\includegraphics[width=0.06\linewidth]{imgs/ao_mri/true_k1.png} & 
\includegraphics[width=0.06\linewidth]{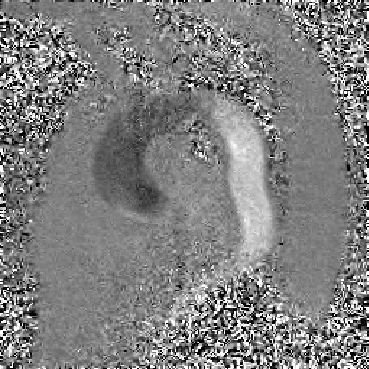} & 
\includegraphics[width=0.06\linewidth]{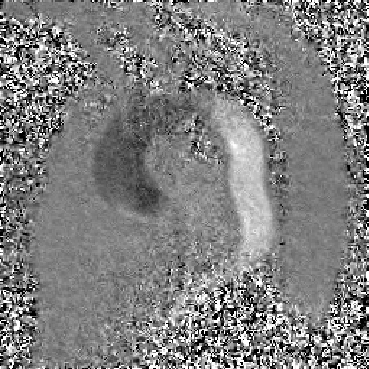} & 
\includegraphics[width=0.06\linewidth]{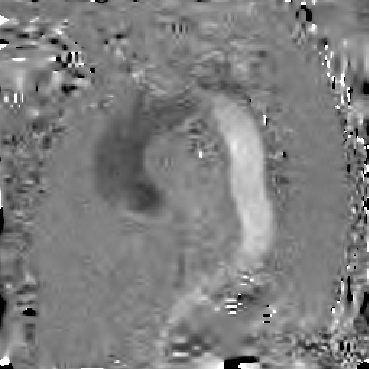}\\
\end{tabular}\label{fig:aortamri_img_A}
}
\subfloat[]{
\begin{tabular}[b]{c | c | c | c}%
\includegraphics[width=0.06\linewidth]{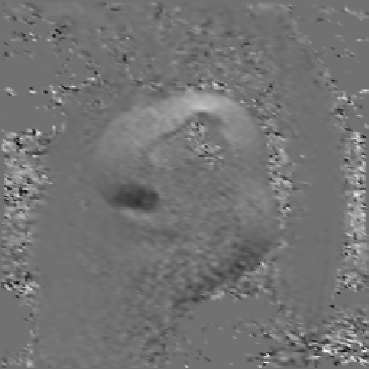} & 
\includegraphics[width=0.06\linewidth]{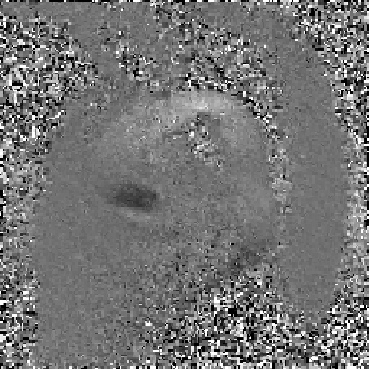} & 
\includegraphics[width=0.06\linewidth]{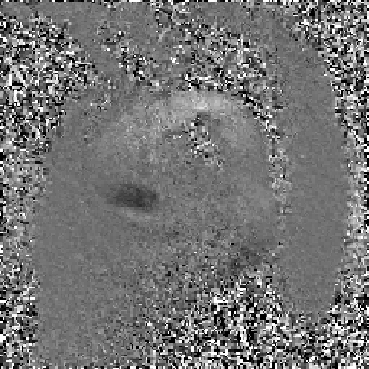} & 
\includegraphics[width=0.06\linewidth]{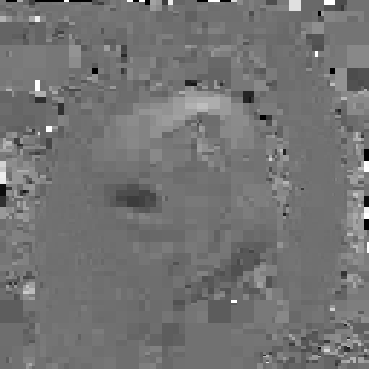}\\
\hline
\includegraphics[width=0.06\linewidth]{imgs/ao_mri/true_k2.png} & 
\includegraphics[width=0.06\linewidth]{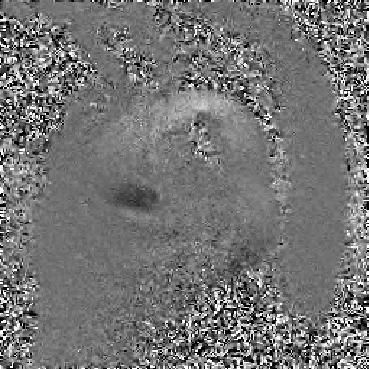} & 
\includegraphics[width=0.06\linewidth]{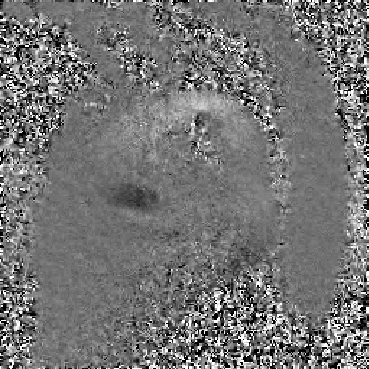} & 
\includegraphics[width=0.06\linewidth]{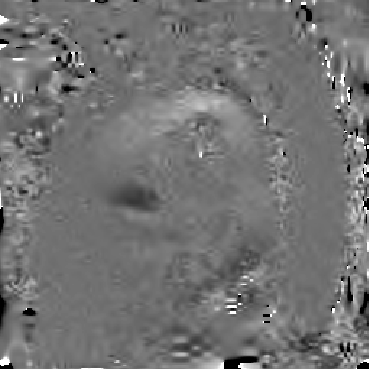}\\
\hline
\includegraphics[width=0.06\linewidth]{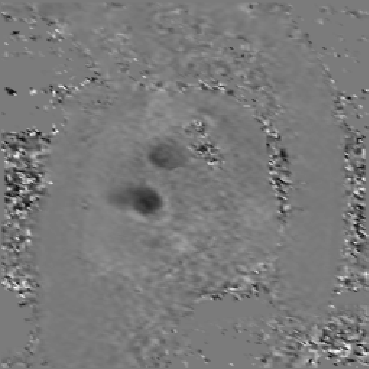} & 
\includegraphics[width=0.06\linewidth]{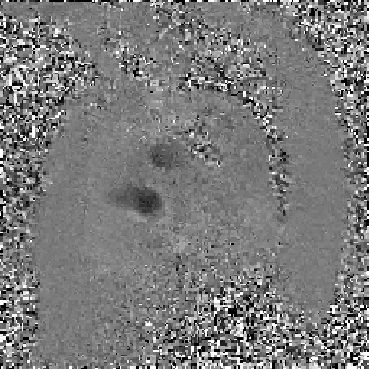} & 
\includegraphics[width=0.06\linewidth]{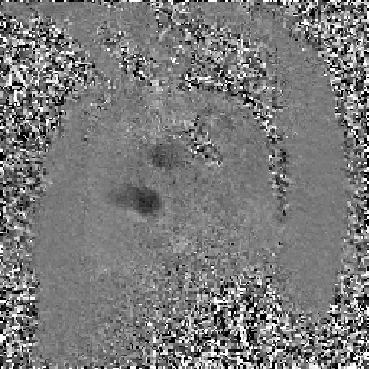} & 
\includegraphics[width=0.06\linewidth]{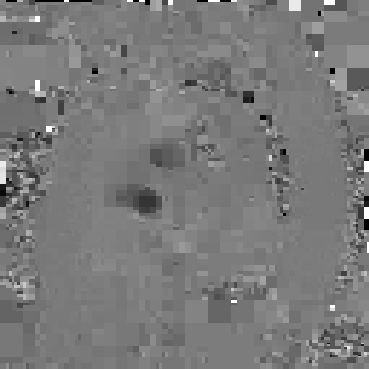}\\
\hline
\includegraphics[width=0.06\linewidth]{imgs/ao_mri/true_k3.png} & 
\includegraphics[width=0.06\linewidth]{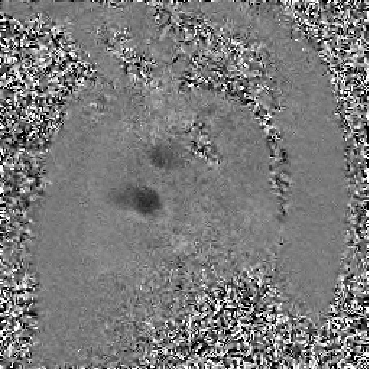} & 
\includegraphics[width=0.06\linewidth]{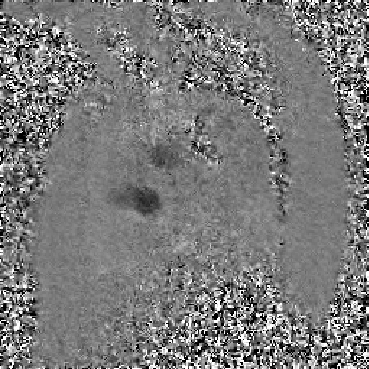} & 
\includegraphics[width=0.06\linewidth]{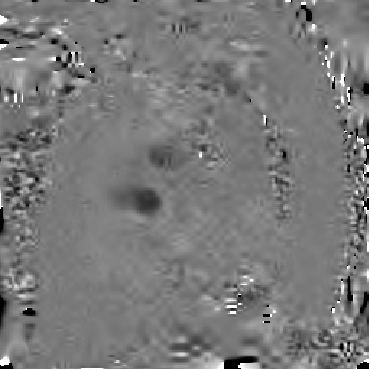}\\
\end{tabular}\label{fig:aortamri_img_B}
}
\subfloat[]{
\begin{tabular}[b]{c | c | c}%
\includegraphics[width=0.06\linewidth]{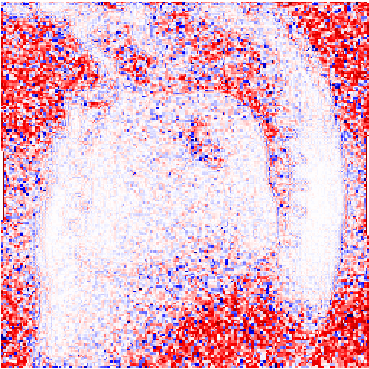} & 
\includegraphics[width=0.06\linewidth]{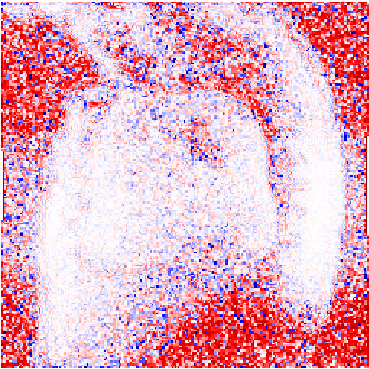} & 
\includegraphics[width=0.06\linewidth]{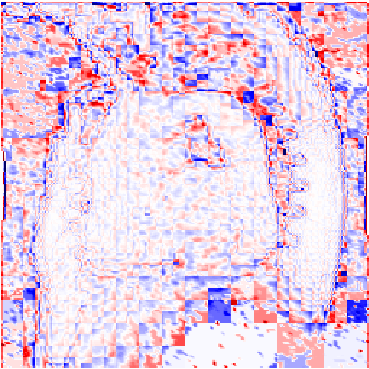}\\
\hline
\includegraphics[width=0.06\linewidth]{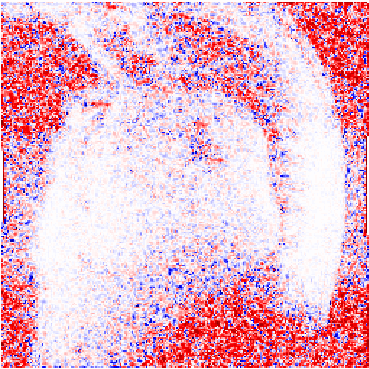} & 
\includegraphics[width=0.06\linewidth]{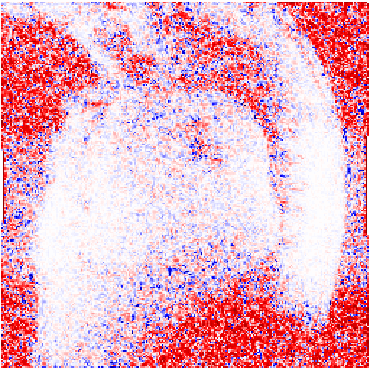} & 
\includegraphics[width=0.06\linewidth]{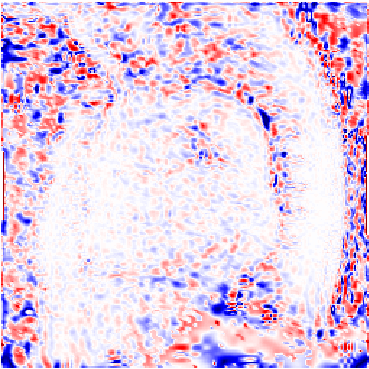}\\
\hline
\includegraphics[width=0.06\linewidth]{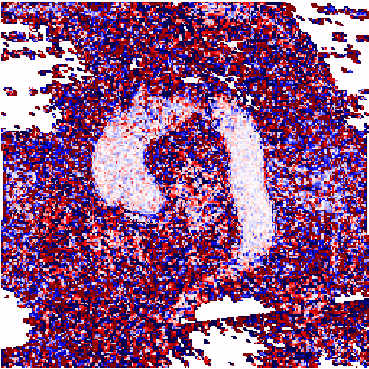} & 
\includegraphics[width=0.06\linewidth]{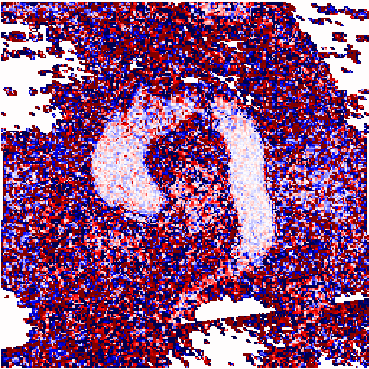} & 
\includegraphics[width=0.06\linewidth]{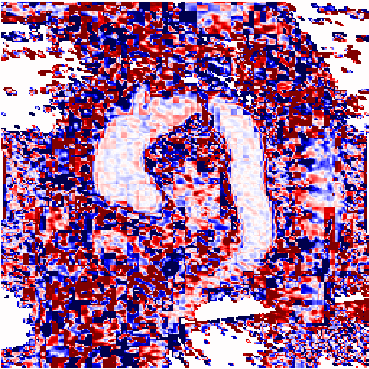}\\
\hline
\includegraphics[width=0.06\linewidth]{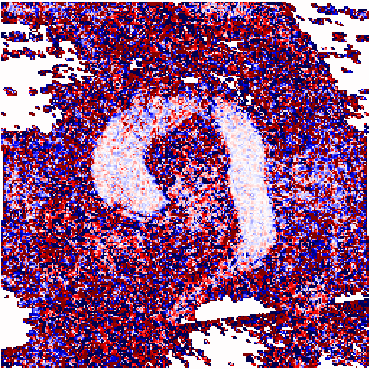} & 
\includegraphics[width=0.06\linewidth]{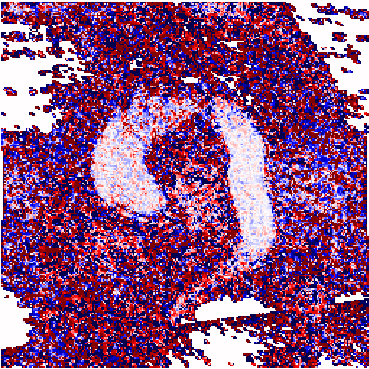} & 
\includegraphics[width=0.06\linewidth]{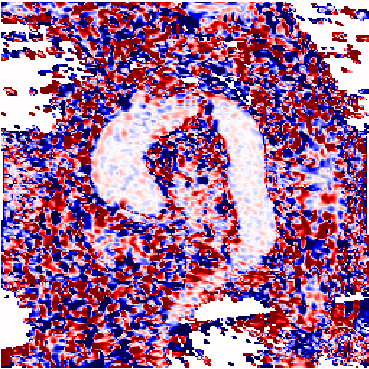}\\
\end{tabular}\label{fig:aortamri_img_C}
}
\subfloat[]{
\begin{tabular}[b]{c | c | c}%
\includegraphics[width=0.06\linewidth]{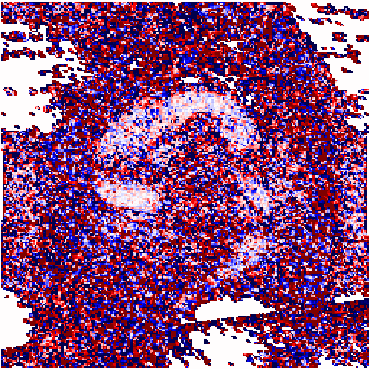} & 
\includegraphics[width=0.06\linewidth]{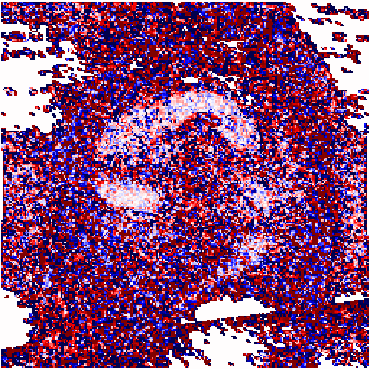} & 
\includegraphics[width=0.06\linewidth]{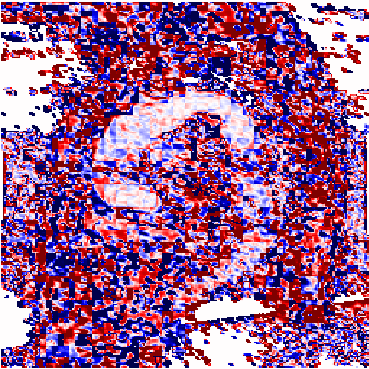}\\
\hline
\includegraphics[width=0.06\linewidth]{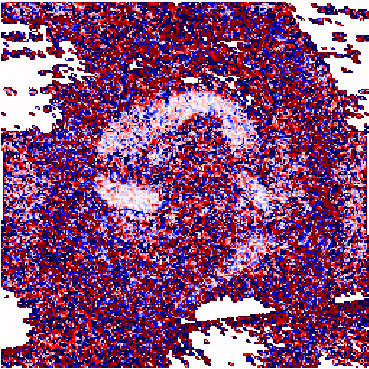} & 
\includegraphics[width=0.06\linewidth]{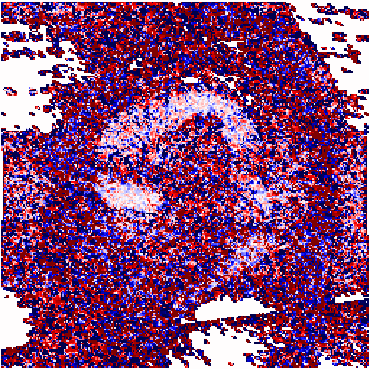} & 
\includegraphics[width=0.06\linewidth]{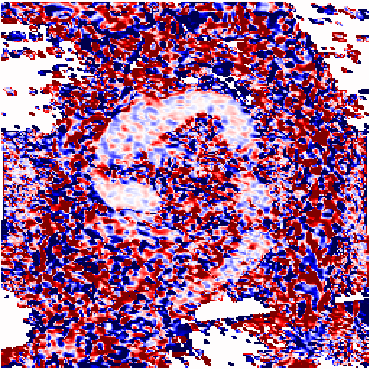}\\
\hline
\includegraphics[width=0.06\linewidth]{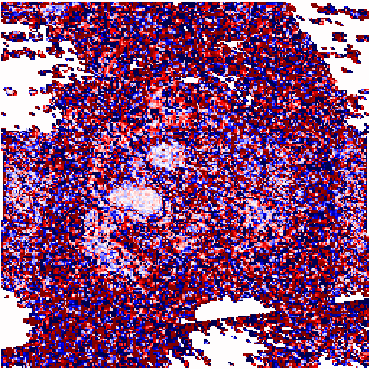} & 
\includegraphics[width=0.06\linewidth]{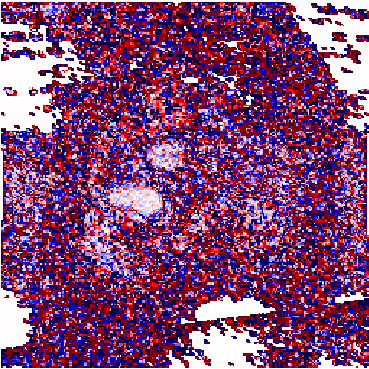} & 
\includegraphics[width=0.06\linewidth]{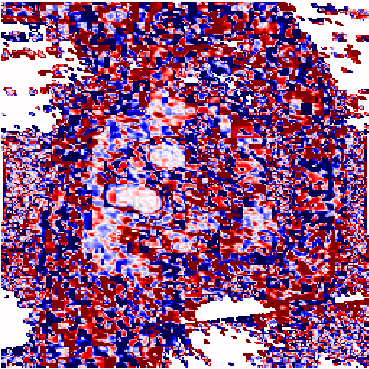}\\
\hline
\includegraphics[width=0.06\linewidth]{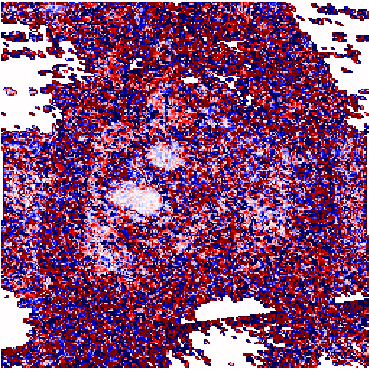} & 
\includegraphics[width=0.06\linewidth]{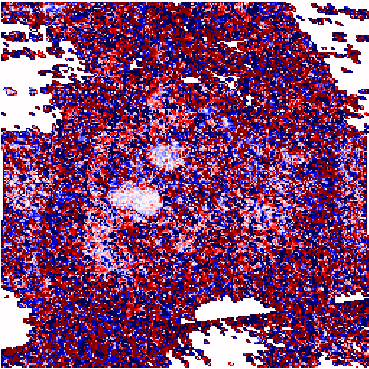} & 
\includegraphics[width=0.06\linewidth]{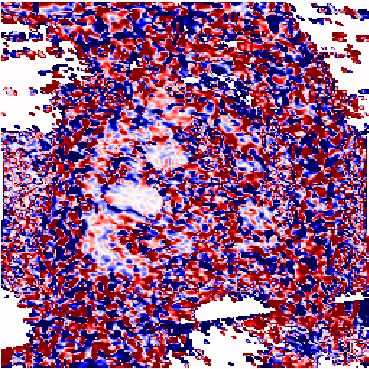}\\
\end{tabular}\label{fig:aortamri_img_D}
}
\caption{Recovered images for the MRI aortic flow test case. Image density and velocities are shown in (a) for $k\in\{0,1\}$ and in (b) for $k\in\{2,3\}$. The columns in (a,b) contain the true image followed by reconstructions generated with CS, CSDEB and stOMP, respectively. The results in the first and third rows were generated using a Fourier-Haar operator, whereas a Db8 frame was used for the second and fourth rows.
Error patterns with respect to the true image are shown in (c) ($k\in\{0,1\}$) and (d) ($k=\{2,3\}$) where again the rows and columns refer to the selected recovery algorithm and wavelet frame. Errors with respect to the average reconstruction contain similar patterns and are therefore not shown.}\label{fig:aortamri_img}
\end{figure}

\begin{figure}[ht!]
\centering
\includegraphics[width=0.8\textwidth]{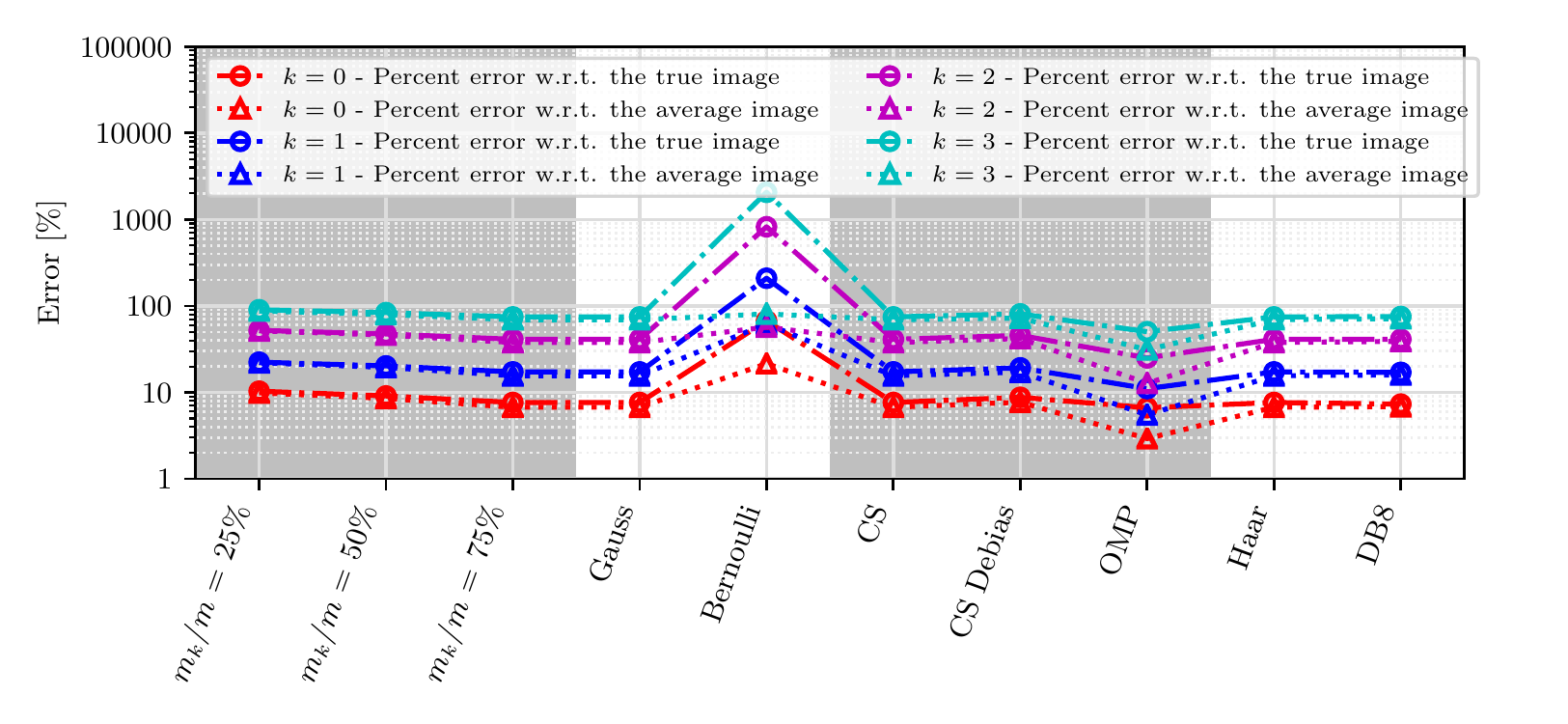}
\caption{Summary of percent errors $\text{PE}_{k,s},\,k\in\{0,3\},\,s\in\{t,a\}$ for the MRI aortic flow test case.}\label{fig:aortamri_pe}
\end{figure}

All recovery techniques produced comparable reconstruction accuracy, as shown in Figure~\ref{fig:aortamri_img}. 
Noise correlations were also in line with the results from the other test cases, with correlation lengths in the order of 2-3 pixels. 
Finally, percent errors in Figure~\ref{fig:aortamri_pe} were found to be one order of magnitude greater than other test cases, due to the higher degree of detail and the reduced smoothness of the underlying flow. 

\begin{figure}[ht!]
\centering
\subfloat[]{\includegraphics[width=0.18\textwidth]{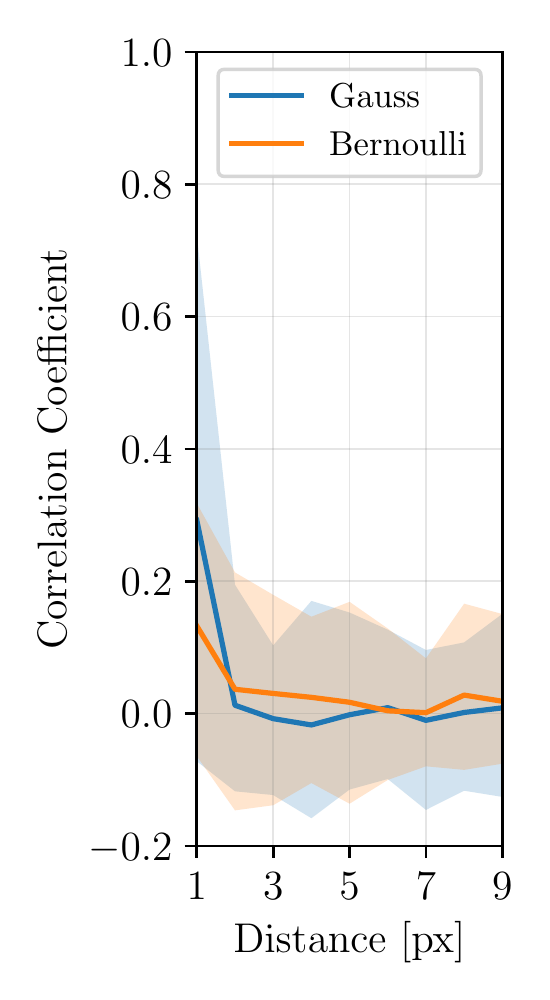}}
\subfloat[]{\includegraphics[width=0.18\textwidth]{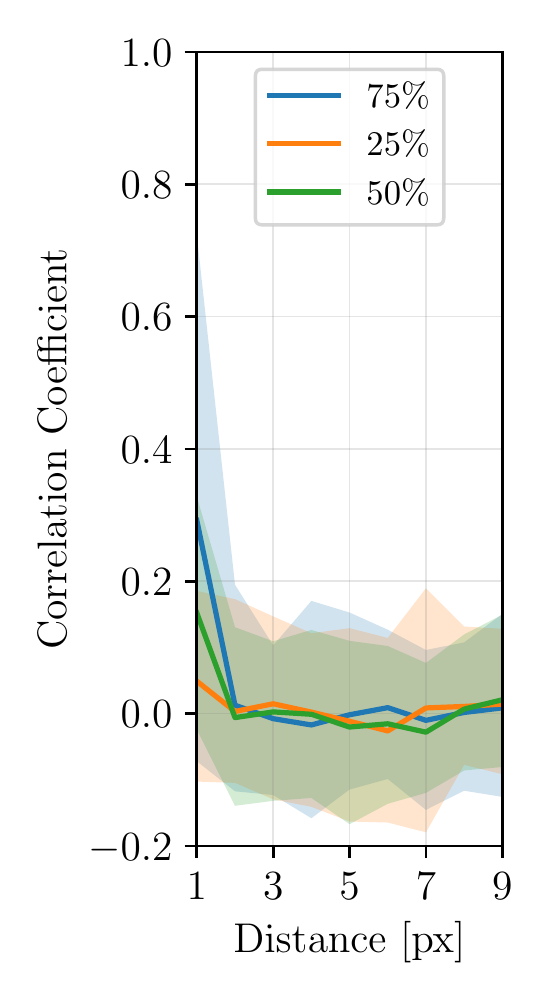}}
\subfloat[]{\includegraphics[width=0.18\textwidth]{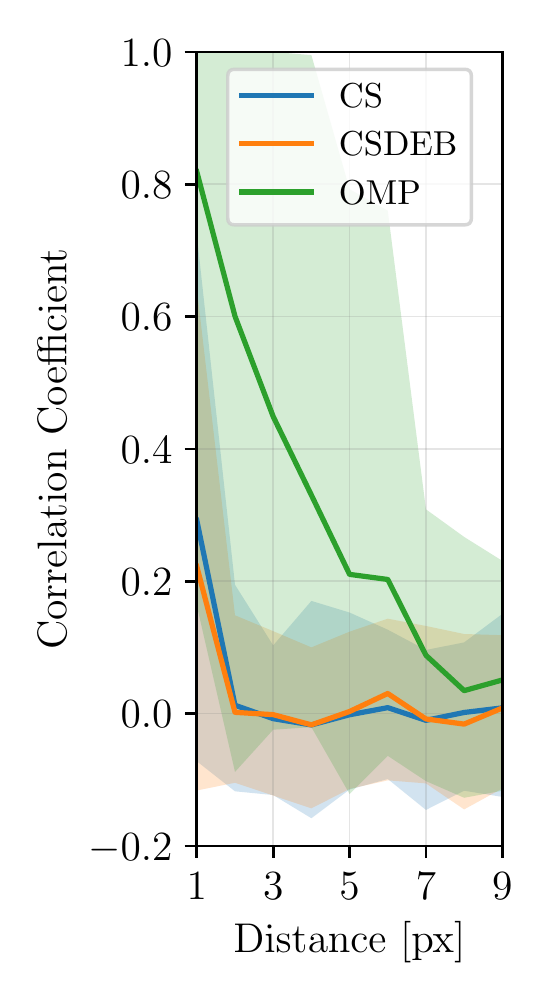}}
\subfloat[]{\includegraphics[width=0.18\textwidth]{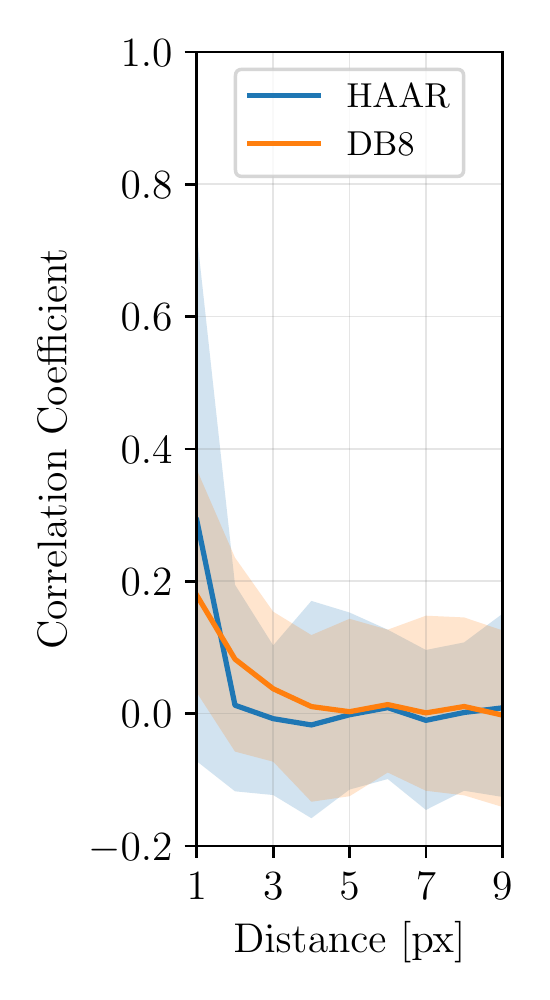}}
\caption{Reconstruction noise correlation for MRI aortic flow obtained by perturbing recovery parameters from baseline conditions. The correlations are only shown for $k=3$, but similar results were obtained for $k\in\{0,1,2\}$.}\label{fig:aortamri_corr}
\end{figure}
%
%

\section{Discussion}\label{sec:discussion}

Our goal is to perform a theoretical an numerical study of the structure of the covariance of the reconstructed density and velocity in 4D flow when the partial measurements in \(k\)-space are corrupted by noise. In particular, we attempted to identify the parameters that have a larger impact on the covariance and thus on the spatial correlations. 

Our numerical results show that reconstructed image values are spatially correlated. The correlation length is short, fluctuating between 2 and 3 pixels. Furthermore, the correlation length seems to be stable for different values of the noise variance. This suggests that there is a reasonable range of values of \(\sigma\) that are well-approximated by the noiseless limit. Hence, it is likely the asymptotic expressions we provide for the covariance can be used in practice to assess the spatial correlations that appear in the reconstructed images. As a consequence, the covariance is similar to that we would obtain for a surrogate affine reconstruction method. 

Another parameter that does not seem to have a noticeable impact on the correlation length is the wavelet basis used for reconstruction. In contrast, correlations between adjacent pixels appear to be sensitive to the amount of undersampling up to 75\%. Additionally, pixels at larger distances become correlated for larger undersampling ratios, i.e., from 85\% to 95\%. This seems to be independent of the distribution of the sampling set. However, differences in the distribution of the sampling set have a large impact on the correlation length. For instance, Bernoulli undersampling removes relevant low-frequency information, leading to high reconstruction noise with large correlation lengths.

Algorithm selection has also an impact on the correlation length. Recovery with a greedy algorithm (stOMP~\cite{donoho2012sparse}) is observed to produce smaller reconstruction noise than other algorithms (those discussed in Section~\ref{sec:recoveryCompressedSensing}) and smaller reconstruction bias up to 75\% undersampling. However, the correlation length is significantly larger, and the quality of the stOMP reconstructions degrades significantly for Gaussian undersampling greater then $\sim$80\% when using Db8 wavelets, due to a significant reduction in the norms of the atoms at the coarser approximation level.

We were surprised by the quality of the linear reconstructions. Even with 4\(\times\) or 8\(\times\) less frequencies, use of a variable density Gaussian mask led to noise intensities and a quality of the reconstructed image comparable to the non-linear reconstruction algorithms. While this performance can be partially attributed to the smoothness of the anatomies and velocity fields analyzed in this study (e.g. see the discussion in Section~\ref{sec:recoveryOrthogonalProjection}), it was also apparent for the aortic MRI scan discussed in Section~\ref{sec:mriAorticFlow}, which is characterized by a realistic spectrum of \(k\)-space frequencies. This suggests how the selection of an appropriate undersampling mask is as important as the development of efficient recovery algorithms.

Finally, we compared the effect of random \(k\)-space noise under a fixed sampling set, and have also varied both the noise and the sampling set for a fixed undersampling ratio. Randomness in the undersampling set does not seem to significantly alter the spatial correlations in the reconstructed images.

Consequently, our results show that a reconstructed density and velocity from undersampled \(k\)-space measurement corrupted by noise do exhibit spatial correlations. Furthermore, these spatial correlations are well-approximated by the expressions found in the noiseless limit. Interestingly, the distribution of the sampling set and the reconstruction method seem to have had the stronger impact on the spatial correlations, whereas \(k\)-space noise variance and the wavelet basis had a limited impact. 

The present study has several limitations mainly related to the synthetic nature of the selected tests cases. Specifically, we neglect artifacts that may arise due, for example, to acquisition protocols, patient movement, field inhomogeneities, multiple coils, and we focus on the artifacts due only to the reconstruction procedure. However, we believe it is an important first step to assess the coherence of visual artifacts due to measurement noise when applying reconstruction methods, and identifying some of the most relevant factors that impact these artifacts.

Future work will focus on a finer analysis of the structure of the covariance. For instance, it would be of interest to assess the visual structure of the artifacts. In addition, our efforts will focus on developing random field models for the noise from undersampled velocity reconstruction for applications in biomedicine.

\section{Acknowledgements}\label{sec:acks}

The authors gratefully acknowledge the support of Notre Dame International through two Luksic Family Collaboration Grant awards in 2018 and 2021 (PI Daniele Schiavazzi), a Luksic Family Collaboration Grant award 2018 (PI Carlos Sing-Long), the grant ANID -- FONDECYT -- 1211643  (PI Carlos Sing-Long), a grant ANID -- Millennium  Science  Initiative  Program -- NCN17\_059  and  ANID  --  Millennium  Science  Initiative Program -- NCN17\_129, and the Open Seed Fund 2020 for collaboration between the Pontificia Universidad Cat\'olica de Chile, School of Engineering and the University of Notre Dame (PI Carlos Sing-Long).

\bibliographystyle{unsrt}
\bibliography{references}  

\appendix
\section{Proofs}
\label{apx:proofs}

\subsection{The noiseless limit}
\label{apx:noiselessLimit}

We consider without loss a continuous function \(f: \C^m \to \C^n\) such that \(f(\vc{0}) = \vc{0}\) and the random variable \(\vf := f(\sigma \vz)\) for \(\vz\sim N(\vc{0},\vI) + iN(\vc{0},\vI)\) and \(\sigma > 0\). We analyze the structure of its first and second-order moments in the asymptotic regime \(\sigma \to 0\). 

Let \(\Sph^m := \set{\vy:\, \nrm{\vy}_2 = 1}\) be the unit sphere in \(\C^m\) endowed with the subspace topology, its Borel \(\sigma\)-algebra, and the uniform measure. Let \(\vo \in \Sph^m\). The one-directional Gateaux derivative exists at \(\vo\) if \(\set{t^{-1} f(t\vo)}_{t > 0}\) has a limit as \(t \to 0\). If this is the case, we let
\begin{equation}
\label{eq:defGateaux}
    df(\vo) := \lim_{t \to 0^{+}}t^{-1} f(t\vo)
\end{equation}
be the one-directional Gateaux derivative along \(\vo\). When the limit exists for almost every \(\vo\in \Sph^{m}\) we let \(df(\vo)\) be equal to the above limit when it exists, and zero otherwise. As a consequence \(df\) is Borel measurable. Finally, recall that for \(h:\Rp \times \Sph^{m} \to \C\) measurable we have
\[
    \ev_{\vz}\bset{h(\nrm{\vz}_2, \ovz)} = \ev_{\rho}\bset{\ev_{\vo}\bset{h(\rho, \vo)}}
\]
where \(\ovz := \vz / \nrm{\vz}_2\), \(\rho \sim \chi_{2m-1}\) and \(\vo\sim \textsc{Uniform}(\Sph^{m})\). 

Since we only assume the one-directional Gateaux derivative exists for almost every \(\vo\) we begin with the following approximation lemma.

\begin{lemma}
    \label{apx:lem:noiselessLimit:residual}
    Let \(\vo \sim \textsc{Uniform}(\Sph^m)\). Suppose that the limit~\eqref{eq:defGateaux} exists for almost every \(\vo\in \Sph^m\) and that \(df\) satisfies
    \[
        \ev_{\vo}\bset{\nrm{df(\vo)}_2^2} < \infty\quad\mbox{and}\quad \lim_{t\to 0^+} \ev\bset{\nrm{df(\vo) - t^{-1}f(t\vo)}_2^2} = 0.
    \]
    In addition, suppose there exists \(C, \alpha, r > 0\) such that \(\nrm{f(\vy)}_2\leq C\nrm{\vy}_2^{\alpha}\) for \(\nrm{\vy}_2 > r\). Then \(\ev\bset{\nrm{f(\sigma\vz)}_2} < \infty\) for any \(\sigma > 0\) and
    \begin{equation}
        \label{eq:noiselessLimit:keyBound}
        R_f(\sigma) := \sigma^{-2}\ev_{\vz}\bset{\nrm{\sigma \nrm{\vz}_2 df(\ovz) - f(\sigma\vz)}_2^2 \ind\set{\sigma\nrm{\vz}_2 \leq r}}
    \end{equation}
    tends to zero as \(\sigma\to 0\).
\end{lemma}

\begin{proof}[Proof of Lemma~\ref{apx:lem:noiselessLimit:residual}]
    First, note that
    \[
        \ev_\vz\bset{\nrm{f(\sigma\vz)}_2^2} \leq \sup_{\nrm{\vy}_2 \leq r/\sigma}\nrm{f(\vy)}_2 \prob\set{\sigma\vz \leq r} + C\sigma^\alpha\ev_\vz\bset{\nrm{\vz}_2^\alpha\ind\set{\sigma \vz > r}} < \infty.
    \]
    Let \(\eps > 0\). By hypothesis, there exists \(t_\eps > 0\) such that 
    \[
        \forall\, t \in (0, t_\eps):\,\, \ev_\vo\bset{\nrm{df(\vo) - t^{-1} f(t\vo)}_2^2} < \frac{1}{2\ev\bset{\rho^2}}\eps.
    \]
    Hence,
    \begin{align*}
        \ev_{\vo}\bset{\nrm{\nrm{\vz}_2 df(\ovz) - \sigma^{-1}f(\sigma\vz)}_2^2 \ind\set{\sigma\nrm{\vz}_2 \leq t_\eps}} &= \ev_{\rho}\Lbset{\rho^2 \ev_{\vo}\Lbset{\Lnrm{df(\vo) - \frac{f(\sigma \rho \vo)}{\sigma\rho}}_2^2}\ind\set{\sigma \rho \leq t_\eps}}\\
            &< \frac{1}{2\ev\bset{\rho^2}}\eps\ev\bset{\rho^2 \ind\set{\sigma \rho \leq t_\eps}}\\
            &\leq \frac{1}{2}\eps.
    \end{align*}
    Hence, from the decomposition
    \begin{align*}
        R_f(\sigma) &= \ev_{\rho,\vo}\bset{\nrm{\rho df(\vo) - f(\sigma\rho\vo)}_2^2 \ind\set{\sigma\rho \leq t_\eps}} + \ev_{\rho,\vo}\bset{\nrm{\rho df(\vo) - f(\sigma\rho\vo)}_2^2 \ind\set{t_\eps < \sigma\rho \leq r}} \\
            &\leq \frac{1}{2}\eps + \ev_{\rho,\vo}\bset{\nrm{\rho df(\vo) - f(\sigma\rho\vo)}_2^2 \ind\set{t_\eps < \sigma\rho \leq r}}
    \end{align*}
    it suffices to bound the second term for sufficiently small \(\sigma\). In particular, if \(r < t_\eps / \sigma\) then
    \begin{align*}
        \ev_{\rho,\vo}\bset{\nrm{\rho df(\vo) - f(\sigma\rho\vo)}_2^2 \ind\set{t_\eps < \sigma\rho \leq r}}&\leq 2\ev_{\vo}\bset{\nrm{df(\vo)}_2^2}\ev_{\rho}\set{\rho^2\ind\set{t_\eps/\sigma < \rho}} \\
            &\quad + 2 C^2\ev_\rho\bset{\rho^{2\alpha - 1}\ind\set{t_\eps / \sigma < \rho}}.
    \end{align*}
    Since both terms tend to zero as \(\sigma \to 0\) we conclude that \(h(\sigma) \to 0\) as \(\sigma\to 0\).
\end{proof}

It follows from the lemma that when the hypotheses hold we also have
\begin{equation}
\label{eq:noiselessLimit:keyBoundLinear}
    \sigma^{-1}\ev\bset{\nrm{\sigma \nrm{\vz}_2 df(\ovz) - f(\sigma\vz)}_2 \ind\set{\sigma\nrm{\vz}_2 \leq r}} = o(\sigma).
\end{equation}

We now provide the limit of the expectation \(\ev_{\vz}\bset{f(\sigma\vz)}\) as \(\sigma\to 0\).

\begin{theorem}
\label{apx:thm:noiselessLimit:expectation}
    Suppose the hypotheses of Lemma~\ref{apx:lem:noiselessLimit:residual} hold. Then
    \[
        \lim_{\sigma\to 0} \sigma^{-1}\ev\bset{f(\sigma\vz)} = \ev_{\rho}\bset{\rho}\ev_{\vo}\bset{df(\vo)}
    \]
    where \(\rho \sim \chi_{2m-1}\) and \(\vo\sim \textsc{Uniform}(\Sph^{m})\). In particular, if \(f\) is differentiable at the origin, the limit is zero.
\end{theorem}

\begin{proof}[Proof of Theorem~\ref{apx:thm:noiselessLimit:expectation}]
    First observe that
    \[
        \sigma^{-1}\ev_\vz\bset{\nrm{f(\sigma\vz)}_2\ind\set{\sigma\nrm{\vz}_2 > r}} \leq C \sigma^{\alpha-1} \ev_\vz\bset{\nrm{\vz}_2^{\alpha}\ind\set{\nrm{\vz}_2 > r/\sigma}}
    \]
    whence the right-hand side tends to zero as \(\sigma\to 0\); in fact, it is of order \(o(\sigma)\). From the decomposition
    \begin{align*}
        \sigma^{-1}\ev_\vz\bset{f(\sigma\vz)} &= \ev_\vz\bset{\nrm{\vz}_2 df(\ovz) \ind\set{\sigma\nrm{\vz}_2 \leq r}} + \ev_\vz\bset{f(\sigma\vz) - \sigma\nrm{\vz}_2 df(\ovz)) \ind\set{\sigma\nrm{\vz}_2 \leq r}} + o(\sigma) \\
        &= \ev_\vz\bset{\nrm{\vz}_2 df(\ovz) \ind\set{\sigma\nrm{\vz}_2 \leq r}} + o(\sigma)
    \end{align*}
    where we used~\eqref{eq:noiselessLimit:keyBoundLinear}. Since
    \[
        \ev_\vz\bset{\nrm{\vz}_2 df(\ovz) \ind\set{\sigma\nrm{\vz}_2 \leq r}} = \ev_\vo\bset{ df(\vo)} \ev_\rho\bset{\rho\ind\set{ \rho\leq r/\sigma}}
    \]
    where \(\rho \sim \chi_{2m-1}\) variable and \(\vo\sim \textsc{Uniform}(\Sph^m)\) the first claim follows. For the second, it suffices to note that \(df(\vo) = Df(\vc{0}) \vo\) when \(f\) is differentiable at the origin, and that \(\ev\bset{\vo} = 0\).
\end{proof}

Note that it is not true in general that \(\ev_{\vz}\bset{f(\sigma\vz)} \to \vc{0}\). We now provide the limit of the covariance as \(\sigma\to 0\).

\begin{theorem}
\label{apx:thm:noiselessLimit:covariance}
    Suppose the hypotheses of Lemma~\ref{apx:lem:noiselessLimit:residual} hold. Then
    \[
        \lim_{\sigma\to 0} \sigma^{-2}\var_\vz\bset{f(\sigma\vz)} = \var_{\vz}\bset{\nrm{\vz}_2df(\ovz)} = \var_\rho\bset{\rho} \ev_{\vo}\bset{df(\vo)df(\vo)^*} +\ev_\rho\bset{\rho}^2 \var_{\vo}\bset{df(\vo)}.
    \]
    where \(\rho \sim \chi_{2m-1}\) and \(\vo\sim \textsc{Uniform}(\Sph^{m})\). In particular, if \(f\) is differentiable at the origin with differential matrix \(Df(\vc{0})\) then
    \[
        \lim_{\sigma\to 0} \sigma^{-2}\var_\vz\bset{f(\sigma\vz)} = 2Df(\vc{0})Df(\vc{0})^*.
    \]
\end{theorem}

\begin{proof}[Proof of Theorem~\ref{apx:thm:noiselessLimit:covariance}]
    First, from the same arguments in the proof of Theorem~\ref{apx:thm:noiselessLimit:expectation} we have
    \begin{align*}
        \nrm{\ev\bset{(\vf - \ev \vf)(\vf - \ev\vf)^*\ind\set{\sigma\nrm{\vz}_2 > r}}}_F &\leq \ev\bset{\nrm{\vf - \ev\vf}_2^2 \ind\set{\sigma\nrm{\vz}_2 > r}}\\
        &\leq 2\ev\bset{\nrm{\vf}_2^2\ind\set{\sigma\nrm{\vz}_2 > r}} + 2\nrm{\ev\bset{\vf}}_2^2\prob\set{\sigma\nrm{\vz}_2 > r}.
    \end{align*}
    By Lemma~\ref{apx:lem:noiselessLimit:residual} the upper bound is \(o(\sigma^2)\). 
    
    Let \(\vh = \sigma\nrm{\vz}_2df(\omega(\vz))\) and define \(\delta \vh = \vh - \ev\bset{\vh}\) and \(\delta \vf = \vf - \ev\bset{\vf}\). Consider the decomposition
    \[
        \delta\vf\delta\vf^*\ind\set{\sigma\nrm{\vz}_2\leq r} = (\delta\vh\delta\vh^* + \delta\vh(\delta\vf - \delta\vh)^* + (\delta\vf -\delta\vh)\delta\vh^* + (\delta\vf - \delta\vh)(\delta\vf - \delta\vh)^*)\ind\set{\sigma\nrm{\vz}_2\leq r}.
    \]
    To bound the terms on the right-hand side, note that
    \[
        \ev\bset{\nrm{\delta\vh}_2\ind\set{\sigma\nrm{\vz}_2 \leq r}} \leq 2\sigma \ev\bset{\nrm{\vz}_2\nrm{df(\omega(\vz))}_2\ind\set{\sigma\nrm{\vz}_2 \leq r}} \leq 2\sigma\ev_\rho\bset{\rho}\ev_{\vo}\bset{\nrm{f(\vo)}_2}
    \]
    whence the term is \(O(\sigma)\). In addition,
    \begin{align*}
        \ev\bset{\nrm{\delta\vf - \delta\vh}_2^2\ind\set{\sigma\nrm{\vz}_2 \leq r}} &\leq 2\ev\bset{\nrm{\vf - \vh}_2^2\ind\set{\sigma\nrm{\vz}_2 \leq r}}+ 2\nrm{\ev\bset{\vf} - \ev\bset{\vh}}_2^2 \prob\set{\sigma\nrm{\vz}_2 \leq r}\\
        &\leq \sigma^2 R_f(\sigma) + 2\sigma^{2} \nrm{\sigma^{-1}(\ev\bset{\vf} - \ev\bset{\vh})}_2^2.
    \end{align*}
    From Lemma~\ref{apx:lem:noiselessLimit:residual} and Theorem~\ref{apx:thm:noiselessLimit:expectation} the upper bound is \(o(\sigma^2)\). In particular, 
    \[
        \ev\bset{\nrm{\delta\vf - \delta\vh}_2\ind\set{\sigma\nrm{\vz}_2 \leq r}} = o(\sigma).
    \]
    Therefore,
    \begin{align*}
        \nrm{\ev\bset{\delta\vh(\delta\vf - \delta\vh)^*}}_F &\leq \ev\bset{\nrm{\delta\vh}_2\nrm{\delta\vf - \delta\vh}_2\ind\set{\sigma\nrm{\vz}_2 \leq r}}\\
        &\leq \ev\bset{\nrm{\delta\vh}_2^2\ind\set{\sigma\nrm{\vz}_2 \leq r}}^{1/2}\ev\bset{\nrm{\delta\vf - \delta\vh}_2^2\ind\set{\sigma\nrm{\vz}_2 \leq r}}^{1/2}
    \end{align*}
    where the upper bound is \(o(\sigma^2)\), and
    \[
        \nrm{\ev\bset{(\delta \vf - \delta\vh)(\delta\vf - \delta\vh)^*\ind\set{\sigma\nrm{\vz}_2 \leq r}}}_F \leq \ev\bset{\nrm{\delta \vf - \delta\vh}_2^2\ind\set{\sigma\nrm{\vz}_2 \leq r}},
    \]
    where the upper bound is also \(o(\sigma^2)\). We conclude that
    \begin{align*}
        \ev\bset{\delta\vf\delta\vf^*} &= \ev\bset{\delta\vf\delta\vf^*\ind\set{\sigma\nrm{\vz}_2 \leq r}} + o(\sigma^2)\\
            &= \ev\bset{\delta\vh\delta\vh^*\ind\set{\sigma\nrm{\vz}_2\leq r}} + o(\sigma^2).
    \end{align*}
    Since \(\ev\bset{\nrm{\delta\vh}_2^2} < \infty\) we have
    \[
        \lim_{\sigma\to 0} \sigma^{-2}\var_{\vz}\bset{f(\sigma\vz)} = \var_{\vz}\bset{\nrm{\vz}_2 df(\ovz)}.
    \]
    Finally, since \(\nrm{\vz}_2 df(\ovz) \stackrel{d}{=} \rho df(\vo)\) we deduce
    \[
        \var_{\vz}\bset{\nrm{\vz}_2 df(\ovz)} = \ev_{\rho,\vo}\bset{(\rho df(\vo) - \ev_\rho\bset{\rho}\ev_{\vo}\bset{df(\vo)}) (\rho df(\vo) - \ev_\rho\bset{\rho}\ev_{\vo}\bset{df(\vo)})^*}
    \]
    from where the theorem follows. 
\end{proof}

Theorem~\ref{thm:noiselessLimit:momentsNoise} follows from Theorem~\ref{apx:thm:noiselessLimit:expectation} and~\ref{apx:thm:noiselessLimit:covariance} applied to \(f(\vy) = \whxO(\mFO\vx + \vy) - \whxO(\mFO\vx)\).

To prove Theorem~\ref{thm:noiselessLimit:momentsSamplingNoise} we need the following theorem.

\begin{theorem}
\label{apx:thm:noiselessLimit:overSamplingSet}
    Let \(\set{f_i}_{j=1}^J\) be a collection of continuous functions satisfying the hypotheses of Lemma~\ref{apx:lem:noiselessLimit:residual} and let \(N\) be a random variable taking values on \(\set{1,\ldots, J}\). Define
    \[
        \bar{f}(\vy) := \ev_N\bset{f_N(\vy)} = \sum_{j=1}^J\prob\set{N = j} f_j(\vy).
    \]
    Then \(\bar{f}\) satisfies the same hypotheses of Lemma~\ref{apx:lem:noiselessLimit:residual} and
    \begin{align*}
        \lim_{\sigma\to 0}\sigma^{-1}\ev_{N,\vz}\bset{f_N(\sigma\vz)} &= \ev_\rho\bset{\rho}\ev_{\vo}\bset{d\bar{f}(\vo)},\\
        \lim_{\sigma\to 0}\sigma^{-2}\var_{N,\vz}\bset{f_N(\sigma\vz)} &= \ev_\rho\bset{\rho^2} \var_{N,\vo}\bset{df_N(\vo)} + \var_{\rho}\bset{\rho} \ev_{\vo}\bset{d\bar{f}(\vo)}\ev_{\vo}\bset{d\bar{f}(\vo)^*},
    \end{align*}
    where \(\rho \sim \chi_{2m-1}\) variable and \(\vo\sim \textsc{Uniform}(\Sph^m)\).
\end{theorem}

\begin{proof}[Proof of Theorem~\ref{apx:thm:noiselessLimit:overSamplingSet}]
    It is clear \(\bar{f}\) satisfies the hypothesis of Lemma~\ref{apx:lem:noiselessLimit:residual} as it involves a finite sum. We only prove the formula for the variance. Write
    \[
        \var_{N,z}\bset{f_N(\sigma\vz)} = \ev_z\bset{\var_N\bset{f_N(\sigma\vz)}} + \var_z\bset{\ev_N\bset{f_N(\sigma\vz)}} = \ev_z\bset{\var_N\bset{f_N(\sigma\vz)}} + \var_z\bset{\bar{f}(\sigma\vz)}.
    \]
    The first term can be expanded as
    \[
        \ev_z\bset{\var_N\bset{f_N(\sigma\vz)}} = \sum_{j=1}^J \prob\set{N=j}\ev_z\bset{(f_j(\sigma\vz) - \bar{f}(\sigma\vz))(f_j(\sigma\vz) - \bar{f}(\sigma\vz))^*}.
    \]
    It suffices to determine the limit for each term in the sum. Since
    \[
        \var_z\bset{f_j(\sigma\vz) - \bar{f}(\sigma\vz)} = \ev_z\bset{(f_j(\sigma\vz) - \bar{f}(\sigma\vz))(f_j(\sigma\vz) - \bar{f}(\sigma\vz))^*} - \ev_z\bset{f_j(\sigma\vz) - \bar{f}(\sigma\vz)}\ev_z\bset{f_j(\sigma\vz) - \bar{f}(\sigma\vz)}^*
    \]
    we conclude from Theorems~\ref{apx:thm:noiselessLimit:expectation} and~\ref{apx:thm:noiselessLimit:covariance} that
    \begin{multline*}
        \lim_{\sigma\to 0} \sigma^{-2}\ev_z\bset{(f_j(\sigma\vz) - \bar{f}(\sigma\vz))(f_j(\sigma\vz) - \bar{f}(\sigma\vz))^*} = \var_{z}\bset{\nrm{\vz}_2 (df_j(\ovz) - d\bar{f}(\ovz))} \\+\: \ev_{\rho}\bset{\rho}^2 \ev_{\vo}\bset{df_j(\vo) - d\bar{f}(\vo)} \ev_{\vo}\bset{df_j(\vo) - d\bar{f}(\vo)}^*.
    \end{multline*}
    The first term in the right-hand side can be written as
    \begin{multline*}
        \var_{z}\bset{\nrm{\vz}_2 (df_j(\ovz) - d\bar{f}(\ovz))} = \var_\rho\bset{\rho} \ev_{\vo}\bset{ (df_j(\vo) - d\bar{f}(\vo)) (df_j(\vo) - d\bar{f}(\vo))^*} \\+\:\ev_\rho\bset{\rho}^2 \var_{\vo}\bset{(df_j(\vo) - d\bar{f}(\vo))}
    \end{multline*}
    whence 
    \[
        \lim_{\sigma\to 0} \sigma^{-2}\ev_z\bset{(f_j(\sigma\vz) - \bar{f}(\sigma\vz))(f_j(\sigma\vz) - \bar{f}(\sigma\vz))^*} = \ev_\rho\bset{\rho^2} \ev_{\vo}\bset{(df_j(\vo) - d\bar{f}(\vo)) (df_j(\vo) - d\bar{f}(\vo))^*}.
    \]
    Therefore, we conclude that
    \begin{align*}
        \lim_{\sigma\to 0} \sigma^{-2}\ev_z\bset{\var_N\bset{f_N(\sigma\vz)}} &= \ev_\rho\bset{\rho^2} \ev_{N,\vo}\bset{(df_N(\vo) - d\bar{f}(\vo)) (df_N(\vo) - d\bar{f}(\vo))^*}\\
        &= \ev_\rho\bset{\rho^2} \ev_{N,\vo}\bset{df_N(\vo)df_N(\vo)^*} - \ev_\rho\bset{\rho^2} \ev_{\vo}\bset{d\bar{f}(\vo)d\bar{f}(\vo)^*}.
    \end{align*}
    By Theorem~\ref{apx:thm:noiselessLimit:covariance} we also have
    \[
        \lim_{\sigma\to 0} \sigma^{-2}\var_{\vz}\bset{\bar{f}(\sigma\vz)} = \var_\rho\bset{\rho} \ev_{\vo}\bset{d\bar{f}(\vo)d\bar{f}(\vo)^*} +\ev_\rho\bset{\rho}^2 \var_{\vo}\bset{d\bar{f}(\vo)}.
    \]
    Hence,
    \[
        \lim_{\sigma\to 0}\sigma^{-2}\var_{N,\vz}\bset{f_N(\sigma\vz)} = \ev_\rho\bset{\rho^2} \var_{N,\vo}\bset{df_N(\vo)} + \var_{\rho}\bset{\rho} \ev_{\vo}\bset{d\bar{f}(\vo)}\ev_{\vo}\bset{d\bar{f}(\vo)^*}. 
    \]
\end{proof}

To deduce Theorem~\ref{thm:noiselessLimit:momentsSamplingNoise} from Theorem~\ref{apx:thm:noiselessLimit:overSamplingSet} we define \(f_{\Omega}(\sigma\vz) := \whx_{\Omega}(\mSO(\mF\vx + \sigma\vz)) - \whx_{\Omega}(\mSO\mF\vx)\). Since \(f_\Omega\) is differentiable by hypothesis, we have
\[
    df_{\Omega}(\vo) = \mSO^* D\whx_{\Omega}(\mSO\mF\vx)\vo
\]
whence \(\ev_{\vo}\bset{df_{\Omega}(\vo)} \equiv 0\). By Theorem~\ref{apx:thm:noiselessLimit:overSamplingSet} we have
\[
    \var_{\Omega,\vz}\bset{df_{\Omega}(\vo)} = \ev_{\Omega,\vz}\bset{df_{\Omega}(\vo)df_{\Omega}(\vo)^*} = \frac{1}{2m}\ev_{\Omega}\bset{\mSO^* D\whx_{\Omega}(\mSO\mF\vx))D\whx_{\Omega}(\mSO\mF\vx))^*\mSO}.
\]
Hence, from Theorem~\ref{thm:noiselessLimit:momentsNoise} we have
\[
    \ev_{\vz}\bset{\whxO(\mSO(\mF\vx + \sigma\vz)} = \whxO(\mSO\mF\vx) + o(\sigma).
\]
As a consequence,
\[
    \lim_{\sigma\to 0}\ev_{\Omega,\vz}\bset{(\whx_{\Omega}(\mSO(\mF\vx + \sigma\vz)) - \ev_{\Omega,\vz}\bset{\whx_{\Omega}(\mSO(\mF\vx + \sigma\vz))})(\whx_{\Omega}(\mSO\mF\vx) - \ev_{\Omega}\bset{\whx_{\Omega}(\mSO\mF\vx)})^*} = \var_\Omega\bset{\whx_{\Omega}(\mSO\mF\vx)}.
\]
Therefore,
\[
    \lim_{\sigma\to 0}\var_{\Omega,\vz}\bset{\whx_{\Omega}(\mSO(\mF\vx + \sigma\vz))} =  \var_{\Omega}\bset{\whx_{\Omega}(\mSO\mF\vx)}.
\]
and the claim follows.

\subsection{The density and velocity in the noiseless limit}
\label{apx:noiselessLimit:4dFlow}

We prove Theorem~\ref{thm:noiselessLimit:4dflow}. We make the following observation. Suppose \(f(\vc{0}) \neq 0\). If \(F:\C^n \to \C^{d}\) is differentiable at the origin, and \(F(f(\vc{0})) = \vc{0}\) then
\[
    d(F\circ f)(\vo) := \lim_{t\to 0} t^{-1} (F\circ f)(t\vo) = DF(f(\vc{0})) df(\vo).
\]
Hence, by Theorem~\ref{apx:thm:noiselessLimit:covariance} we have
\[
    \lim_{\sigma\to 0} \sigma^{-2}\var_\vz\bset{(F\circ f)(\sigma\vz)} = DF(f(\vc{0}))\var_{\vz}\bset{\nrm{\vz}_2 df(\ovz)} DF(f(\vc{0}))^*.
\]
Hence, to compute the \((i,j)\)-the entry of the right-hand side we need to compute the rows of \(DF(f(\vc{0}))\). To use this strategy to prove Theorem~\ref{thm:noiselessLimit:4dflow} we need to ensure the differentiability of \(\Tfinv\). It is apparent \(\Tfinv\) is not complex-differentiable. Hence, we leverage the Wirtinger calculus~\cite[Ch.~1]{Remmert1991}. We let
\begin{align*}
    f(\vz_0,\ldots,\vz_3) &= (\whx_{\Omega_0}(\mFO\vx_0 + \vz_0),\ldots, \whx_{\Omega_3}(\mFO\vx_3 + \vz_3),\ol{\whx}_{\Omega_0}(\mFO\vx_0 + \vz_0),\ldots, \ol{\whx}_{\Omega_3}(\mFO\vx_3 + \vz_3))\\
    F(\vx_0,\ldots,\vx_3,\ol{\vx}_0,\ldots,\ol{\vx}_3) &= \Tfinv(\vx_0,\ldots,\vx_3,\ol{\vx}_0,\ldots,\ol{\vx}_3) - \Tfinv(\whvx_0,\ldots,\whvx_3,\ol{\whvx}_0,\ldots,\ol{\whvx}_3)
\end{align*}
where \(\whvx_k = \whx_{\Omega_k}(\mFO\vx_k)\) and \(\ol{\vx}\) denotes the vector with the conjugate of the entries of \(\vx\); in this case \(\Tfinv\) is given by the expressions
\[
    \wh{\rho}(\vec{r}) = |x_0(\vec{r})| = \sqrt{x_0(\vec{r})x_0(\vec{r})^*}
\]
and
\begin{align*}
    \wh{\theta}_0(\vec{r}) &= \frac{1}{2i} (\log(x_0(\vec{r})) - \log(x_0(\vec{r})^*))\\
    \wh{v}_k(\vec{r}) &= \frac{v_{\text{enc}}}{2\pi i} (\log(x_k(\vec{r})) - (\log(x_k(\vec{r})^*) - \log(x_0(\vec{r})) - \log(x_0(\vec{r})^*))
\end{align*}
which are all Wirtinger-differentiable away from the non-positive real axis. 

\begin{proof}[Proof of Theorem~\ref{thm:noiselessLimit:4dflow}]
    First, note that the only non-zero derivatives of \(\rho(\vec{r})\) are
    \[
        \frac{\partial \rho(\vec{r})}{\partial x_0(\vec{r})} = \frac{1}{2}\sqrt{\frac{x_0(\vec{r})^*}{x_0(\vec{r})}} = \frac{1}{2}e^{-i\whth_0(\vec{r})}\quad\mbox{and}\quad \frac{\partial \rho(\vec{r})}{\partial x_0(\vec{r})^*} = \frac{1}{2}\sqrt{\frac{x_0(\vec{r})}{x_0(\vec{r})^*}} = \frac{1}{2}e^{i\whth_0(\vec{r})}.
    \]
    Hence
    \begin{align*}
        \lim_{\sigma\to 0} \cov(\wh{\rho}(\vecp), \wh{\rho}(\vecq)) &= \frac{1}{4}e^{-i\whth_0(\vecp) + i\whth_0(\vecq)} \cov^0(\whx_0(\vecp),\whx_0(\vecq)) + \frac{1}{4}e^{i\whth_0(\vecp) - i\whth_0(\vecq)} \cov^0(\whx_0(\vecp)^*,\whx_0(\vecq)^*)\\
        &\quad +\:\frac{1}{4}e^{i\whth_0(\vecp) + i\whth_0(\vecq)} \cov^0(\whx_0(\vecp)^*,\whx_0(\vecq)) + \frac{1}{4}e^{-i\whth_0(\vecp) - i\whth_0(\vecq)} \cov^0(\whx_0(\vecp),\whx_0(\vecq)^*).
    \end{align*}
    The velocity components have non-zero derivatives with respect to the components of \(\vx_0\) given by
    \[
        \frac{\partial v_k(\vec{r})}{\partial x_0(\vec{r})} = -\frac{v_{\text{enc}}}{2\pi i}\frac{1}{x_0(\vec{r})}\quad\mbox{and}\quad \frac{\partial v_k(\vec{r})}{\partial x_0(\vec{r})^*} = \frac{v_{\text{enc}}}{2\pi i}\frac{1}{x_0(\vec{r})^*}.
    \]
    This implies 
    \begin{align*}
        \lim_{\sigma\to 0} \cov(\wh{\rho}(\vecp), \wh{v}_k(\vecq)) &= \frac{\venc}{4\pi i }\frac{e^{-i\whth_0(\vecp)}}{\whx_0(\vecq)^*} \cov^0(\whx_0(\vecp),\whx_0(\vecq)) - \frac{\venc}{4\pi i }\frac{e^{i\whth_0(\vecp)}}{\whx_0(\vecq)} \cov^0(\whx_0(\vecp)^*,\whx_0(\vecq)^*) \\
        &\quad+\: \frac{\venc}{4\pi i}\frac{e^{i\whth_0(\vecp)}}{\whx_0(\vecq)^*} \cov^0(\whx_0(\vecp)^*,\whx_0(\vecq)) - \frac{\venc}{4\pi i }\frac{e^{-i\whth_0(\vecp)}}{\whx_0(\vecq)} \cov^0(\whx_0(\vecp),\whx_0(\vecq)^*),
    \end{align*}
    Finally, the velocity component \(v_k\) depends only on the components of \(\vx_k\) whence its non-zero derivatives are
    \[
        \frac{\partial v_k(\vec{r})}{\partial x_k(\vec{r})} = \frac{v_{\text{enc}}}{2\pi i}\frac{1}{x_k(\vec{r})}\quad\mbox{and}\quad \frac{\partial v_k(\vec{r})}{\partial x_k(\vec{r})^*} =- \frac{v_{\text{enc}}}{2\pi i}\frac{1}{x_k(\vec{r})^*}.
    \]
    whence
    \begin{align*}
        \lim_{\sigma\to 0} \cov(\whv_k(\vecp), \whv_\ell(\vecq)) &= \frac{\venc^2}{4\pi^2}\left(\frac{\cov^0(\whx_0(\vecp),\whx_0(\vecq))}{\whx_0(\vecp)\whx_0(\vecq)^*} + \frac{\cov^0(\whx_0(\vecp)^*,\whx_0(\vecq)^*)}{\whx_0(\vecp)^*\whx_0(\vecq)} \right) \\
        &\quad -\:\frac{\venc^2}{4\pi^2}\left(\frac{\cov^0(\whx_0(\vecp)^*,\whx_0(\vecq))}{\whx_0(\vecp)^*\whx_0(\vecq)} + \frac{\cov^0(\whx_0(\vecp),\whx_0(\vecq)^*)}{\whx_0(\vecq)\whx_0(\vecq)^*} \right),
    \end{align*}
    for \(k\neq \ell\) and
    \begin{align*}
        \lim_{\sigma\to 0} \cov(\whv_k(\vecp), \whv_k(\vecq)) &= \frac{\venc^2}{4\pi^2}\left(\frac{\cov^0(\whx_0(\vecp),\whx_0(\vecq))}{\whx_0(\vecp)\whx_0(\vecq)^*} + \frac{\cov^0(\whx_0(\vecp)^*,\whx_0(\vecq)^*)}{\whx_0(\vecp)^*\whx_0(\vecq)} \right) \\
        &\quad -\:\frac{\venc^2}{4\pi^2}\left(\frac{\cov^0(\whx_0(\vecp)^*,\whx_0(\vecq))}{\whx_0(\vecp)^*\whx_0(\vecq)} + \frac{\cov^0(\whx_0(\vecp),\whx_0(\vecq)^*)}{\whx_0(\vecq)\whx_0(\vecq)^*} \right) \\
        &\quad+\:\frac{\venc^2}{4\pi^2}\left(\frac{\cov^0(\whx_k(\vecp),\whx_k(\vecq))}{\whx_k(\vecp)\whx_k(\vecq)^*} + \frac{\cov^0(\whx_k(\vecp)^*,\whx_k(\vecq)^*)}{\whx_k(\vecq)^*\whx_k(\vecq)} \right) \\
        &\quad-\:\frac{\venc^2}{4\pi^2}\left(\frac{\cov^0(\whx_k(\vecp)^*,\whx_k(\vecq))}{\whx_k(\vecp)^*\whx_k(\vecq)} + \frac{\cov^0(\whx_k(\vecp),\whx_k(\vecq)^*)}{\whx_k(\vecq)\whx_k(\vecq)^*} \right).
    \end{align*}
\end{proof}

\subsection{Proof of Proposition~\ref{prop:l2NormRecovery:covariance:samplingNoise}}
\label{apx:l2NormRecovery:proof}

Write first
\[
    \whvx_\Omega - \ev_{\Omega,z}\bset{\whvxO} = \mF^*(\mSO^*\mSO - \vc{D}_\mu)\mF\vx + \sigma\mF^*\mSO^*\mSO\vz.
\]
Then
\[
    \var_{\Omega,z}\bset{\whvx_\Omega(\vc{S}_\Omega(\mF\vx + \sigma\vz)))} = 2\sigma^2 \mF^*\vc{D}_\mu\mF + \mF^*\ev_{\Omega}\bset{(\mSO^*\mSO - \vc{D}_\mu)\mF\vx\vx^*\mF^* (\mSO^*\mSO - \vc{D}_\mu)} \mF.
\]
Let \(\bar{B}(\vec{\xi}) = B(\vec{\xi}) - \mu(\vec{\xi})\) so that \(\mSO^*\mSO - \vc{D}_\mu = \diag(\bar{B}) = :\bar{\vc{B}}\). Then, if we define \(\vc{X} = \mF\vx \vx^*\mF^*\), we deduce
\[
    \ev_\Omega\bset{\bar{B}_k X_{k,\ell} \bar{B}_\ell} = \begin{cases}
        \mu_k(1 - \mu_k)X_{k,k} & k = \ell\\
        0 & k\neq \ell
    \end{cases}
\]
and the matrix is diagonal. It follows that the diagonal components are the magnitude squared of the components of the vector \(\vc{D}_s \mF\vx\) where \(s_k = \sqrt{\mu_k(1 - \mu_k)}\). This proves the proposition.

\end{document}